\documentclass[twoside,11pt]{article}
\usepackage{./mcite}
\usepackage{graphicx}
\usepackage{xspace}
\usepackage{amssymb}
\usepackage{setspace}

\newcommand{\be}{\begin{equation}}
\newcommand{\ee}{\end{equation}}
\newcommand{\bea}{\begin{eqnarray}}
\newcommand{\eea}{\end{eqnarray}}

\newcommand{\comment}[1]{}

\ifnum 1=1
\fi 

\def\and{\&}

\def\Cerenkov{Cherenkov\xspace}

\def\deg{$^\circ$\xspace}
\def\etal{et al.\xspace}
\def\knee{{\sl knee}\xspace}

\def\lnA{\langle\ln A\rangle}
\def\lg{{\rm lg}}

\def\modell{poly-gonato model\xspace}
\def\gcm2{g/cm$^2$\xspace}

\def\Xmax{$X_{\rm max}$\xspace}
\def\xn{$X_R$\xspace}
\def\ga{\raisebox{-0.2em}{\,$\stackrel{\scriptstyle>}{\scriptstyle\sim}$\,}}

\def\lleft{{\sl left}\xspace}
\def\LLeft{{\sl Left panel}\xspace}
\def\rright{{\sl right}\xspace}
\def\RRight{{\sl Right panel}\xspace}
\def\TTop{{\sl Top panel}\xspace}
\def\ttop{{\sl top}\xspace}
\def\BBottom{{\sl Bottom panel}\xspace}
\def\bbottom{{\sl bottom}\xspace}
\def\eref#1{(\ref{#1})\xspace}
\def\fref#1{Fig.\,\ref{#1}\xspace}

\def\Fref#1{Figure~\ref{#1}\xspace}
\def\sref#1{Sec.\,\ref{#1}\xspace}

\def\rref#1{Ref.\,\cite{#1}\xspace}

\def\lg{\log_{10}}

\def\breite{0.80\textwidth}

\def\line{---}
\def\dashed{-\,-\,-}
\def\dotted{$\cdot\cdot\cdot$}
\def\dashdot{-$\cdot$-$\cdot$}

\def\Caption#1{{\centering\begin{minipage}{0.9\textwidth}\caption{#1}\end{minipage}\\}}
\def\Section#1{\section{#1}}

\begin{document}

\title{ \vspace{1cm} Cosmic Rays from the Knee to the Highest Energies
\thanks{Invited review, submitted to Progress in Particle and Nuclear Physics.}
}
\author{Johannes Bl\"umer$^1$, Ralph Engel$^1$, and J\"org R. H\"orandel$^2$\\
\\
$^1$Karlsruhe Institute of Technology (KIT)\footnote{KIT is the cooperation of
    University Karlsruhe and Forschungszentrum Karlsruhe},
    Institut f\"ur Kernphysik, \\
    P.O. Box 3640, 76021 Karlsruhe, Germany\\
$^2$Radboud University Nijmegen, Department of Astrophysics,\\
    P.O. Box 9010, 6500 GL Nijmegen, The Netherlands}

\maketitle
\begin{abstract} 
  This review summarizes recent developments in the understanding of
  high-energy cosmic rays. It focuses on galactic and presumably
  extragalactic particles in the energy range from the knee
  ($10^{15}$~eV) up to the highest energies observed ($>10^{20}$~eV).
  Emphasis is put on observational results, their interpretation, and the
  global picture of cosmic rays that has emerged during the last
  decade.
\end{abstract}

\tableofcontents


\Section{Introduction\label{intsec}}

Cosmic rays are ionized atomic nuclei reaching the Earth from outside the
Solar System. Although already discovered in 1912, their sources and
propagation mechanisms are still subject of intense research. During the last
decade significant progress has been made and a consistent picture of cosmic-ray observations begins to evolve. This review describes recent progress in the
exploration of the sources and propagation of high-energy cosmic rays, focusing
on observational results and the emerging global picture.

Many reviews of cosmic-ray theory and observations are available in
the literature. Most of these reviews concentrate on different aspects
and energy ranges of high-energy cosmic rays which, in turn, are
presented in much more detail than possible here. For example,
covering the knee energy region, experimental data are compiled in
\cite{Wiebel-Sooth99a,*Swordy:2002df,Hoerandel:2002yg} and a comparison
of observations and model predictions can be found in
\cite{Hoerandel:2004gv}.  The data in the range between the knee and
the ankle, and their interpretation are discussed in
\cite{Giller:2008zza,*DeDonato:2008wq}. Focusing on the upper end of
the cosmic-ray spectrum, measurement techniques and observations are
reviewed in
\cite{Sokolsky:1992rz,Nagano:2000ve,Bertou:2000ip,*Engel:2004ui,*Cronin:2004ye,*Bergman:2007kn,Kampert:2008pr}.
More theoretical and phenomenological aspects of the physics of ultra
high-energy cosmic rays are subject of the reviews
\cite{Olinto:2000sa,*Sigl:2002yk,*Anchordoqui:2002hs,*Protheroe:2003vc}
and different source scenarios are discussed in depth in
\cite{Hillas:1985is,Torres:2004hk} (acceleration scenarios) and
\cite{Bhattacharjee:1998qc} (non-acceleration scenarios).  An
exhaustive compilation of experimental results of the full cosmic-ray
energy range can be found in \cite{Grieder:2001ct} and a recent
review, emphasizing measurement and analysis techniques, is given in
\cite{Haungs:2003jv}.

In this article, we discuss high-energy cosmic-ray measurements
covering the energy range from the knee to the highest energies.
By concentrating mainly on observational results of the last decade and their
implications for our overall understanding of cosmic rays, this review
is complementary to the other articles.

The exploration of cosmic rays is mainly driven by new experimental findings.
Hence, we begin this review with a short historical overview, followed by an
introduction to the physics of high-energy cosmic rays (\sref{intsec}).

In the energy region of interest cosmic rays are measured indirectly with large
detector installations below the atmosphere, registering secondary particles
produced in extensive air showers, initiated by high-energy cosmic rays.  In
\sref{detsec}, basic properties of air showers are introduced and major
detection techniques are discussed.

Recent experimental results concerning the flux of cosmic rays, their elemental
composition, and studies of anisotropies in their arrival directions are
presented in sections \ref{energysec} to \ref{anisosec}. The global picture
evolving from these measurements and their impact on the present understanding
of the origin of high-energy cosmic rays is emphasized in \sref{astrosec}.  The
importance of the understanding of high-energy hadronic interactions for the
interpretation of air shower data is underlined in \sref{accelsec}. Concluding
remarks (\sref{concsec}) complete the review.

\subsubsection*{Historical Overview}
Cosmic rays were discovered in the year 1912 by V.F. Hess during several
ascends with hydrogen-filled balloons up to altitudes of 5~km \cite{hess}.  He
measured the ionization rate of air as function of altitude. 
Electrometers served as standard
devices to measure ionizing radiation at this time \cite{wulf}. 
Hess found an increase of ionizing radiation with increasing
height and he concluded that radiation penetrates from outer space into the
atmosphere.  For the discovery of the cosmic radiation, V.F.~Hess has been
awarded the Nobel Price in 1936.  During the subsequent years W. Kolh\"orster
made further ascends with improved electrometers, measuring the altitude
variation of the ionization up to heights of 9~km \cite{kolhoerster}.

In 1929 W.~Bothe and W.~Kolh\"orster measured coincident signals in two
Geiger-M\"uller counters \cite{bothekolhoerster}.  Placing absorber material in
between the two counters they also measured the absorption characteristics of
the radiation.  They concluded that the ``H\"ohenstrahlung'' (or cosmic
radiation) is of corpuscular nature, i.e.\  consists of charged particles.
Similar conclusions were drawn from measurements by J. Clay, who showed that
the intensity of cosmic rays depends on the (magnetic) latitude of the
observer \cite{clay}. This was a clear indication that a large fraction of
cosmic radiation consists of charged particles.

Kolh\"orster continued his work with Geiger-M\"uller tubes operated in
coincidence.  In February 1938 he reported the discovery of coincident signals
between two tubes with distances as far as 75~m
\cite{kolhoersterschauer}.  He concluded that the tubes were hit by secondary
particles or showers generated by cosmic rays in the atmosphere. 
In the late 1930s P.~Auger undertook investigations of cosmic radiation at
the Jungfraujoch, Switzerland at 3500~m.a.s.l.  He used Wilson chambers and
Geiger-M\"uller tubes separated by large distances and operated in coincidence
\cite{augerschauer}.  Similar to Kolh\"orster, Auger concluded that the
registered particles are secondaries generated in the atmosphere, originating
from a single primary cosmic ray.

In the 1940s the origin of the primary radiation could be revealed with
measurements on balloons at high altitudes.  M. Schein showed that the
positively charged primary particles were mostly protons \cite{schein}.  Cloud
chambers and photographic plates were carried into the stratosphere and it was
found that cosmic rays are made up of fully ionized atomic nuclei moving at
speeds closely to that of light \cite{brandtpeters}.  Many nuclei of the
periodic table up to $Z\approx40$ were found and their relative abundances
determined.  Hydrogen and helium occur most frequently, and the distribution in
mass of the heavier nuclei appeared to be similar to that in the solar system.
Elements more massive than iron or nickel were found to be very rare.

Since the mid 1940s large detector arrays were installed to measure extensive
air showers.  For most investigations detectors with a large surface and a
short time resolution were required.  Early detectors comprised Geiger-M\"uller
counters, progress in the development of photomultipliers lead to the
application of scintillation counters and the newly available \Cerenkov
detectors.
It was found that the energy spectrum of cosmic rays follows 
a power law $dN/dE\propto E^{\gamma}$ over a wide range in energy.
In 1958 G.V.~Kulikov and G.B.~Khristiansen measured the integral electron
number spectrum in air showers using an array of hodoscope counters
\cite{kulikov}. They recognized a kink in the spectrum around $6\times10^5$
particles, corresponding to primary energies of several PeV ($10^{15}$\,eV).  This structure
is now known as the ``knee'' in the energy spectrum.  Since that time there is an
ongoing debate about the origin of this structure.

In the 1960s the air shower array of the M.I.T. group at Volcano Ranch,
New Mexico was the largest cosmic-ray detector.  The set-up
comprised 20 stations equipped with scintillation counters, set up on
a triangular grid, covering a total area of 12~km$^2$.  In 1962 the
first event with an energy of about $10^{20}$~eV has been recorded
with this array \cite{Linsley:1963km}. Bigger air shower arrays were
built (SUGAR \cite{Bell74a}, Haverah Park \cite{Edge73a}, 
Yakutsk \cite{Afanasiev93a}, and AGASA \cite{Chiba:1992nf}) and, after some 
initial attempts, the first successful
fluorescence light detector, called Fly's Eye, was set up in Utah
\cite{Baltrusaitis:1985mx}. With these detectors, another feature of
the cosmic-ray flux -- first discussed in \cite{Linsley:1963m1} -- was firmly established in the early 1990s
and is now known as the ``ankle'' \cite{Bird:1993yi,*Lawrence:1991cc,Nagano:1992jz}.

Finally it should be mentioned that, in the early years, particle
physics was done mainly by studying cosmic rays.  In the
1930s investigations of the cosmic radiation lead to the discovery of
new elementary particles such as the positron \cite{dispositron} or
the muon \cite{dismuon}. The pion was discovered exposing nuclear
emulsions to cosmic radiation at mountain altitudes in 1947
\cite{dispion}. 
New unstable hadrons were found in cosmic-ray interactions in
balloon-borne emulsion chambers \cite{Niu:1971xu} in 1971 which were
later, after the discovery of charm particles, identified as D mesons
\cite{Gaisser:1975xk}. Also a number of exotic phenomena were observed
\cite{Lattes:1980wk,*Slavatinsky:1997th}, none of which could be
confirmed in accelerator experiments.

Over time a standard description of cosmic rays evolved, which is briefly
sketched in the following section.

\subsubsection*{Nature and Origin of Cosmic Rays}

\begin{figure}[t] \centering
 \includegraphics[width=\breite]{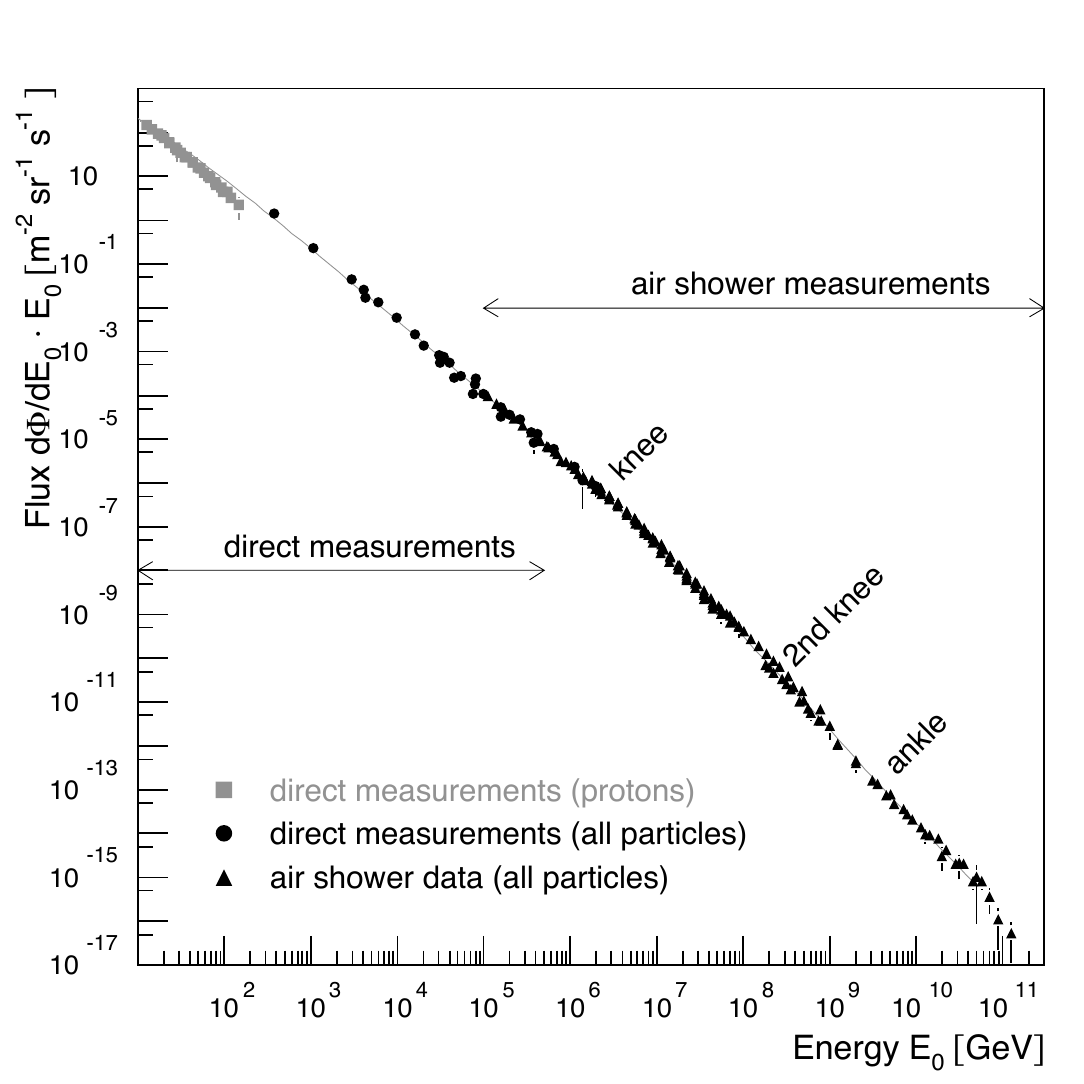}
 \Caption{All-particle energy spectrum of cosmic rays as measured directly
          with detectors above the atmosphere and with air shower detectors.
          At low energies, the flux of primary protons is shown.
          \label{espeksteil}}
\end{figure}

\begin{figure}[t] \centering
 \includegraphics[width=0.95\textwidth]{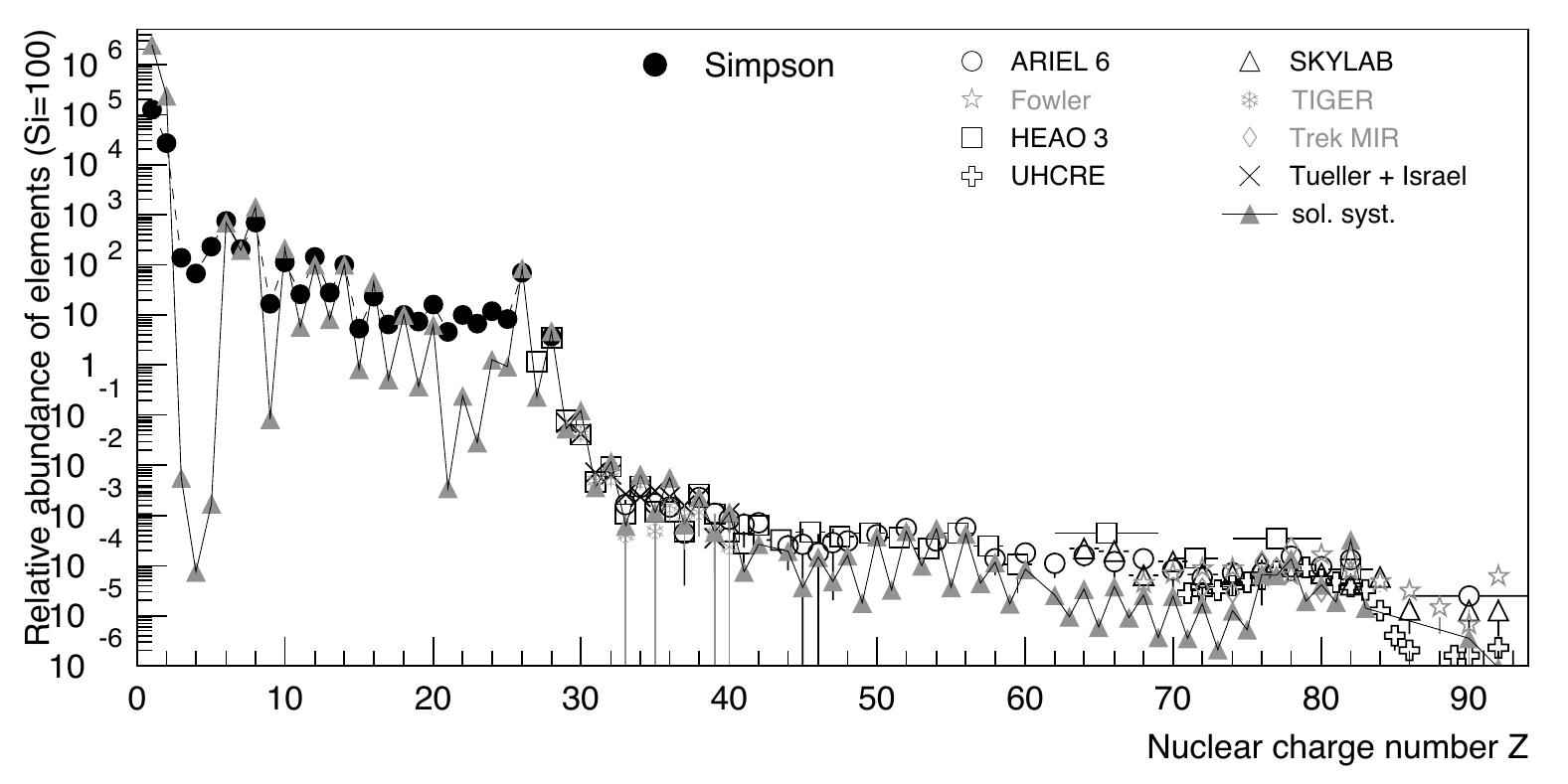}
 \Caption{Abundance of elements in cosmic rays as function of their nuclear
          charge number $Z$ at energies around 1~GeV/n, 
          normalized to $Si=100$ \protect\cite{jrhcospar06}.
          Abundance for nuclei with $Z\le28$ according to \protect\cite{simpson}.
          Heavy nuclei as measured by 
          ARIEL~6 \protect\cite{ariel},
          \protect\cite{fowler},
          HEAO 3 \protect\cite{heao}, 
          SKYLAB \protect\cite{skylab},
          TIGER \protect\cite{tiger},
          TREK/MIR \protect\cite{trek}, 
          \protect\cite{tueller}, as well as
          UHCRE \protect\cite{uhcre}.
          In addition, the abundance of elements in the solar system is
          shown according to \protect\cite{lodders}.}
 \label{abundance}	  
\end{figure}

The solar system is permanently exposed to a flux of highly energetic
ionized atomic nuclei -- the cosmic rays. Their energies extend from the
MeV range to at least $10^{20}$~eV.  The differential energy spectrum of all
cosmic-ray particles is depicted in \fref{espeksteil}.  It falls steeply as
function of energy, decreasing by about a factor 500 per decade in energy.  The
flux decreases from more than 1000 particles per second and square meter at GeV
energies to about one particle per m$^2$ and year at a PeV, and further to less than
one particle per km$^2$ and century above 100~EeV.

The strong decrease in flux poses a big experimental challenge and our knowledge
about the particles and their origin is more and more limited with increasing energy (and
decreasing flux).  At sub-GeV energies individual isotopes are measured with
small detectors in outer space and individual elements can be resolved with
balloon-borne detectors in the TeV regime. At energies exceeding 100~TeV large
detection areas are required to collect a suitable number of particles in a
reasonable time. At present, such detectors are realized at ground level only
and secondary particles generated in the atmosphere (the extensive air
showers) are registered. At PeV energies, groups of elements could be resolved,
while at the highest energies even a classification into ``light'' and ``heavy''
particles becomes already an experimental challenge.

The energy spectrum follows a power law $dN/dE\propto E^\gamma$ over a wide
energy range, indicating non-thermal acceleration processes.  The spectrum
is rather featureless as can be inferred from \fref{espeksteil}.  However,
small structures become clearly visible when the ordinate is multiplied with some power
of the particle energy, as shall be discussed below (see e.g.
\fref{allparticle}).  The spectral index is $\gamma\approx-2.7$ at
energies up to several PeV. Then a steepening is observed, the so-called {\sl
knee}, with $\gamma\approx-3.1$ at higher energies. A further steepening, the
{\sl second knee}, occurs around $4\times 10^{17}$\,eV. 
Finally, at about $4\times 10^{18}$\,eV, at the {\sl
ankle}, the spectrum flattens again.

The abundance of elements in cosmic rays is shown in \fref{abundance} as
function of the nuclear charge number. All elements of the periodic table have
been found in cosmic rays. For the relatively abundant elements up to nickel,
energy spectra for individual elements have been measured
\cite{Wiebel-Sooth99a,Hoerandel:2002yg}.  Abundances as obtained by several
experiments at about 1~GeV/n are depicted.  The cosmic-ray composition is
compared to the abundance of elements in the solar system.  Overall, both
distributions look very alike.  However, there exist certain differences, which
reveal information on the acceleration and propagation of cosmic rays.

The light elements lithium, beryllium, and boron as well as the elements below
iron ($Z=26$) and below lead ($Z=82$) are more abundant in cosmic rays than in
the solar system.  They are assumed to be produced in spallation processes of
the more abundant particles of the CNO, iron, and lead groups during the
journey of cosmic rays through the Galaxy. Hence, they are frequently referred
to as secondary cosmic rays.  As the spallation cross section of the relevant
nuclei is known at GeV energies, the ratio of secondary to primary cosmic rays
is used to infer the propagation path length of cosmic rays in the Galaxy.  An
example is the boron-to-carbon ratio which has been measured as function of
energy \cite{stephens}. The ratio decreases as function of energy which is
frequently explained in Leaky Box models by a rigidity-dependent 
\footnote{Rigidity is defined as particle momentum divided by its charge
$R\,[\mbox{V}]=p/z$.}
decrease of
the path length of cosmic rays in the Galaxy $\Lambda(R) = \Lambda_0
(R/R_0)^{-\delta}$.  Typical values are $\Lambda_0\approx10 - 15$~g/cm$^2$,
$\delta\approx0.5 - 0.6$, and $R_0\approx4$~GV as reference rigidity.

Cosmic-ray particles are assumed to propagate in a diffusive process
through the Galaxy, being deflected many times by the randomly oriented
magnetic fields ($B\sim3$~$\mu$G).  The nuclei are not confined to the galactic
disc, they propagate in the galactic halo as well. 
The scale height of the halo has been estimated with
measurements of the $^{10}$Be/$^9$Be-ratio by the ISOMAX detector
\cite{simon-height} to be a few kpc.  The abundance of radioactive
nuclei in cosmic rays measured with the CRIS instrument yields a residence time in the
Galaxy of about $15\times10^6$~years for particles with GeV energies
\cite{cris-time}.

The energy density of cosmic rays amounts to about
$\rho_{cr}\approx1$~eV/cm$^3$, a value comparable to the energy density of the
visible star light $\rho_{sl}\approx0.3$~eV/cm$^3$, the galactic magnetic
fields $B^2/2\mu_0 \approx0.25$~eV/cm$^3$, or the microwave background
$\rho_{3K}\approx0.25$~eV/cm$^3$.  The power required to sustain a constant
cosmic-ray intensity can be estimated as $L_{cr}=\rho_{cr}
V/\tau_{esc}\approx10^{41}$~erg/s, where $\tau_{esc}$ is the residence time of
cosmic rays in a volume $V$ (the Galaxy and the galactic halo).  With a rate of
about three supernovae per century in a typical Galaxy, the energy required
could be provided by a small fraction ($\approx10\%$) of the kinetic energy
released in supernovae. This had been realized already in 1934 by Baade and
Zwicky \cite{baadezwicky}.  The actual mechanism of acceleration remained
mysterious until Fermi \cite{fermi} proposed a process that involved
interaction of particles with large-scale magnetic fields in the Galaxy.
Eventually, this lead to the currently accepted model of cosmic-ray
acceleration by the first-order Fermi mechanism that operates in strong shock
fronts which are powered by supernova explosions and propagate from a
supernova remnant (SNR) into the interstellar medium
\cite{axford,*krymsky,*bell,*Blandford:1978ky,*Blandford:1987pw}.

Diffusive, first-order shock acceleration works by virtue of the fact that
particles gain an amount of energy $\Delta E\propto E$ at each cycle, where a
cycle consists of a particle passing from the upstream (unshocked) region to
the downstream region and back. At each cycle, there is a probability that the
particle is lost downstream and does not return to the shock. Higher energy
particles are those that have remained longer in the vicinity of the shock and so have had
time to achieve higher energy. After a time $T$ the maximum energy achieved is
$E_{\rm max}\sim Ze\beta_s\cdot B \cdot TV_s$, where $\beta_s=V_s/c$ refers to the
velocity of the shock.  This results in an upper limit, assuming a minimal
diffusion length equal to the Larmor radius of a particle of charge $Ze$ in the
magnetic fields $B$ behind and ahead of the shock.  Using typical values of
Type II supernovae exploding in an average interstellar medium yields $E_{\rm
max}\approx Z \cdot 10^{14}$\,eV \cite{lagagemax}. More recent estimates give a
maximum energy up to one order of magnitude larger for some types of supernovae
\cite{berezhkomax}.  It has also been suggested that the cosmic rays itself
interact with the magnetic fields in the acceleration region, yielding to an
amplification of the fields, which in turn results in much higher energies that
can be reached during the acceleration process \cite{Lucek2000a}. With this
mechanism cosmic rays are supposedly accelerated up to $10^{17}$~eV.

Information on the composition at the source can be obtained from measurements
of the abundance of refractory nuclei. They appear to have undergone minimal
elemental fractionation relative to one another.  Comparing the derived
abundance at the source to the abundance in the solar system reveals that the
two samples exhibit a striking similarity over a wide range
\cite{cris-abundance}.  When uncertainties are taken into account, the
abundances of particular isotopes are consistent with being within 20\% of the
solar values. This indicates that cosmic rays are accelerated out of a sample
of well mixed interstellar matter. Hence, cosmic rays are ``regular'' matter, but
accelerated to extremely high energies.

The understanding of the origin of the knee in the energy spectrum is commonly
thought to be a cornerstone for the understanding of the origin of (galactic)
cosmic rays. Many approaches are discussed in the literature
\cite{Hoerandel:2004gv}.  A popular explanation is that the knee is associated
with the upper limit of acceleration by galactic supernovae, while the ankle is
associated with the onset of an extragalactic population that is less intense
but has a harder spectrum that dominates at sufficiently high energy.  Another
popular explanation is leakage of particles from the Galaxy.  At energies in
the GeV regime measurements indicate a decreasing path length of cosmic rays in
the Galaxy. Extrapolating this to higher energies indicates that above a
certain energy cosmic rays are not contained in the Galaxy (or the galactic
halo) anymore.  In a simple picture, this can be understood since the Larmor
radius of a proton in the galactic magnetic field 
\begin{equation} \label{larmorradeq}
 r_L=1.08~\mbox{pc} \, \frac{E/\mbox{PeV}}{Z\cdot B/\mu\mbox{G}}
\end{equation}
becomes with increasing energy comparable to and finally exceeds the thickness of the galactic disk.

If the knee is caused by the maximum energy attained during the acceleration
process or due to leakage from the Galaxy the energy spectra for individual
elements with charge $Z$ would exhibit a cut-off (or a knee) at an energy
$E_c^Z=Z\cdot E_c^p$, with the cut-off energy $E_c^p$ for protons.  The sum of
the flux of all elements with their individual cut-offs makes up the
all-particle spectrum.  In this picture the knee in the all-particle spectrum
is related to the cut-off for protons and the steeper spectrum above the knee
is a consequence of the subsequent cut-offs for all elements, resulting in a
relatively smooth spectrum above the knee.  \footnote{ Such a scenario has been
pointed out first by Peters \cite{peters} and it has been suggested to call
such a behavior a {\sl Peters cycle} \cite{gaissererice}.}  Since the abundance
of ultra-heavy nuclei (see \fref{abundance}) at GeV energies is very low as
compared to iron the end of the galactic component is often assumed to be at
energies around $30\times E_c^p$.  However, recently it has been proposed that
ultra-heavy elements may play an important role at high energies
\cite{Hoerandel:2002yg} which yields a value of $92\times E_c^p$ for the end of
the galactic component, coinciding with the second knee at $4\times
10^{17}$\,eV.

\begin{figure}[t]
 \includegraphics[width=0.51\textwidth]{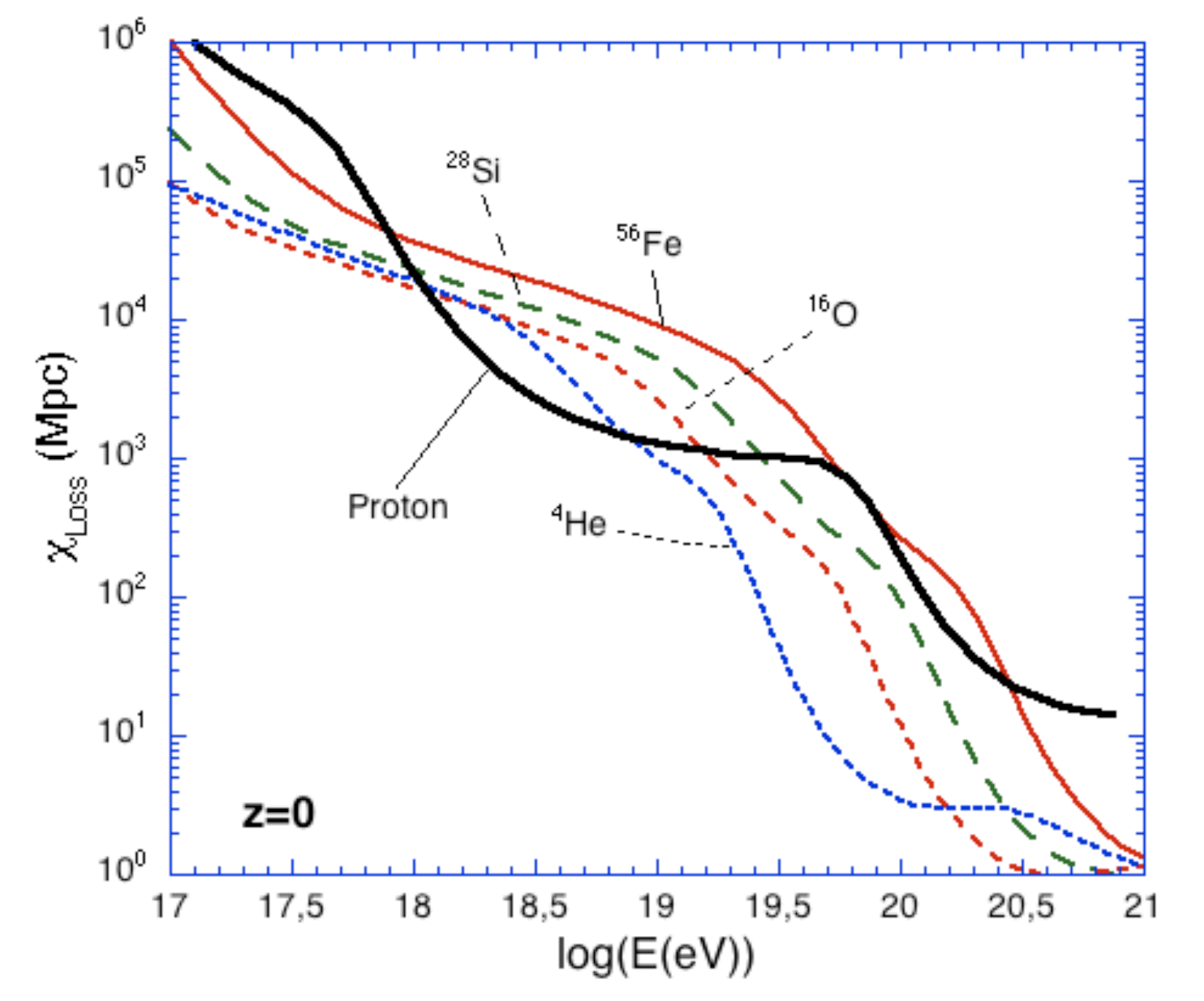}s\hspace*{\fill}
 \includegraphics[width=0.48\textwidth]{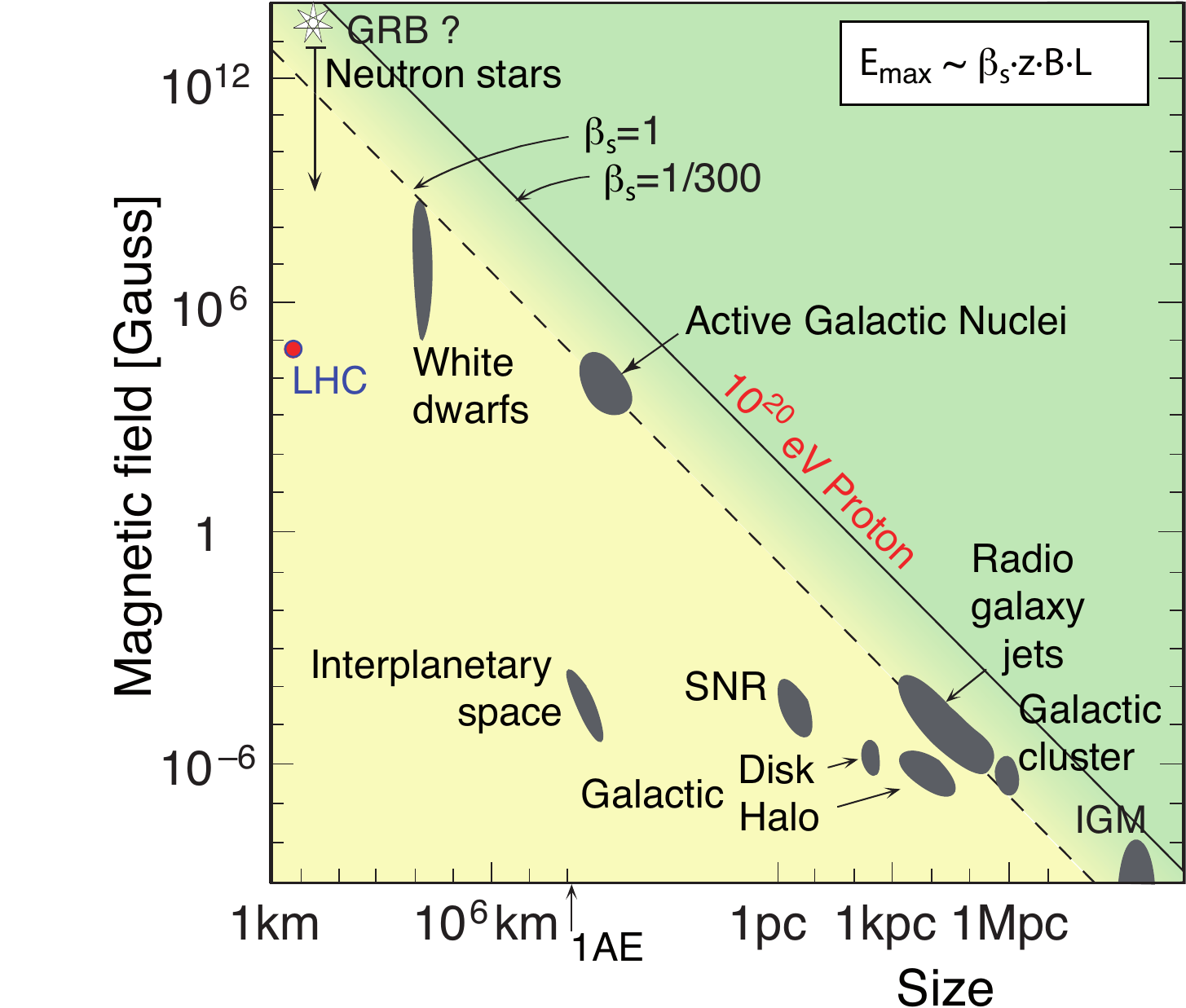}
\Caption{
\LLeft: Energy loss lengths of protons and nuclei
calculated for a redshift $z=0$ \protect\cite{Allard:2006mv}.
\RRight: Hillas plot \protect\cite{Hillas:1985is}  of astrophysical objects in which cosmic rays
could be accelerated (adapted from \protect\cite{Blumer:2000ci}).
\label{fig:GZKPlots}
}
\end{figure}

Another interesting question is that of a natural end of the
spectrum at high energy. Already 40 years ago it has been
realized that interactions of cosmic rays with photons of the
cosmic microwave background (CMB) would result in a cut-off of the spectrum
above $6\times 10^{19}$~eV \cite{Greisen:1966jv,*Zatsepin66}.  All hadronic
particles suffer significant energy losses during propagation above
this energy. Protons interact with background photons forming mainly a
$\Delta^+(1232)$ resonance
\cite{Hill:1983mk,*Berezinsky:1988x1,*Yoshida:1993pt,*Stanev:2000fb} and nuclei are broken
up due to photodisintegration
\cite{Stecker:1998ib,*Yamamoto:2003tn,*Khan:2004nd}.  A compilation of
energy loss lengths of protons and various nuclei is shown in
Fig.~\ref{fig:GZKPlots} (\lleft) \cite{Allard:2006mv}.  The energy
loss length of photons depends on the flux of the universal radio
background (URB) which is not well known (see, for example, discussion
in \cite{Fodor:2002qf}). Depending on the assumptions on the URB, the
energy loss length is significantly smaller than or comparable to that
of hadronic particles.

The short energy loss lengths indicate that cosmic rays with energies above
$10^{20}$\,eV should come from sources within a $\sim 100$\,Mpc sphere.
Astrophysical sources within our Galaxy are disfavored. Even though rapidly
spinning young neutron stars could be thought of accelerating particles to the
highest energies observed \cite{Blasi:2000xm}, it would be difficult to explain
the apparent isotropic arrival direction of UHECRs to energies beyond
$10^{19}$~eV. As cosmic rays of energy greater than $10^{18}$\,eV are no
longer confined by galactic magnetic fields, it is natural to assume that they
are produced by extragalactic sources. 

Considering diffusive shock acceleration, which is 
thought to accelerate cosmic rays in SNRs, the magnetic field strength $B$ in the source and the
size $R$ of the source region are related to the maximum acceleration energy by \cite{Hillas:1985is}
\begin{equation}
E_{\rm max} 
 \simeq  10^{18}\,{\rm eV}~Z~\beta_s~\left(\frac{R}{\rm kpc}\right)
\left(\frac{B}{\rm \mu G}\right),
\end{equation}
where $\beta_s$ is the shock velocity in units of $c$ and $Z$ is the
particle charge. This relation is shown in Fig.~\ref{fig:GZKPlots}
(\rright) for various astrophysical objects. The list of the very few
viable candidate sources includes active galactic nuclei (AGN)
\cite{Protheroe:1992qs,Berezhko:2008xn,Biermann:2008yv}, radio lobes of FR II galaxies
\cite{Rachen:1992pg,Hardcastle:2008jw}, and gamma-ray bursts (GRBs)
\cite{Waxman:1995vg,*Vietri:1995hs,*Murase:2008mr} (for a review of
astrophysical sources, see \cite{Torres:2004hk}).
 
Many alternative, non-acceleration scenarios -- called top-down models -- have
been proposed. In these models, UHECRs are produced in decays of super-heavy
objects such as super-heavy dark matter
\cite{Berezinsky:1997hy,*Aloisio:2006yi}, cryptons \cite{Ellis:2005jc}, or
topological defects \cite{Hill:1982iq}. All of these models postulate new
particle physics and predict typically high gamma-ray fluxes at ultra-high
energy \cite{Protheroe:1995ft,*Berezinsky:1998ft,*Sarkar:2001se}.  Finally there are propagation model scenarios
in which the GZK energy loss processes are evaded or shifted to higher
energies. Examples are violation of Lorentz invariance
\cite{Coleman:1998ti,*Ellis:2000sf,*Jankiewicz:2003sm,*Scully:2008jp,*Bietenholz:2008ni},
the $Z$-burst model \cite{Weiler99a,*Fargion99a,*Ahlers:2005zy} or postulation of new particles with
properties similar to protons \cite{Albuquerque:1998va,Kachelriess:2003yy}.  Reviews of the
different scenarios can be found in \cite{Bhattacharjee:1998qc,Kachelriess:2004ax}.

Measurements of the arrival direction distribution, primary mass
composition and flux will be the key ingredients to solving the puzzle of
UHECR sources. Given an expected angular deflection of only a few
degrees for particles above $10^{19.5}$\,eV in our Galaxy
\cite{Stanev:1996qj,*Tinyakov:2001ir,*Alvarez-Muniz:2001vf} and the
existence of large cosmic voids with negligible magnetic fields
\cite{Kronberg94a,*Sigl:2004yk,*Dolag:2004kp}, high statistics
measurements should finally allow cosmic-ray astronomy and reveal
correlations with sources or source regions.  Similarly, knowing the
composition of UHECRs will restrict the classes of source models. For
example, a mixed composition would exclude top-down
models. Another very important source of complementary information is given by
secondary particle fluxes, i.e.\ gamma-rays and neutrinos, produced in
UHECRs sources and during propagation (see, for example,
\cite{Semikoz:2003wv,Seckel:2005cm,Gelmini:2005wu,Allard:2006mv,Allard:2005cx,Becker:2007sv}).


\Section{Detection Techniques\label{detsec}}

\subsection*{Shower Properties\label{sec:ShowerProperties}}

In the following we will introduce some general properties of extensive air
showers that are employed in cosmic-ray measurements.  Detailed presentations
of this subject can be found in
\cite{Gaisser90a,*Stanev:2004ys,*Lemoine:2001zz}.

\subsubsection*{Electromagnetic Showers}

Showers initiated by photons or electrons (called em.\ showers
henceforth) are governed mainly by the particle production processes
(i) brems\-strahlung of electrons\footnote{In the following we will use
  the term electron to refer to both electrons and positrons.} and (ii)
pair production of electrons by photons. In addition to radiative
energy losses, electrons are subject to ionization energy loss. The
total energy loss $dE/dX$ of electrons can be written as
\begin{equation}
\frac{dE}{dX} = - \alpha(E) - \frac{E}{X_R},
\label{eq:EnergyLoss}
\end{equation}
where $\alpha(E)$ is the ionization energy loss given by the
Bethe-Bloch formula which depends logarithmically on energy. 
The {\it radiation length}, $X_R$, depends on the
material the shower evolves in and is $X_R \approx 37$\,g/cm$^2$ in air.
Particle multiplication and ionization energy loss are competing
processes in showers. The energy at which an electron loses the same
energy due to ionization and particle production is called 
{\it critical energy}, $E_c = X_R \langle \alpha(E_c) \rangle \approx 86$\,MeV.

Some properties of em.\ showers can already be understood within the
very simple model of Heitler \cite{Heitler44}.  In this model it
is assumed that the incoming particle interacts in the atmosphere
after traveling a depth distance $\lambda_{\rm em}$ and produces two
new particles with half the energy of the initial particle. These two
new particles again interact at a distance $\lambda_{\rm em}$ from
their production point. After $n$ generations of successive
interactions the number of particles is $2^n$. The number of particles
as function of depth $X$ can be written as $N(X) = 2^{X/\lambda_{\rm em}}$. The
production of new particles continues until the energy of the
secondary particles is smaller than the critical energy $E_c$, below
which absorption processes dominate over further particle
multiplication. The maximum number of particles in the shower is
$N_{\rm max} = E_0/E_c$ and the depth of shower maximum is given by
$X_{\rm max} = \lambda_{\rm em} \log_2(E_0/E_c)$, with $E_0$ being the
primary particle energy. Heitler's model is certainly an over-simplified
picture of an air shower but illustrates two important features. The
number of particles at shower maximum is approximately proportional to
the primary energy and the depth of shower maximum increases
logarithmically with energy.

Approximate formulae for the longitudinal shower size profile and the lateral 
particle distribution of em.\ showers have been derived from cascade 
theory \cite{Rossi41a,Kamata58,*Nishimura65}. Considering only shower particles of energy $E$, 
the depth of shower maximum is given by
$X_{\rm max} \approx X_R \ln(E_0/E)$. Accounting for the energy distribution of particles in a shower  
this expression becomes (here written for electrons in a photon-induced shower)
\begin{equation}
X_{\rm max} \approx X_R \left[ \ln\left( \frac{E_0}{E_c} \right) + \frac{1}{2}\right] .
\end{equation}
The increase of the depth of shower maximum per decade in energy is
called {\it elongation rate}, $D_{10}$. The elongation rate of em.\
showers is energy-independent, $D_{10}^{\rm em} = \ln(10) X_R \approx
85$\,g/cm$^2$.  For energies larger than $E_c$, the energy spectrum of
secondary particles in a shower follows approximately a power law
$ dN/dE \sim E^{-(1+s)} $ with
where $s$ denotes the shower age parameter. The shower age is often phenomenologically defined as 
$s \approx 3 X/(X + 2 X_{\rm max})$. Based on the solutions of
the one-dimensional cascade equations given in \cite{Rossi41a}, Greisen
developed a compact and still often used parametrization of the mean
longitudinal shower size profile (number of charged particles)
\cite{Greisen65}
\begin{equation}
N_e(X) = \frac{0.31}{\sqrt{\ln E_0/E_c}} \exp\left\{\left(1 - \frac{3}{2}\ln s\right) \frac{X}{X_R} \right\}.
\label{eq:long-prof-em-shower}
\end{equation}
A recent derivation of this expression is given in \cite{Lipari:2008td}. 
In data analysis, a function proposed by Gaisser and Hillas \cite{Gaisser77a} 
is often used to fit measured shower profiles
\begin{equation}
N(X) = N_{\rm max}\left(\frac{X-X_0}{X_{\rm max}-X_0}\right)^{(X_{\rm max} - X)/\Lambda}
\exp\left(\frac{X_{\rm max}-X}{\Lambda}\right)\ .
\label{eq:GaisserHillas}
\end{equation}

The particle density in dependence on the distance to the shower core
(lateral distance) is mainly determined by multiple
Coulomb scattering of electrons. For electrons of low energy
$E$, the increase of variance of the effective scattering angle per traversed
depth $\delta X$ is approximately given by $\langle \delta
\theta^2\rangle \approx (E_s/E)^2/X_R \delta X$, with $E_s \approx
21$\,MeV \cite{Moliere48a}.
Results of detailed calculations of the lateral particle distribution by Nishimura and Kamata
\cite{Kamata58,*Nishimura65} were parametrized by Greisen \cite{nkggreisen} as
\begin{equation}
\frac{dN_e}{r dr d\varphi} = C(s) N_e(X)\left(\frac{r}{r_1}\right)^{s-2} 
\left(1+\frac{r}{r_1}\right)^{s-4.5},
\label{eq:NKG}
\end{equation}
with $C(s) = \Gamma(4.5-s)/[ 2\pi r_1^2 \Gamma(s)\Gamma(4.5-2s)]$
being a normalization constant.  Eq.~(\ref{eq:NKG}) is called
Nishimura-Kamata-Greisen (NKG) function. The lateral distribution of
the shower particles scales with the Moli\`ere unit $r_1 = X_R E_s/E_c \approx
9.3$\,g/cm$^2$ and hence depends on the local air density. The effect
of varying atmospheric density along the shower track can approximately be
taken into account by calculating the Moli\`ere unit not at
observation hight but 2-3 radiation lengths higher up in the
atmosphere \cite{Greisen65}.

At very high energy, two additional processes become important and
change the characteristics of em.\ showers.
First of all, the amplitudes of subsequent interactions of photons or
electrons, which are independent at low energy, have to be added
coherently \cite{Landau:1953um,*Migdal:1956tc} as the interaction
length becomes comparable to the separation of subsequent
interactions. The resulting effect is known as
Landau-Pomeranchuk-Migdal (LPM) effect and leads to the suppression of
new particle production in certain kinematic regions
\cite{Stanev:1982au,*Klein:1998du}. In air, the LPM effect typically becomes
important at energies above $10^{18}$\,eV: Shower-to-shower
fluctuations of em.\ showers increase drastically and the depth of
maximum is shifted deeper into the atmosphere.

Another important ultra high-energy process is that of magnetic pair
production and brems\-strahlung
\cite{Erber:1966vv,*Stanev:1996ux}. Gamma-rays of energies exceeding
$10^{19.5}$\,eV can interact with the geomagnetic field of the Earth.
Such interactions typically take place a thousand kilometers above the
atmosphere.  Mainly due to magnetic brems\-strahlung, a shower of more
than 100 secondary photons and a few electrons is formed, which
interact in the atmosphere simultaneously. Simulations of this
effect can be found in
\cite{Bertou:2000x1,Vankov:2002cb,Homola:2003ru,Homola:2006wf,*Risse:2007sd}. As the primary
energy is shared by many secondary particles, the LPM effect hardly
influences such showers. Due to the superposition of many lower-energy
em.\ showers, shower-to-shower fluctuations of converted primary
photons are significantly reduced. The dependence of the geomagnetic
pre-shower effect on the local arrival direction can be used to search
for photons in a model-independent way, see \cite{Bertou:2000x1,Homola:2006wf,*Risse:2007sd}.

\subsubsection*{Hadron-initiated Showers}

Most of the differences between photon and hadron-initiated showers
are related to the fact that hadronic showers develop a significant
muon component whereas there are very few muons in purely em.\ showers. 
Furthermore, hadronic multiparticle production is characterized by 
large event-to-event fluctuations. 

Again a model similar to that from Heitler is useful to understand the
basic properties of hadronic showers (see, for example,
\cite{Matthews:2005sd,Risse:2004ac,Pierog:2006qu,jrherice06}). For simplicity
we assume that a hadronic interaction of a particle with energy $E$ produces
$n_{\rm tot}$ new particles with energy $E/n_{\rm tot}$, two thirds of which
being charged particles $n_{\rm ch}$ (charged pions) and one third being
neutral particles $n_{\rm neut}$ (neutral pions).  Neutral particles decay
immediately into em.\ particles ($\pi^0 \rightarrow 2 \gamma$).  Charged
particles re-interact with air nuclei if their energy is greater than some
typical decay energy $E_{\rm dec}$ or decay otherwise.  The number of
generations of hadronic interactions, $n$, follows from $E_{\rm
dec}=E_0/(n_{\rm tot})^n$.

Supposing that one muon is produced in the decay of each charged
particle, one gets
\begin{equation}
N_{\mu}=(n_{\rm ch})^n=\left( \frac{E_0}{E_{\rm dec}} \right)^\alpha,
\label{eq:HeitlerNmu}
\end{equation} 
with $\alpha=\ln{n_{\rm ch}}/\ln{n_{\rm tot}}\approx 0.86 \dots 0.93$. The numerical
values for $\alpha$ and $E_{\rm dec}$ depend on the muon energy threshold and are given in
\cite{Alvarez-Muniz:2002ne} for different hadronic interaction models.
The number of muons produced in an air
shower increases almost linearly with primary energy and depends on
the air density (through $E_{\rm dec}$) and the charged and total
particle multiplicities of hadronic interactions.

Also the energy transferred to the em.\ shower component can be
estimated within this simple model.  In each hadronic interaction,
a certain fraction of the initial energy is transferred to the 
em.\ shower component. After
$n$ generations the energy in the hadronic and em.\  components is
given by
\begin{equation} 
E_{\rm had}  =  \left(\frac{2}{3}\right)^n  E_0, \hspace*{2cm}
E_{\rm em}  =  E_0 - E_{\rm had}.
\label{eq:HeitlerMissingEnergy} 
\end{equation}
Simulations show that the number of generations of charged pions is typically about 5
to 6 \cite{Meurer:2005dt} and increases slightly with primary shower energy.
Correspondingly the fraction of energy transferred to the em.\ component increases from 
about $70-80$\% at $10^{15}$\,eV to $90-95$\% at $10^{20}$\,eV.

The depth of shower maximum of a hadron-induced shower is given by that
of the em.\ shower component, $X^e_{\rm max}$. The first hadronic
interaction produces em.\ particles of energy $\sim E_0/n_{\rm tot}$.
Therefore one can write in lowest order approximation
\begin{eqnarray}
X_{\rm max}(E_0) &\sim & \lambda_{\rm had}+X^e_{\rm max}(E_0/n_{\rm tot}) \nonumber\\
               &\sim & \lambda_{\rm had}+X_R\cdot\ln\left( \frac{E_0}{n_{\rm tot}E_c}\right),  
\label{eqxmax}
\end{eqnarray}
where $\lambda_{\rm had}$ is the hadronic interaction length. From
Eq.~(\ref{eqxmax}) follows the {\it elongation rate theorem}, stating
that hadronic showers have always an elongation rate equal to or smaller
than that of em.\ showers \cite{Linsley:1981gh}
\begin{eqnarray}
D_{10}^h & \approx & (1-B_\lambda -B_n) \ln(10) X_R \ln (E_0/E_c) \nonumber\\
         & \approx &  (1-B_\lambda -B_n) D_{10}^{\rm em}.
\label{eq:ERTheorem}
\end{eqnarray}
The coefficients are $B_\lambda = -d\lambda_{\rm had}/d\ln E$ and $B_n
= d\ln(n_{\rm tot})/d\ln E$ (see also discussion in
\cite{Alvarez-Muniz:2002ne} for the next higher order term).

If the primary particle is a nucleus, one can use the {\it
  superposition model} to deduce the main shower characteristics \cite{Engel:1992vf}. In
this model, a nucleus with mass $A$ and energy $E_0$ is considered as
$A$ independent nucleons with energy $E_{\rm h}=E_0/A$. The
superposition of the individual nucleon showers yields
\begin{eqnarray}
N^A_{\rm max} & \approx & A\cdot\frac{E_{\rm h}}{E_c}=\frac{E_0}{E_c}=N_{\rm max} \label{nmaxa}\\
X^A_{\rm max} & \approx & X_{\rm max}(E_0/A) \label{xmaxeq}\\
N^A_{\mu} & \approx & A\cdot \left( \frac{E_0/A}{E_{\rm dec}}
                            \right)^\alpha=A^{1-\alpha}\cdot N_{\mu}\label{nmua}.
\end{eqnarray}
To a good approximation, while the number of charged 
particles at shower maximum is independent of the primary hadron, both, the number of muons and the depth of maximum do depend on the mass of the primary particle. The heavier the shower-initiating particle the more muons are expected for a given primary energy. In addition, the superposition of $A$ independent showers naturally explains why the shower-to-shower fluctuations are smaller for shower initiated by nuclei as compared to proton showers.

Detailed shower simulations confirm the basic shower properties
discussed so far. Quantitative predictions depend on the details of
modeling particle production and transport \cite{Knapp96a,Knapp:2002vs}.  Whereas em.\
interactions are rather well understood within perturbative QED,
hadronic multiparticle production cannot be calculated within QCD from
first principles. Phenomenological models have to be used to describe
the final states of hadronic interactions.  The parameters of these
models are determined by comparing the model predictions with
accelerator measurements. Reflecting the different methods for
describing data, low- and high-energy interaction models are
distinguished. The former ones are typically based on the picture of 
intermediate resonance formation and decay as well as parametrizations of data. The
latter ones are involving the production of color strings and their
fragmentation.

For high energy interactions 
($E_{\rm lab} \gtrsim 100$~GeV), the hadronic interaction models often used in air shower simulations are
DPMJET II.55 and III \cite{Ranft:1994fd,Roesler01a}, 
EPOS \cite{Werner:2005jf,Pierog:2006qv},
QGSJET~01~\cite{Kalmykov89a-e,*Kalmykov:1993qe,*Kalmykov:1997te} and 
QGSJET~II~\cite{Ostapchenko:2005nj,*Ostapchenko:2006vr}, 
as well as SIBYLL 2.1 \cite{Engel:1992vf,Fletcher:1994bd,Engel:1999db}. 
These models reproduce accelerator data reasonably well but
are characterized by different extrapolations above 
a center-of-mass energy $E_{\rm cms}\sim 1.8$ TeV
($E \sim 10^{15}$\,eV), leading to very different shower predictions at 
high energy \cite{Knapp96a,Knapp:2002vs,Engel:2004is,Heck08CORSIKASchool}.

The situation is different at low
energy where more measurements from fixed target experiments are
available. There, one of the main problems is the
extrapolation of measurements to the very forward phase space region
close to the beam direction and the lack of measurements of
pion-induced interactions with light nuclei \cite{Meurer:2005dt}.  At low energy, models based on data
parameterizations and/or microscopic models such as FLUKA \cite{Ballarini:2005tg},
GHEISHA \cite{Fesefeldt85a}, UrQMD \cite{Bleicher99a}, or TARGET \cite{Engel01c} are
used. Differences of the predictions of these models are important
for the number of muons \cite{Hillas:1997tf,*Engel:1999vy,*Drescher:2002vp,*Heck:2003br,*Drescher:2003gh}, 
in particular at large lateral distances.

\begin{figure}[t]
 \includegraphics[width=0.52\textwidth]{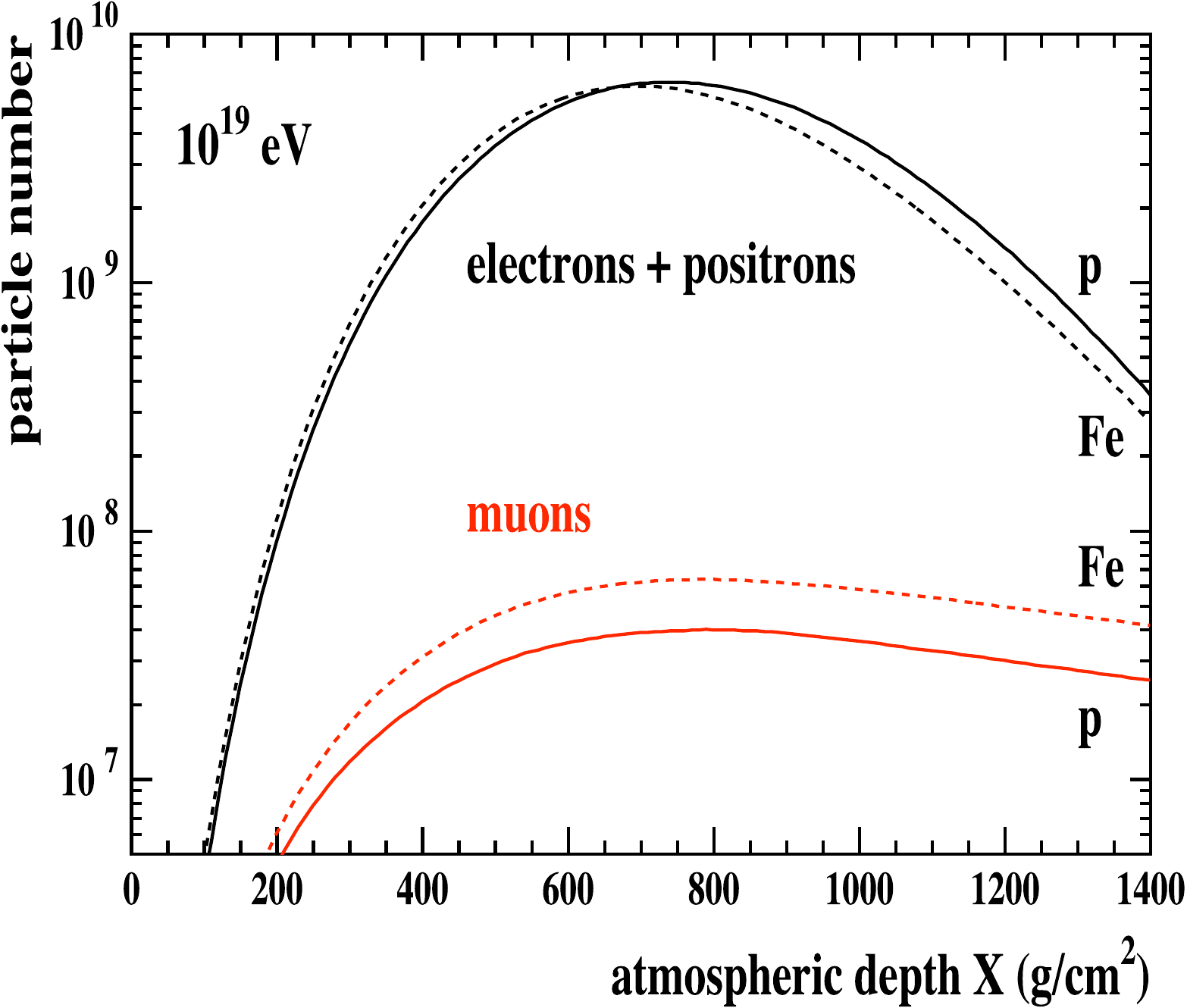}\hspace*{\fill}
 \includegraphics[width=0.47\textwidth]{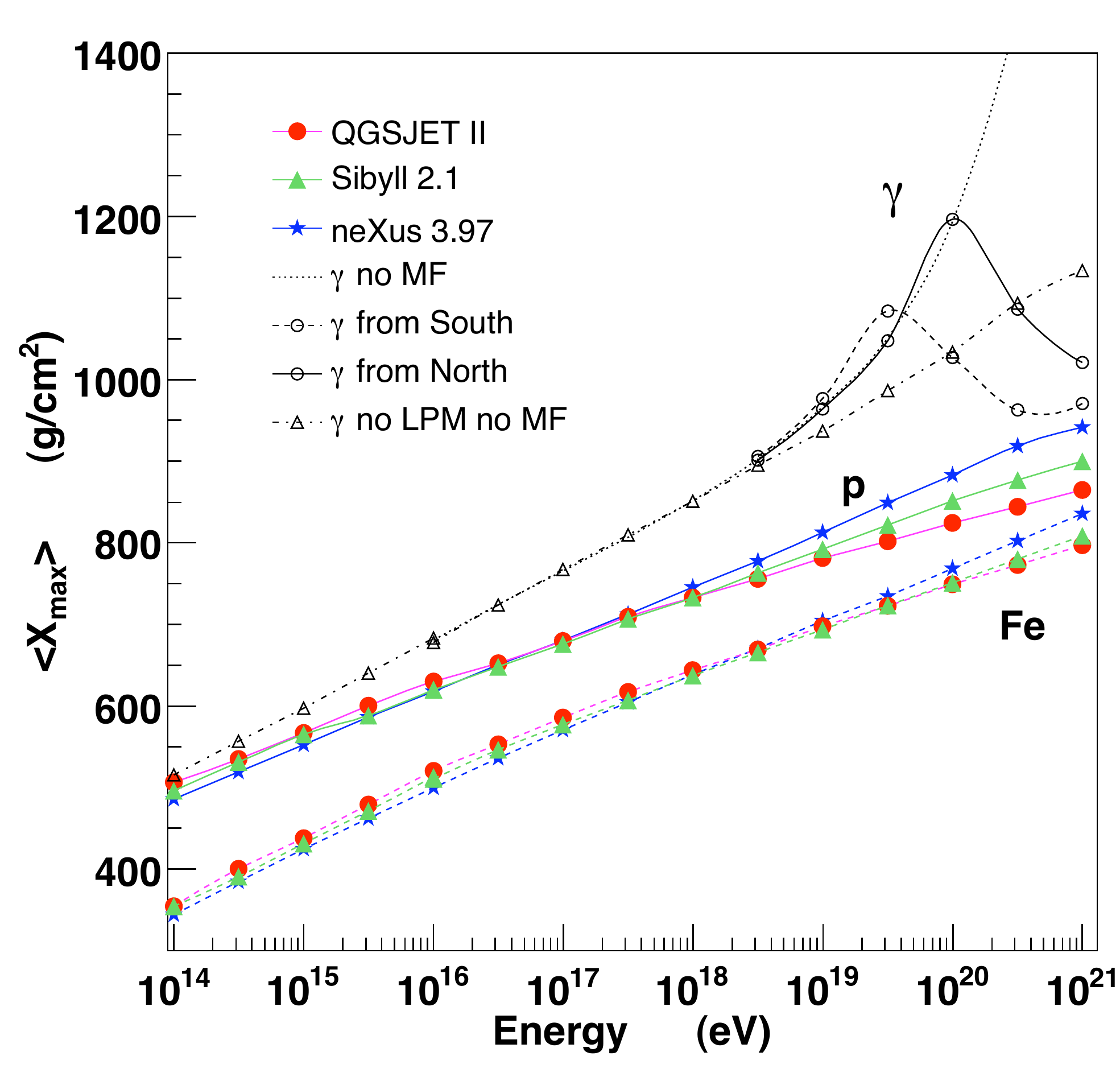}
\Caption{\LLeft: Longitudinal shower profile. Shown are the
  number of electrons with $E_{\rm kin}> 250$\,keV and muons with $E_{\rm
    kin}> 250$\,MeV (from \protect\cite{Risse:2004ac}).  
 \RRight: Mean depth of shower maximum calculated with
   different hadronic interaction models and CONEX (from \protect\cite{Pierog:2005aa}, modified). 
\label{fig:longi}
}
\end{figure}

\begin{figure}[t]\centering
 \includegraphics[width=\breite]{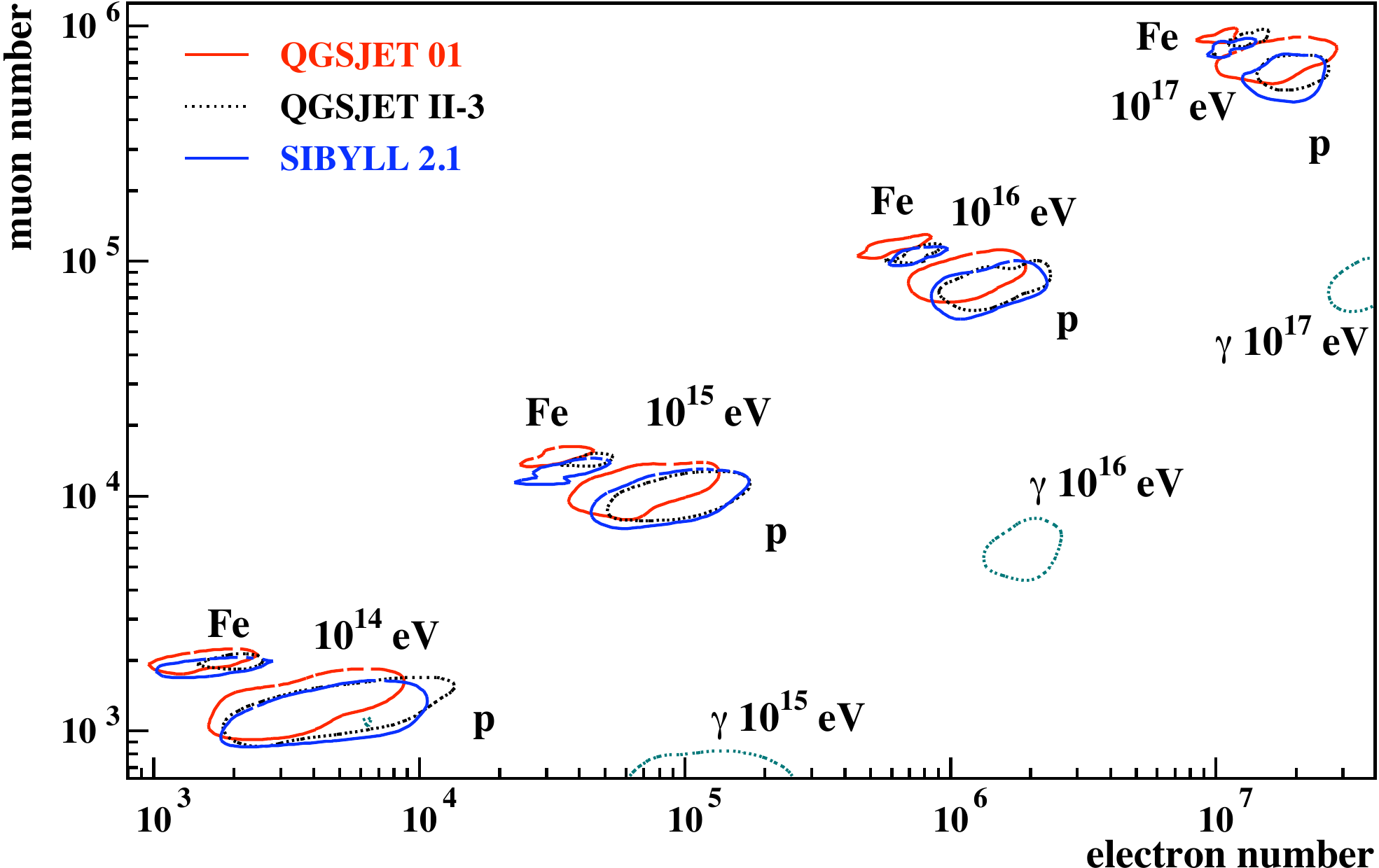}
\Caption{ Expected number of muons and electrons in vertical showers
  at sea level.  The curves show the FWHM of the distributions for
  different primary particles and energies, as obtained with QGSJET
  01, QGSJET-II, and SIBYLL 2.1 
  \protect\cite{Heck08CORSIKASchool,Heck2006private}.  \label{fig:NeNmuModels}
}
\end{figure}

In Figs.~\ref{fig:longi}-\ref{fig:NeNmuModels}, we illustrate some
shower properties discussed only qualitatively so far by showing the
results of detailed Monte Carlo simulations done with CORSIKA
\cite{Heck98a} and CONEX \cite{Bergmann:2006yz}.

The longitudinal profile of a typical proton shower of $10^{19}$\,eV
is shown in Fig.~\ref{fig:longi} (\lleft). In the region of the
shower maximum, less than 1\% of the charged shower particles are muons.
The electromagnetic component is absorbed much faster than
the muonic one. 

As muons are mainly produced in hadronic interactions, their number and lateral
distribution can be used as composition-sensitive observables. 
The predicted muon distributions depend on the assumptions on hadron
production in air showers. This is demonstrated in Fig.~\ref{fig:NeNmuModels}
where the expected number of muons and electrons is shown for showers initiated
by proton, iron, and gamma-ray primaries, as calculated with different
interaction models \cite{Heck08CORSIKASchool}. The energy dependence of the
total muon number and ratio between the muon numbers of proton and iron showers is in
agreement with the expectations from the superposition model.  At high energy,
the discrimination power of electron-muon number measurements is subject to
large systematic uncertainties due to our limited understanding of hadronic
multiparticle production.

In Fig.~\ref{fig:longi} (\rright), we show the mean depth of
shower maximum, $\langle X_{\rm max}\rangle$, for different primary particles. The difference between
proton and iron showers is in agreement with the predictions of the
superposition model.  Whereas the expected mean depth of maximum
depends again sensitively on the chosen hadronic interaction model,
the fluctuations are rather model-independent at a given energy.
The elongation rate of em.\ showers obtained
with detailed model simulations coincides with that predicted in em.\
cascade theory \cite{Rossi41a}. The increase of the em.\ elongation rate at high
energy is caused by the LPM effect. Also, at ultra high-energy, photon
interaction with the geomagnetic field can lead to a negative elongation
rate.  The elongation rates found for hadronic showers within different
models ($D_{10}^{\rm had} \approx 50 \dots 60 $\,g/cm$^2$ at $10^{19}$\,eV) can be
qualitatively explained with the elongation rate theorem
\cite{Linsley:1981gh,Alvarez-Muniz:2002ne}. An increase of the observed $\langle
X_{\rm max} \rangle$ faster than or comparable to the elongation rate
of em.\ showers is a model-independent signature for a transition to a
lighter composition. However, over a limited range in energy, drastic changes
in the characteristics of hadronic multiparticle production can also lead
to an elongation rate comparable to that of em.\ showers
\cite{Pajares:2000sn,*DiasdeDeus:2005sq,*AlvarezMuniz:2006ye}.

Finally, it should be mentioned that the em.\ shower component
exhibits a number of universality features that are independent of the
primary hadron type and also rather insensitive to the primary
energy. For example, near the shower core, the electron energy
distribution is a universal function of shower age and the angular
distribution of electrons depends almost exclusively on the electron
energy and only slightly on shower age
\cite{Hillas:1982vn,Giller:2005qz,Nerling:2006yt}.  Furthermore, the
longitudinal shower profile (particle flux) at a given lateral
distance can be well parametrized by an universal function of the
depth of shower maximum and shower size at maximum \cite{Chou:2005yq,*Schmidt:2007vq}. 
This reflects the universality of the lateral distribution of 
electromagnetic particles if the lateral distance is measured in 
Moliere units \cite{Giller:2005gp,*Lafebre:2009en}.

\subsection*{Measurement of Charged Secondary Particles}

\subsubsection*{Detection Techniques}

The classical set-up to measure air showers is an array of scintillation
detectors, registering charged particles from the shower.  In each detector the
density of charged particles (mostly electrons, positrons, 
but also some converted photons) is measured. From this information the lateral
distribution of the electromagnetic component is inferred.  This yields
information on the position of the shower core and the total number of
particles in the shower.  Measurements of the arrival times of the particles
with a resolution of a few ns allow to reconstruct the orientation of the
shower plane and thus, perpendicular to it, the arrival direction of the
primary cosmic ray.  Due to the large number of secondary particles 
it is usually sufficient to cover only a small fraction of the
total area with detectors.  Typical values range from 1.2\,\% for the KASCADE
array to $5\times10^{-6}$ for the Auger array.

Examples for arrays in the knee region are the EAS-TOP experiment
\cite{eastop}, where 37 stations of scintillation detectors were distributed
over an area of $\sim 10^5$~m$^2$,  located above the Gran Sasso underground
laboratory at an altitude of 2005~m, or the KASCADE experiment
\cite{kascadenim}, where 252 detector stations are placed on a rectangular 13~m
grid, covering an area of $200\times200$~m$^2$. Detectors that operate at
higher energy are the scintillator array of the KASCADE-Grande experiment
\cite{grande}, or the AGASA experiment, which covered an area of about
100~km$^2$ with 111 scintillation counter stations \cite{Chiba:1992nf}.

Another technique utilized to measure charged particles are water \Cerenkov
detectors, like the surface array of the Pierre Auger Observatory with its 1600
detectors distributed over an area of 3000~km$^2$ \cite{Abraham:2004dt}. They
are relatively deep (typically $\sim1$~m) compared to scintillators with
typical thicknesses of the order of several cm. Consequently they have a larger
response to showers with large zenith angles, e.g. the FWHM of the declination
distributions were $\sim40$\deg and $\sim75$\deg for the Volcano Ranch
(scintillators) \cite{Linsley:1961kt} and Haverah Park (water \Cerenkov) \cite{Edge73a} experiments respectively.
This results in a much larger sky coverage of a water \Cerenkov experiment
compared to a scintillator array. They are also suitable for the
measurement of horizontal showers that can be used to detect neutrinos at
the highest energies.

Various techniques are applied for the detection of muons in air showers.
Frequently, particle counters are covered with absorbers of lead, iron, or soil
with a thickness of a few tens \xn to suppress the electromagnetic component,
as in the KASCADE \cite{kascadenim} or AGASA \cite{Chiba:1992nf} experiments.
Muons can also be identified via their trajectories in tracking devices, like
the HEGRA CRT detectors \cite{hegracrt} or the muon tracking detector of
KASCADE \cite{mtdnim}.  In underground laboratories, well shielded by rock,
soil, water, or ice absorbers with a thickness corresponding to several
1000~m\,w.e. (meter water equivalent), muons with thresholds in the TeV regime
are registered \cite{macro1,*macro2}.

The energy of hadrons is usually determined with calorimeters.  The principle
idea is to absorb an incoming particle and to measure the dissipated energy.
Examples are the hadron calorimeters of the EAS-TOP \cite{eastopkal} or KASCADE
experiments \cite{kalonim}.

\subsubsection*{Measured Parameters}

The direction of the shower axis and hence of the primary cosmic ray is
obtained by measurements of the arrival time of the shower front in the
detector plane. The shower direction is the normal to the reconstructed,
slightly curved shower front. Typical detectors have a time resolution in the
range from 0.5~ns up to a few ns.  The angular resolution (68\% value) of the
KASCADE detector field decreases from 0.55\deg for small showers with
$N_e\approx10^4$  to  0.1\deg for showers with $N_e\approx10^6$
\cite{kascadenim}.  The angular resolution of the Auger surface detectors
decreases from values around 2.2\deg for showers with energies below 4~EeV to
values from 0.5\deg to 1\deg, depending on the zenith angle, for higher
energies ($>10$~EeV) \cite{Bonifazi:2009ma}.

To obtain the position of the shower core and the number of particles, the
particle densities are measured and an appropriate lateral distribution
function is fitted to the data. The number of particles is calculated by
integration of the measured lateral distribution.
Historically, the choices of parameterizations of the electron and muon 
lateral distributions were influenced by a review by Greisen \cite{nkggreisen}.
The Nishimura-Kamata-Greisen (NKG) function became the standard function 
to describe the particle density $\rho$ for a shower
with the number of electrons $N_e$ and shower age $s$ as function of distance
$r$ to the shower, see Eq.~\eref{eq:NKG}
\cite{Kamata58,nkggreisen}.
Traditionally, a fixed Moli\`ere unit $r_1=79$~m is used and the parameter
$s$ is limited to $0.5<s<1.5$.  To parameterize the lateral distribution of
muons, Greisen suggested the function \cite{nkggreisen}
\begin{equation}
 \rho_\mu(r,N_\mu)= C N_\mu \left(\frac{r}{r_G}\right)^{-\beta}
   \left(1+\frac{r}{r_G}\right)^{-2.5}
\end{equation}
to describe the density of muons $\rho_\mu$ as function of distance $r$ to the
shower axis for a shower with a total muon number $N_\mu$. The Greisen radius
is $r_G=320$~m.

The KASCADE group found that the lateral distributions of all three major
shower components (electromagnetic, muonic, and hadronic) can be parameterized
using the NKG function \cite{kascadelateral}.  Fitting simultaneously the
parameters $N_e$, $r_1$, and $s$, the measured lateral distributions of the
electromagnetic component can be reproduced with an accuracy of about 1\%,
yielding $r_1^e\approx20-30$~m. The experimental muon distributions are
described with an accuracy of 5\% using $r_1^\mu=420$~m.  Finally, for the
lateral distribution of hadrons with energies above 50~GeV a value
$r_1^h\approx10$~m has been found.

The NKG function has been analytically developed to describe pure
electromagnetic showers. For hadron-induced air showers it exhibits
shortcomings describing the measured electron lateral distributions, most
obvious at large core distances.  Investigations of the KASCADE group
showed that the measured electron lateral distributions for showers with
energies up to $10^{17}$\,eV and core distances up to 200~m can be described better
using a modified NKG function \cite{kascademodnkg}
\begin{equation}
 \rho_{NKG}^{mod}=N_e\cdot c(s)\cdot \left(\frac{r}{r_0}\right)^{s-\alpha}
    \left(1+\frac{r}{r_0}\right)^{s-\beta} \mbox{~with~}
    c(s)=\frac{\Gamma(\beta-s)}{2\pi r_0^2 \Gamma(s-\alpha+2)
    \Gamma(\alpha+\beta-2s-2)} .
\end{equation}
Optimizing the parameters with Monte Carlo data, the values $\alpha=1.5$ and
$\beta=3.6$ have been obtained, when $r_0=40$~m is used for the scale
parameter.

For showers in the EeV energy range, the AGASA group uses the
parameterization
\begin{equation}
 \rho(r) = C \left(\frac{r}{r_M}\right)^{-1.2}
 \left(1+\frac{r}{r_M}\right)^{-(\eta-1.2)}
 \left(1.0+\left(\frac{r}{1000~\mbox{m}}\right)^2\right)^{-\delta}
\end{equation}
to describe the lateral distribution of charged particles up to distances of
several km from the shower axis \cite{agasalat}, inspired by a function
suggested by Linsley \cite{linsleylat}.  The parameters are $\eta=3.8$,
$\delta=0.6\pm0.1$ and a Moli\`ere unit $r_M=91.6$~m for near vertical
showers with $\sec\theta<1.2$.

Another parameterization for the electron lateral distribution has been
suggested by the Haverah Park Collaboration \cite{haverahlat}. It is also applied
in the Auger Observatory \cite{Barnhill:2005pr}. The signal $S$ in a water
\Cerenkov detector is parameterized as function of distance $r$ to the shower
core as $S(r)=kr^{-(\eta+r/r_s)}$ for $r<800$~m and $S(r)=(1/800)^\delta k
r^{-(\eta+r/r_s)+\delta}$ at larger distances.  The shape parameter $\eta$
varies with zenith angle, while the parameter $\delta$ and the scale radius
$r_s=4000$~m are fixed.

The position of the shower core is determined by the KASCADE group with an
uncertainty of 5~m for showers with an electron number $N_e\approx10^4$
improving to less than 1~m for large showers ($N_e\approx10^{6.5}$)
\cite{kascadenim}. The error in the reconstructed number of electrons decreases
in the same electron number interval from 18\% to less than 4\%.
Similar values were obtained in the EAS-TOP experiment \cite{eastop}. In the
electron number range from $N_e=10^{4.8}$ to $10^{6.8}$, the error in the
position of the shower core improves from 7.5 to 2~m and the uncertainty in the
number of electrons decreases from $\Delta N_e/N_e=28\%$ to about 10\%.

\subsubsection*{Energy Estimators}
One of the most important parameters to characterize a shower is the energy of
the primary particle. Various methods are discussed in the literature to obtain
this value.

For KASCADE, it has been found that, at sea level, the number of muons with
energy above 230~MeV in the range from 40~m to 200~m from the shower axis is
a good measure for the primary energy independent of the mass of the primary particle
\cite{wwtestjpg}. With its relatively high detector density the lateral
distribution of muons is measured very well in the radius range from 40 to
200~m. Extrapolating beyond these limits would introduce uncertainties related
to the (less well known) shape of the lateral distribution. Hence, the number
of muons is reconstructed using a distance range only, in which detectors are
present.

Another method is to use the correlation between the number of electrons and
muons reconstructed. For example the CASA-MIA group uses the relation $E_0 =
0.8~\mbox{GeV} (N_e + 25N_\mu)$ to estimate the primary energy \cite{casae}.
Similarly, the KASCADE-Grande experiment obtained the relation
\begin{equation}
 \lg\left(\frac{E_0}{\rm GeV}\right)=0.313\,\lg N_e + 0.666 \,\lg N_\mu 
      + 1.24/\cos\theta + 0.580
\end{equation}
to estimate the primary energy as function of the observed number of electrons
($E_e>3$~MeV) and muons ($E_\mu>300$~MeV) at sea level for showers with zenith
angle $\theta$ \cite{glasstetterpune}.

A similar method has been applied by the AGASA group
\cite{Takeda:2002at}. Here, the particle density as measured 600~m
from the shower core is used to estimate the primary energy. At this
distance the fluctuations in the lateral distribution are found to be
relatively small \cite{hillass600,Dai:1988bd}. The measured value is
corrected for attenuation due to the zenith angle and converted to a
value $S_0(600)$ for vertically incident showers. The conversion to
primary energy via the relation $E[\mbox{eV}]=2.23\times10^{17}\cdot
S_0(600)^{1.02}[\mbox{m}^2]$ yields at $10^{20}$~eV a difference
between proton and iron induced showers of about 10\% and a similar
difference is obtained using two different codes to describe hadronic
interactions in the atmosphere, namely QGSJET and SIBYLL
\cite{agasa01}. 

The methods described so far depend on simulations of the shower development in
the atmosphere. To avoid this uncertainty another method is being applied in
the Auger Observatory \cite{Abraham:2008ru}. It
makes use of the constant intensity method, which relies on the fact that
primary cosmic rays arrive isotropically.  The value $S(1000)$ of the measured
signal in a water \Cerenkov detector at a distance of 1000~m from the shower
axis is used. The dependence of this parameter on the depth in the atmosphere
(which varies as $\sec\theta$ of the shower zenith angle $\theta$) is obtained
from the measured data.  Using this dependence, the actually measured signal is
converted to the value $S_{38}$, representing the signal for a shower with a
zenith angle of 38\deg.  Finally, the primary energy is estimated as $E[{\rm
EeV}]=0.149\cdot S_{38}^{1.08}$ using the fluorescence telescopes as optical
calorimeters, which define the energy scale in a direct way (see
\sref{sec:Fluorescence}) \cite{Abraham:2008ru}.

\subsubsection*{Composition Estimators}

To estimate the mass of the shower-inducing primary particle the
following array observables are used: the electron-to-muon number ratio, the
arrival time distribution of the particles, the curvature of the
shower front, and the slope of the lateral distribution.

The method applied most frequently is the measurement of the electron-to-muon
ratio at ground level.  Plotting the number of electrons $N_e$ and muons
$N_\mu$ in a plane as shown in \fref{fig:NeNmuModels}, we find an energy axis
(in the direction of the main diagonal) and (almost perpendicular to it) a mass
axis.  Using a Heitler cascade model the ratio of electrons to muons at shower
maximum can be estimated, yielding the relation \cite{jrherice06}
\begin{equation} \label{nenmeq}
 \frac{N_e}{N_\mu}= \left(\frac{E_0}{A\cdot1~\mbox{PeV}}\right)^{0.15}
    \mbox{~or~} \lg\left(\frac{N_e}{N_\mu}\right)=C-0.065\ln A .
\end{equation}
This illustrates that $\lg N_e/N_\mu$ is a function of $\ln A$.
It can be estimated that an uncertainty in $\lg(N_e/N_\mu)$
of about 16\% results in an uncertainty of about one unit in $\ln A$.

Also the shape of the shower front or the arrival time distribution of the
particles at ground level is utilized as an indirect estimate of the depth of
the shower maximum. For heavy nuclei, muons are produced earlier in the
shower development and reach the ground also earlier as compared to the
electromagnetic component whose particles follow the shower core and branch off
to large lateral distances only very late before reaching the detector array.
It is understood that the narrower pulse profiles correspond to higher
production altitudes, which in turn lead to similar arrival times in the
detectors (see e.g.\ \cite{Ave:2002gc,Ave:2003ab}).

\subsection*{Measurement of \Cerenkov Light}

Many particles in the shower disc travel with relativistic velocities through
the atmosphere. Approximately one third of the charged particles 
emit \Cerenkov light in the forward direction \cite{Giller:2004cf}, the
\Cerenkov angle in air at sea level amounts to 1.3\deg only.  Electrons (and
positrons) are the most abundant charged species in air showers. Due to their
relatively low \Cerenkov threshold (21~MeV at sea level), they contribute
mostly to the \Cerenkov light in air showers.

At present, for the detection of \Cerenkov light, two techniques are applied:
integrating detectors, in principle consisting of arrays of photomultipliers
inside light collecting cones, looking upwards in the sky, and imaging
detectors or telescopes, composed of large area collection mirrors and a camera
with segmented read-out.
Optical detectors such as \Cerenkov detectors and fluorescence detectors
(described in the next section) can only be operated during clear moon-less
nights to obtain reliable data. This restricts their duty cycle to about 10\%.

\subsubsection*{Light Integrating Detectors}

The basic idea of integrating detectors is to measure the lateral distribution
of the \Cerenkov light with an array of photomultipliers distributed over a
large area on ground level. To enlarge the collection area, the PMTs are
installed inside light collecting cones (Winston cones).  Such observations
yield the lateral density distribution of \Cerenkov photons.  It can be
parameterized by the empirical relation 
\begin{equation} \label{cerlatfun}
 C(r)=\left\{ \begin{array}{ll}
     C_{120}\cdot \exp(s[120-r/\mbox{m}]);&   30~\mbox{m}<r\le120~\mbox{m} \\
     C_{120}\cdot (r/120~\mbox{m})^{-\beta};&   120~\mbox{m}<r\le350~\mbox{m},\\
              \end{array}\right.
\end{equation}
a combination of an exponential and a power law function \cite{blanca}.
$C_{120}$ is the \Cerenkov light intensity at 120~m distance from the shower
core, $s$ the exponential inner slope, and $\beta$ the outer slope.  The energy
of the primary particle is strongly correlated with the photon density at
120~m, $C_{120}$ grows approximately as $E^{1.07}$. The average depth of the
shower maximum \Xmax is approximately linearly related to the exponential
slope $s$. Hence, from the \Cerenkov measurements both, energy and mass of the
primary particle can be derived. The latter through the dependence of the
average depth of the shower maximum on the primary particle mass.

This technique was pioneered by the AIROBICC experiment on La Palma island
\cite{hegraairobic}.  Another example is the BLANCA instrument, which was
located at the Dugway Proving Ground in Utah, USA and operated in coincidence
with the CASA experiment at the same site \cite{blanca}. BLANCA consisted of
144 angle-integrating detectors which recorded the lateral distribution of air
shower \Cerenkov light.  A \Cerenkov detector array is also installed in the
Tunka valley close to lake Baikal in Siberia
\cite{tunka13,*tunka,tunka04}, consisting of 25 detectors that cover an
area of 0.1~km$^2$.  It is planned to extend this installation with 133 optical
detectors covering an area of about 1~km$^2$ \cite{tunka133}.  

\subsubsection*{Imaging \Cerenkov Detectors}
Cosmic-ray events within the field of view of an imaging atmospheric \Cerenkov
telescope produce a focal plane image which corresponds to the direction and
intensity of \Cerenkov light coming from the air shower.  When the direction of
the air shower core and the distance of the shower axis from the telescopes are
known, simple geometry can be used to reconstruct the light received from each
altitude of the shower. The amount of \Cerenkov light produced is strongly
correlated with the number of electrons in the shower and is used to estimate
the shower size as a function of depth in the atmosphere from which the
location of the shower maximum can be determined. This procedure is essentially
geometrical and has the advantage of being almost independent of numerical
simulations expect for the calculation of the angular distribution of
\Cerenkov light around the shower axis.

Large \Cerenkov telescopes are used to reconstruct air showers initiated by
primary gamma rays in TeV $\gamma$-ray astronomy.  The telescopes and analysis
procedures are designed to effectively suppress the much more abundant (up to a
factor of 1000) hadron induced showers. The technique was established by the
pioneering work of the WHIPPLE telescope \cite{whipple}. Among the presently
largest installations are the H.E.S.S. \cite{hessexp}, MAGIC \cite{magic}, and
VERITAS \cite{veritas} telescopes.

An example for a \Cerenkov telescope, optimized for the reconstruction of
hadron induced showers was the Dual Imaging \Cerenkov Experiment (DICE)
\cite{dice}. It was located inside the CASA-MIA array in Dugway, Utah (USA) and
comprised two telescopes, each equipped with a 2~m diameter spherical mirror
viewed by an array of 256 PMTs.

\subsection*{Measurement of Fluorescence Light \label{sec:Fluorescence} }

At very high energy ($E \gtrsim 10^{17}$\,eV), the fluorescence light
technique can be used to measure directly the longitudinal profile of
air showers. This technique is based on the detection of fluorescence light
emitted by nitrogen molecules that are excited by charged particles
traversing the atmosphere.

\begin{figure}[t]
 \includegraphics[width=0.49\textwidth]{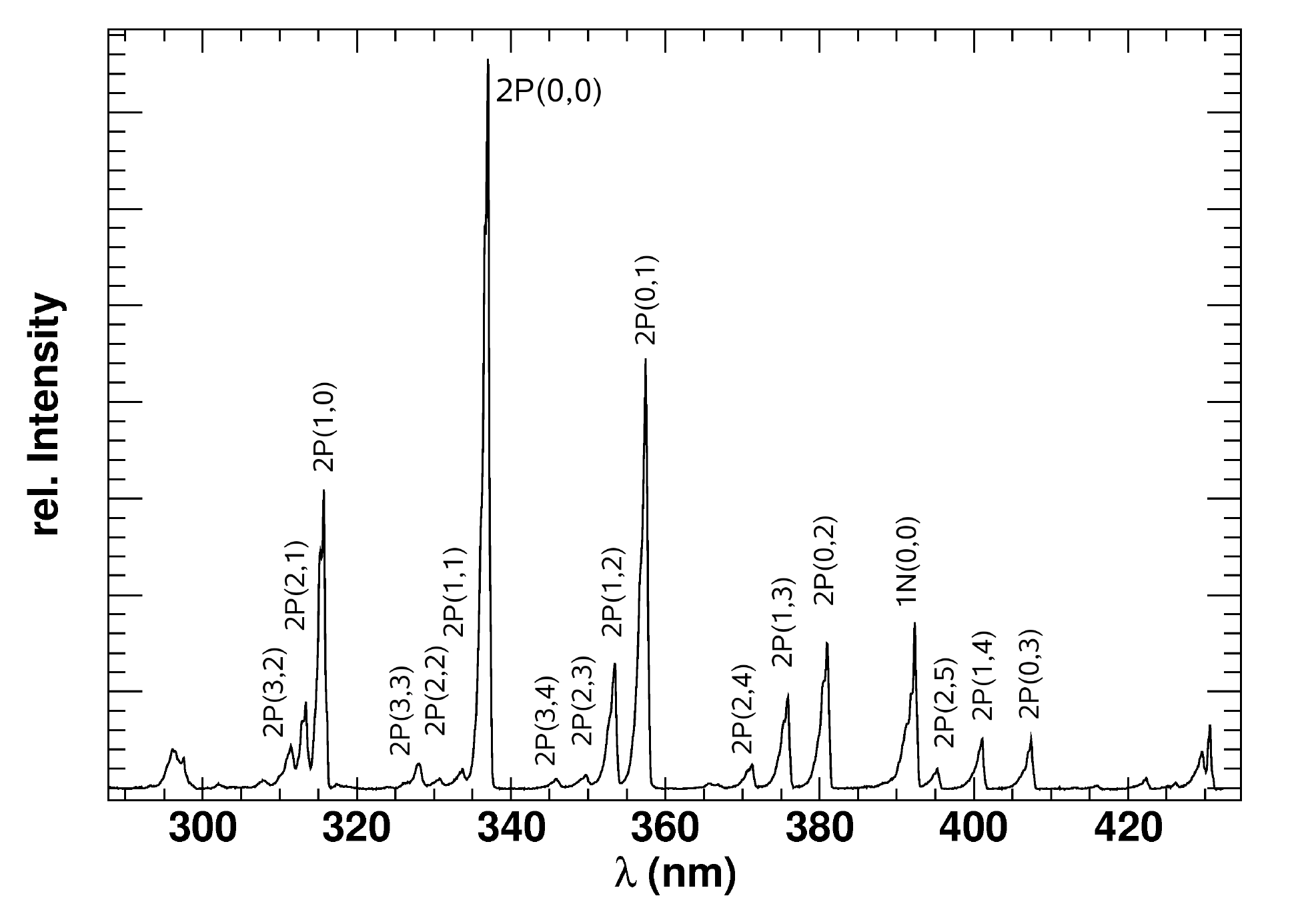}\hspace*{\fill}
 \includegraphics[width=0.49\textwidth]{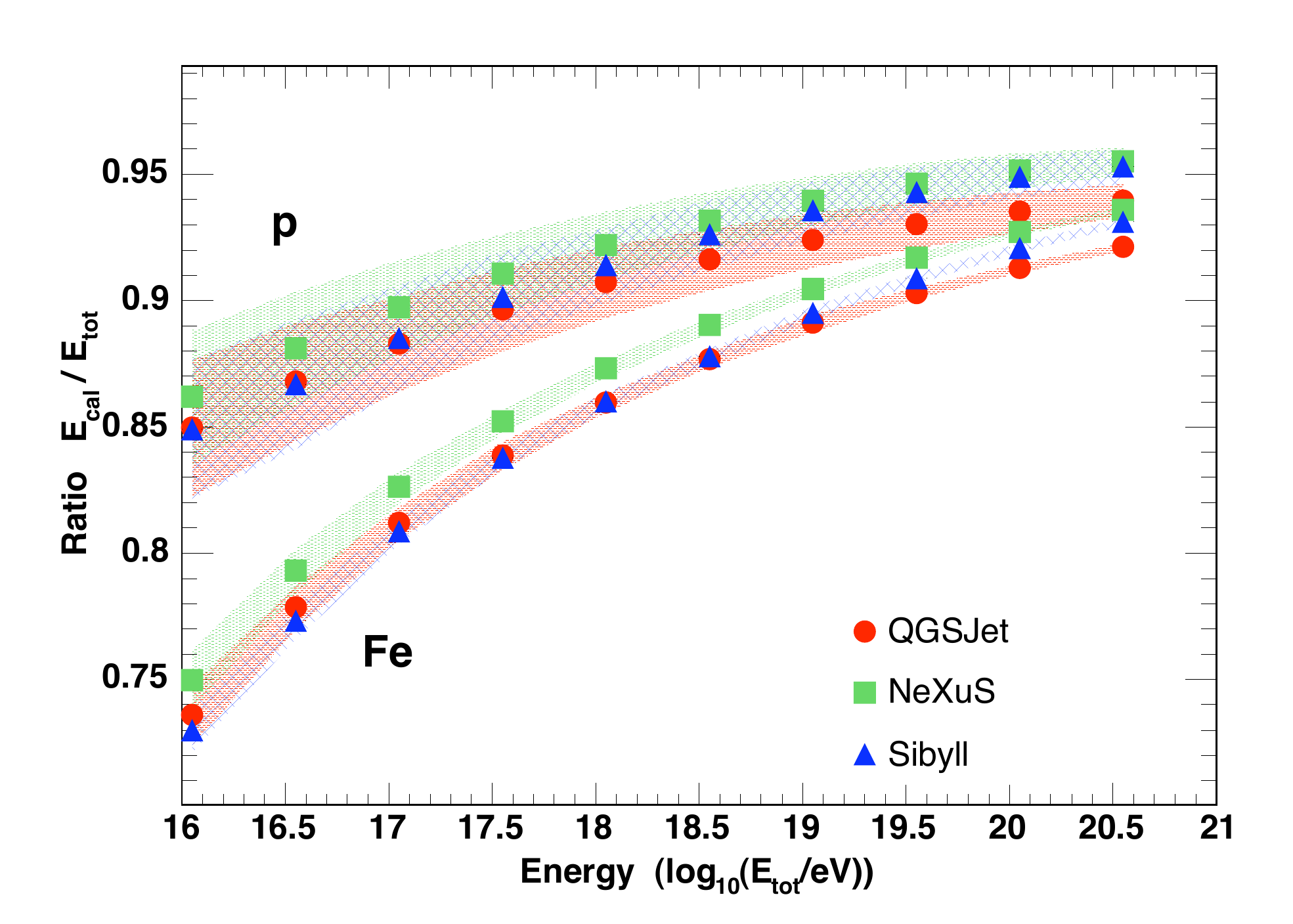}
\Caption{ \LLeft: Fluorescence light spectrum of air as measured
  by the AIRFLY experiment \protect\cite{Arciprete:2006pb,airfly5fwp} with an
  electron beam of 3 MeV. The measurement was done at $20^\circ$ C and a
  pressure of 800\,hPa.  
\RRight: Ratio of calorimetric to total
  shower energy \protect\cite{Pierog:2005aa}. Shown are predictions calculated
  with CONEX \protect\cite{Bergmann:2006yz} and different high-energy
  interaction models. The symbols present the mean values and 
  $1\sigma$ fluctuations are indicated by the shaded
  bands.  \label{fig:FYSpectrum}
}
\end{figure}

There are two transitions of electronic states of the nitrogen molecule,
called 2P and 1N for historical reasons, that lead -- in combination with
the change of the vibrational and rotational states of the molecule -- to
several fluorescence emission bands. A recent spectrometer measurement
\cite{Arciprete:2006pb,airfly5fwp} of these bands is shown in
Fig.~\ref{fig:FYSpectrum}. The bands are labeled with the electronic
transition type (2P or 1N) and the change of the vibration quantum
number $(\nu^\prime \rightarrow \nu^{\prime\prime})$, for example,
2P($\nu^\prime,\nu^{\prime\prime}$). Most of the fluorescence light
emission is found in the wave length range between 300 and 400\,nm
(near UV).  The lifetime of the excited states of nitrogen is of the
order of 10\,ns and the fluorescence light is emitted isotropically.

The fluorescence yield, the number of photons produced per deposited
energy, depends on the gas mixture in the atmosphere and atmospheric
conditions.  Collisions between molecules give rise to competing
de-excitation processes (collisional quenching, see, for example,
discussion in \cite{Keilhauer:2005nk}).  The importance of quenching
increases with pressure and almost cancels the density dependence of
the energy deposit per unit length of particle trajectory. This
results in a weakly height dependent rate of about 4-5 fluorescence
photons produced per meter and charged shower particle at relevant
altitudes.  During the last years several experiments have been set up
for measuring the fluorescence yield under different atmospheric
conditions
\cite{Kakimoto:1996pr,Nagano:2003zn,Belz:2005pv,Colin:2006nj,%
Arciprete:2006pb,Waldenmaier:2007am,Ave:2007xh,Abbasi:2007am}.
Recent progress is reviewed in \cite{Proc5FW}.  Still, the fluorescence
yield is currently known only to a precision of about 15\%
\cite{Nagano:2003zn} as end-to-end calibrated data are not yet
available.

Recent measurements confirm the expectation that the fluorescence
yield is independent of the energy of the exciting particle. Only at
very low energy, deviations are expected
\cite{Arqueros:2007rw}. Therefore, if the atmospheric dependence of
the fluorescence yield is taken into account, the fluorescence
technique allows a calorimetric measurement of the energy deposited in
the atmosphere. Simulations show that about 90\% of the total shower
energy is converted to ionization energy and, hence, is accessible for
detection \cite{Risse:2003fw,Barbosa:2003dc,Pierog:2005aa}.  The
average ratio between the energy deposited in the atmosphere and the
primary particle energy is shown in Fig.~\ref{fig:FYSpectrum}
(\rright).  It depends on the primary particle type and energy and, to
some extent, also on the model used in the simulations. However, as
most of the shower energy is transferred to em.\ particles, the model
dependence corresponds to an uncertainty of only a few percent of the
total energy. In case of a gamma-ray as a primary, about 99\% of the
energy is deposited in the atmosphere.

A complete reconstruction of a shower profile with the fluorescence
technique requires the determination of the geometry of the shower
axis, the determination of the \Cerenkov light fraction, and the
correction for the wavelength dependent atmospheric absorption of light.

In shower observations with one fluorescence telescope (monocular
observation), a shower-detector plane is given by the image of the
light track. The orientation of the shower within this plane
follows from the time sequence of the 
PMT signals \cite{Baltrusaitis:1985mx,Kuempel:2008ba}. 
The angular uncertainty of the orientation of the shower-detector plane
depends on the resolution of the fluorescence camera and the length of the
measured track. Typically a resolution of the order of $1^\circ$ is obtained.
In general, the reconstruction resolution of the angle within the shower-detector plane
is much worse and varies
between $4.5^\circ$ and $15^\circ$ (for example, see \cite{Abbasi:2005mk}). The
reconstruction accuracy can be improved considerably by measuring showers
simultaneously with two telescopes (stereo observation).  Showers observed in
stereo mode can be reconstructed with an angular resolution of about
$0.6^\circ$ \cite{Abbasi:2005mk}. A similar reconstruction quality is achieved
in hybrid experiments that use surface detectors to determine the arrival time
of the shower front at ground \cite{Fick:2003qp,Bonifazi:2009ma}.

In fluorescence measurements, the \Cerenkov light signal of air showers can be
considered as a highly asymmetric background contribution but can also
be exploited as
an independent signal \cite{Unger:2008uq}.  Knowing the longitudinal shower profile, the \Cerenkov
light contribution to the detected signal can be estimated using parametrized
electron energy distributions and models for the angular distribution of the
emitted \Cerenkov light
\cite{Baltrusaitis:1985mx,Giller:2005qz,Nerling:2006yt,Giller:2009x1}.

In general, fluorescence detectors require continuous monitoring of
atmospheric conditions, in particular the measurement of the
wavelength dependent Mie scattering length and detection of clouds
\cite{Abbasi:2005gi,Abraham:2004dt,BenZvi:2006xb}.  The temporal
variations of the density profile of the atmosphere can lead to
additional systematic uncertainties of shower reconstruction, in
particular of the depth of shower maximum \cite{Keilhauer:2004jr}.

The first fully functional air shower fluorescence detector was the Fly's Eye
experiment in Utah that began taking data in 1982 and was operated for
10 years \cite{Baltrusaitis:1985mx}.  The Fly's Eye detector was a
setup of two stations, Fly's Eye I (67 spherical mirrors of 1.5 m
diameter, viewed by 880 PMTs in total) and Fly's Eye II (8 mirrors
viewed by 120 PMTs). The Fly's Eye I detector had a total field of view (FoV)
of $360^\circ$ in azimuth and $90^\circ$ in zenith.  With the smaller FoV
 of about $90^\circ$ in azimuth and an elevation range from
$2^\circ$ to $38^\circ$ degrees, the Fly's Eye II station was designed
to measure showers in coincidence with Fly's Eye I. In October 1991 the shower 
of the highest energy measured so far, $E = (3.2 \pm 0.9) \times 10^{20}$\,eV, was 
detected with Fly's Eye I \cite{Bird:1995uy}. 

The successor to the Fly's Eye experiment, the High Resolution Fly's Eye
(HiRes), took data from 1997 (HiRes I) and 1999 (HiRes II) to 2006. The largest
data set of HiRes is that of monocular observations with HiRes I, a telescope
consisting of a ring of mirrors with a FoV from $3^\circ$ to $16^\circ$ in
elevation and full azimuth. HiRes II is built up of two rings of mirrors
covering elevation angles up to $30^\circ$. The optical resolution of the HiRes
detectors is with $1^\circ \times 1^\circ$ per PMT much higher than that of
Fly's Eye.

The Pierre Auger Observatory combines the observation of fluorescence
light using imaging telescopes with the measurement of particles
reaching ground level in a "hybrid approach" \cite{Abraham:2004dt}.
The southern Auger Observatory (near Malargue, Argentina) is the
world's largest air shower detector and comprises 1600 polyethylene
tanks set up in an area covering 3000~km$^2$. Each water \Cerenkov
detector has 3.6~m diameter and is 1.55~m high, containing 12~m$^3$ of
high-purity water. A radio system is used to provide communication
between each station and a central data acquisition system.  Four
telescope systems overlook the surface array.  A single telescope
system comprises six telescopes, overlooking separate volumes of
air. Each telescope has a camera with 440 PMT pixels, whose field of
view is approximately 1.5\deg.  One camera overlooks a total field of
view of 30\deg azimuth $\times$ 28.6\deg elevation.

In the northern hemisphere, the Telescope Array (TA)  is located in Millard
County, Utah, USA \cite{Kawai:2008zza,*TA}. It covers an area
of 860~km$^2$ and comprises 576 scintillator stations and three fluorescence
detector sites on a triangle with about 35~km separation, each equipped with
twelve fluorescence telescopes.

\subsection*{Measurement of Radio Emission from Air Showers}
\label{radiosect}

An independent measurement technique to observe air showers is provided by
means of detection of radio-frequency electromagnetic waves emitted from
showers.  Coherent radio emission generated by extensive air showers was
theoretically predicted by Askaryan in 1961 \cite{askaryanradio} and
experimentally discovered by Jelly \etal in 1965 at a frequency of 44 MHz
\cite{jelleynature}.  Over a period of time this phenomenon has been considered
as an interesting alternative to traditional methods of detection of
high-energy cosmic rays with energy greater than $10^{17}$eV. In the 1960s and
1970s the experimental and theoretical efforts in this direction had only limited
success \cite{allanrev}.  Modern experiments, such as CODALEMA \cite{Ardouin:2005xm}
and LOPES \cite{lopesspie}, aim at studies of radio emission from air showers
using modern, improved instruments.  First break-throughs have been achieved
\cite{Falcke:2005tc,Ardouin:2006nb}.  At present, also activities are under way
to install prototype antenna systems at the site of the Pierre Auger
Observatory in Argentina to investigate the possibility for radio detection of
air showers at the highest energies \cite{vandenBerg:2007tc}. For a review of recent
developments, see \cite{Falcke:2008qk}.

In addition to experimental difficulties there remain questions
concerning the quantitative radio emission theory.  Several mechanisms
of radio emission generation in air have been identified after the
pioneering work of Askaryan where the coherent \Cerenkov radiation of
the charge-excess was put forward \cite{askaryanradio}. This radiation
is very strong for showers developing in dense media
\cite{Zas:1991jv,*saltzberg,*Gorham:2006fy}. In the case of air showers there is also an
alternative radiation due to the acceleration of charged shower
particles in the Earth's magnetic field. It is called geosynchrotron
mechanism and has been recently investigated in detail
\cite{Huege:2003up,*Huege:2006kd,*Huege:2008tn}.  The interrelation
between these two essential mechanisms is not clear at present.
Hence, also combined efforts are in progress, performing accurate
radio emission calculations within the framework of a unified approach
\cite{Engel:2006fs,*Scholten:2007ky,*Werner:2007kh,*Gousset:2009zz}.

\Section{Energy Spectra\label{energysec}}

\begin{figure}[t] \centering
 \includegraphics[width=\breite]{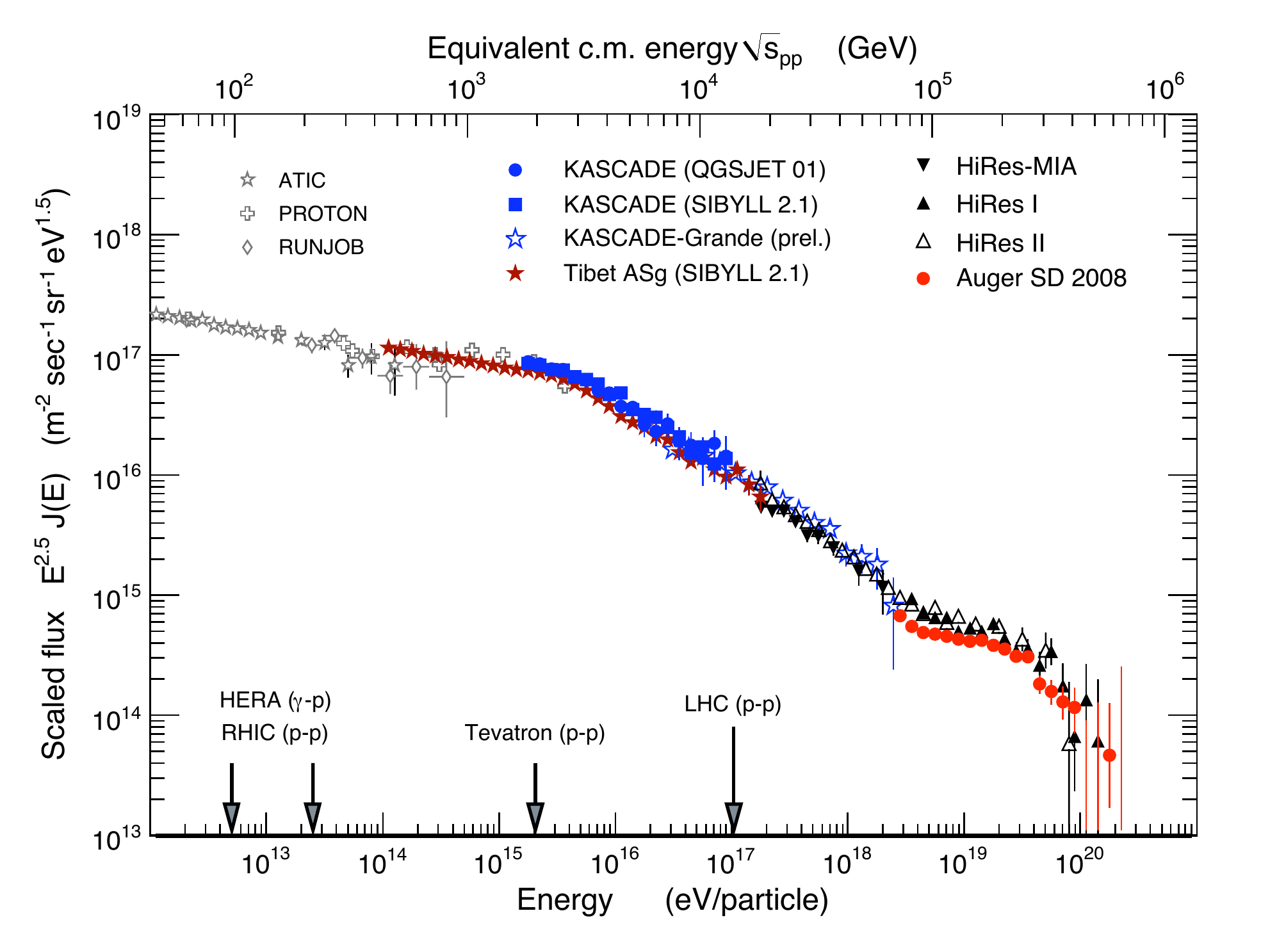}
 \Caption{All-particle cosmic-ray energy spectrum as obtained by
   direct measurements above the atmosphere by the
   ATIC~\protect\cite{Ahn:2003de,wefelpriv},
   PROTON~\protect\cite{Grigorov70a,*Grigorov71a}, and RUNJOB~\protect\cite{runjob05}
   as well as results from air shower experiments. Shown are Tibet
   AS$\gamma$ results obtained with SIBYLL 2.1 \protect\cite{Amenomori:2008jb}, KASCADE
   data (interpreted with two hadronic interaction models)
   \protect\cite{Antoni:2005wq}, preliminary KASCADE-Grande
   results~\protect\cite{ArteagaVelazques:2008xx}, and Akeno
   data~\protect\cite{Nagano:1984db,Nagano:1992jz}. The measurements at high
   energy are represented by
   HiRes-MIA~\protect\cite{AbuZayyad:2000ay,AbuZayyad:1999xa}, HiRes I and
   II~\protect\cite{Abbasi:2007sv},
   and
   Auger~\protect\cite{Abraham:2008ru}.
   \label{allparticle} }
\end{figure}

The all-particle energy spectrum extending from $10^{12}$~eV up to the highest
energies is shown in \fref{allparticle}.  The flux as obtained from direct
measurements above the atmosphere (represented in the figure through results
from ATIC, PROTON, and RUNJOB) extends smoothly to high energies in the air
shower detection regime. The all-particle spectrum can be approximated by a
broken power law $\propto E^\gamma$ with a spectral index $\gamma=-2.7$ below
$E_k\approx4\times10^{15}$~eV. At this energy, the {\sl knee}, the spectral
index changes to $\gamma\approx-3.1$. 

In the following we consider in more detail two energy regions: galactic cosmic
rays up to energies of about $10^{17}$ to $10^{18}$~eV and the extragalactic
component at higher energies. \footnote{The exact energy of the transition from
galactic to extragalactic cosmic rays is presently not known, however, it is
generally assumed to be in the energy range indicated, see also
\sref{astrosec}.}

\subsubsection*{Galactic Cosmic Rays}

\begin{figure}[t] \centering
 \includegraphics[width=\breite]{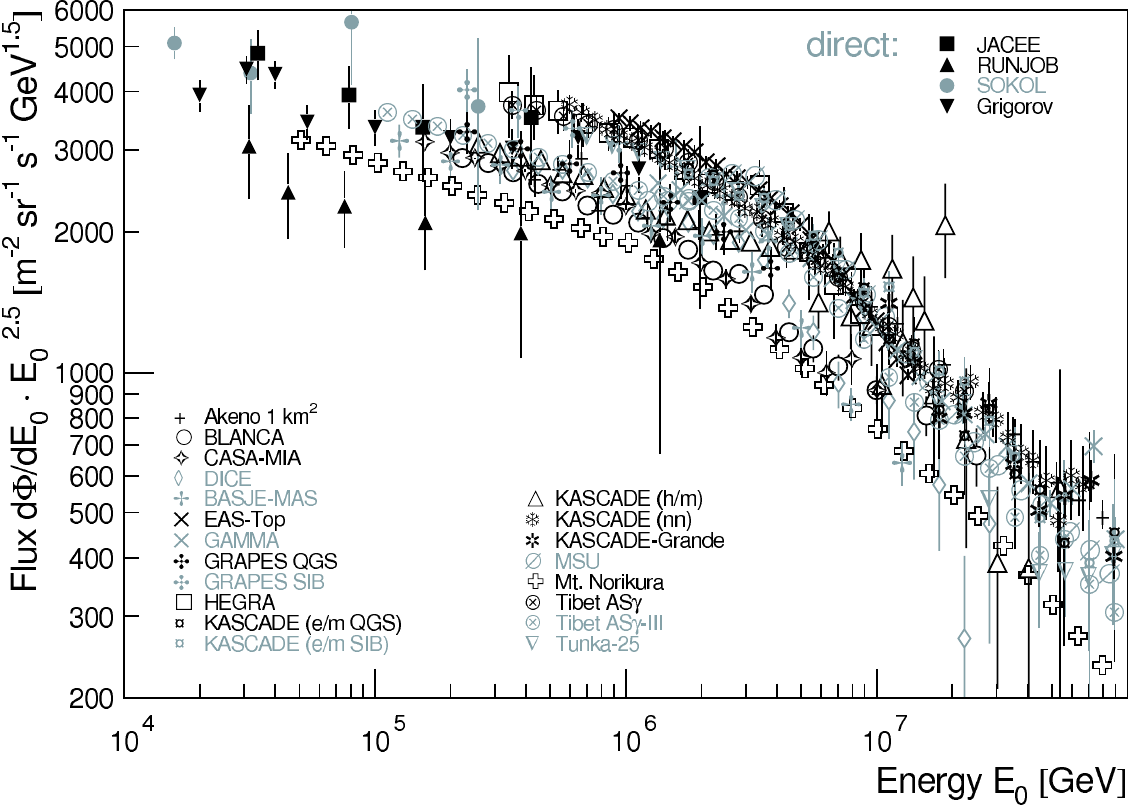}
 \Caption{All-particle energy spectra in the knee region. 
   Results from direct measurements by
   Grigorov \etal \protect\cite{Grigorov70a,*Grigorov71a},
   JACEE \protect\cite{jaceefe},
   RUNJOB \protect\cite{runjob05}, and
   SOKOL \protect\cite{sokol}
   as well as from the air shower experiments
   Akeno 1~km$^2$ \protect\cite{Nagano:1984db},
   BASJE-MAS \protect\cite{basjemas},
   BLANCA \protect\cite{blanca},
   CASA-MIA \protect\cite{casae},
   DICE \protect\cite{dice},
   EAS-TOP \protect\cite{eastope},
   HEGRA \protect\cite{hegraairobic},
   KASCADE electrons and muons interpreted with two hadronic interaction models
        \protect\cite{Antoni:2005wq},
        hadrons \protect\cite{hknie}, and a neural network analysis combining
        different shower components \protect\cite{rothnn},
   MSU \protect\cite{msu},
   Mt.~Norikura \protect\cite{mtnorikura},
   Tibet AS$\gamma$ \protect\cite{tibetasg00} and
         AS$\gamma$-III \protect\cite{Amenomori:2008jb}, as well as
   Tunka-25 \protect\cite{tunka04}.
   \label{knee-zoom}}
\end{figure}

Many groups published results on the all--particle energy spectrum from
indirect measurements in the knee region ($\approx 10^{15}$~eV).
The spectra obtained are compiled in \fref{knee-zoom}. The ordinate has been
multiplied by $E_0^{2.5}$. The individual measurements agree within a factor
of two in the flux values and a similar shape can be recognized for all
experiments with a \knee at energies of about 4~PeV.  Also shown are results
for the all-particle flux as obtained by direct observations above the
atmosphere approaching energies up to 1~PeV.  In the region of overlap, the
results from direct and indirect measurements are in reasonable agreement.
Typical values for the systematic uncertainties of the absolute energy scale
of air shower experiments are about 15 to 25\%.  Renormalizing the energy
scales of the individual experiments to match the all-particle spectrum
obtained by direct measurements in the energy region up to almost a PeV
requires correction factors of the order of 10\% \cite{Hoerandel:2002yg}.  A
remarkable result, indicating that behind an absorber of 11 hadronic
interaction lengths or 30 radiation lengths the energy of the primary particle
is determined with an absolute error of the order of 10\%.  One should keep in
mind that the experiments investigate different air shower components,  are
situated at different atmospheric depths, and use different interaction models
to interpret the observed data. Nevertheless, the systematic differences are
relatively small and the all-particle spectrum seems to be well known.

\begin{figure}[tp] \centering
 \includegraphics[height=0.94\textheight]{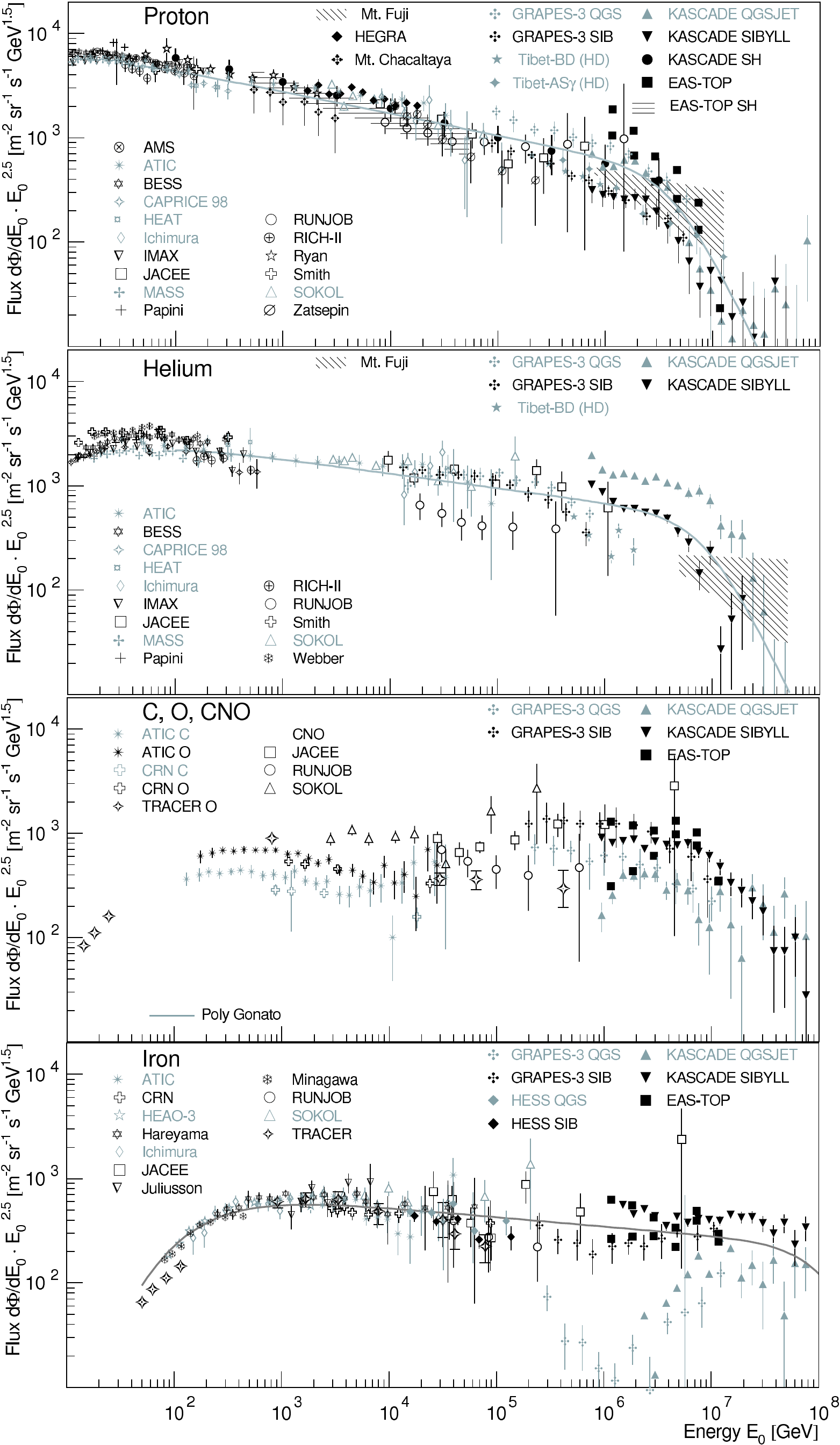}
 \Caption{Energy spectra for elemental groups, caption see next
          page.\label{elementspek}}
\end{figure} 

\begin{figure}[t] 
 {\centering\begin{minipage}{0.9\textwidth}\Fref{elementspek}:
  Cosmic-ray energy spectra for four groups of elements, from top to bottom:
  protons, helium, CNO group, and iron group.
  \newline
   {\bf Protons:}
   Results from direct measurements above the atmosphere by 
   AMS \protect\cite{amsp},
   ATIC \protect\cite{atic05},
   BESS \protect\cite{bess00},
   CAPRICE \protect\cite{caprice98},
   HEAT \protect\cite{heat01},
   \protect\cite{ichimura},
   IMAX \protect\cite{imax00},
   JACEE \protect\cite{jaceephe},
   MASS \protect\cite{mass99},
   \protect\cite{papini},
   RUNJOB \protect\cite{runjob05},
   RICH-II \protect\cite{rich2},
   \protect\cite{ryanp},
   \protect\cite{smith},
   SOKOL \protect\cite{sokol},
   \protect\cite{zatsepinp},
   and 
   fluxes obtained from indirect measurements by
   KASCADE electrons and muons for two hadronic interaction models
   \protect\cite{Antoni:2005wq} and single hadrons \protect\cite{kascadesh},
   EAS-TOP (electrons and muons) \protect\cite{eastopspec} and single hadrons
   \protect\cite{eastopsh},
   GRAPES-3 interpreted with two hadronic interaction models \protect\cite{grapes05},
   HEGRA \protect\cite{hegrap},
   Mt.\ Chacaltaya \protect\cite{chacaltayap},
   Mts.\ Fuji and Kanbala \protect\cite{mtfujip},
   Tibet burst detector (HD) \protect\cite{tibetbdp} and AS$\gamma$ (HD) 
   \protect\cite{tibetasgp}.
\newline
  {\bf Helium:}
    Results from direct measurements above the atmosphere by 
   ATIC \cite{atic05},
   BESS \cite{bess00},
   CAPRICE \cite{caprice98},
   HEAT \cite{heat01},
   \cite{ichimura},
   IMAX \cite{imax00},
   JACEE \cite{jaceephe},
   MASS \cite{mass99},
   \cite{papini},
   RICH-II \cite{rich2},
   RUNJOB \cite{runjob05},
   \cite{smith},
   SOKOL \cite{sokol},
   \cite{webberhe},
   and 
   fluxes obtained from indirect measurements by
   KASCADE electrons and muons for two hadronic interaction models
   \cite{Antoni:2005wq},
   GRAPES-3 interpreted with two hadronic interaction models \cite{grapes05},
   Mts.\ Fuji and Kanbala \cite{mtfujip}, and
   Tibet burst detector (HD) \cite{tibetbdp}.
\newline
   {\bf CNO group:}
   Results from direct measurements above the atmosphere by 
   ATIC (C+O) \cite{atic06},
   CRN (C+O) \cite{crn},
   TRACER (O) \cite{tracer05},
   JACEE (CNO) \cite{jaceemasse},
   RUNJOB (CNO) \cite{runjob05},
   SOKOL (CNO) \cite{sokol},
   and 
   fluxes obtained from indirect measurements by
   KASCADE electrons and muons \cite{Antoni:2005wq},
   GRAPES-3 \cite{grapes05}, 
   the latter two give results for two hadronic interaction models,
   and
   EAS-TOP \cite{eastopspec}.
\newline
 {\bf Iron:}
   Results from direct measurements above the atmosphere by 
    ATIC \cite{atic06},
    CRN \cite{crn},
    HEAO-3 \cite{heao3},
    \cite{juliusson},
    \cite{minagawa},
    TRACER \cite{tracer05}
    (single element resolution) and
    \cite{hareyama},
    \cite{ichimura},
    JACEE \cite{jaceefe},
    RUNJOB \cite{runjob05},
    SOKOL \cite{sokol} 
    (iron group),
   as well as
   fluxes from indirect measurements (iron group) by
   EAS-TOP \cite{eastopspec},
   KASCADE electrons and muons \cite{Antoni:2005wq}, 
   GRAPES-3 \cite{grapes05}, and
   H.E.S.S. direct \Cerenkov light \cite{Aharonian:2007zja}.
   The latter three experiments give results according to
   interpretations of the measured air-shower data with two hadronic
   interaction models, namely QGSJET and SIBYLL.
\newline
   The gray solid lines indicate spectra according to the \modell
   \cite{Hoerandel:2002yg}.
   \end{minipage}}
\end{figure}   

Up to about a $10^{15}$\,eV direct measurements have been performed with
instruments above the atmosphere. As examples, results for primary protons,
helium, and iron nuclei are compiled in \fref{elementspek}.  Recently, also
indirect measurements of elemental groups became possible, as discussed below
and the results of the KASCADE and EAS-TOP experiments are shown in the figures
as well.  Also results from other air shower experiments are shown.  HEGRA used
an imaging \Cerenkov telescope system to derive the primary proton flux
\cite{hegrap}.  Spectra for protons and helium nuclei are obtained from
emulsion chambers exposed at Mts.\ Fuji and Kanbala \cite{mtfujip}.  The Tibet
group performs measurements with a burst detector as well as with emulsion
chambers and an air shower array \cite{tibetbdp,tibetasg03}.  GRAPES-3 uses the
correlation between the registered number of electrons and muons to derive
energy spectra for mass groups \cite{grapes05}.  
The H.E.S.S. \Cerenkov telescope system derived for the first time an energy
spectrum measuring direct \Cerenkov light \cite{Aharonian:2007zja}.  This light is emitted
by the primary nuclei in the atmosphere before its first interaction, i.e.\
before the air shower begins \cite{directc}. Results for iron nuclei are shown.

Over the wide energy range depicted, the flux as obtained by direct
measurements is smoothly continued to higher energies with the results of air
shower measurements. Despite of the experimental uncertainties and systematic
differences between different experiments and different interpretations of air
shower data using various air shower models, a clear picture of the spectra for
elemental groups is evolving.
It is evident that the \knee in the all-particle spectrum is caused by
a depression of the flux of light elements.
The measurements follow power laws with a cut-off at high energies.
The spectra according to the \modell are indicated in the figures as lines. It
can be recognized that the measured values are compatible with cut-offs at
energies proportional to the nuclear charge $\hat{E}_Z=Z\cdot4.5$~PeV
\cite{Hoerandel:2002yg,cospar06}.

\subsubsection*{Extragalactic Cosmic Rays} \label{extrag}

There are several detector installations that were designed to measure
cosmic rays at the highest energies. At the high-energy end, the total
aperture and observation time determine the statistics of expected
events. At low energy, the acceptance range is given by the employed
detection technique and typically the distance between the detectors.

\begin{figure}[t]\centering
\includegraphics[width=\breite]{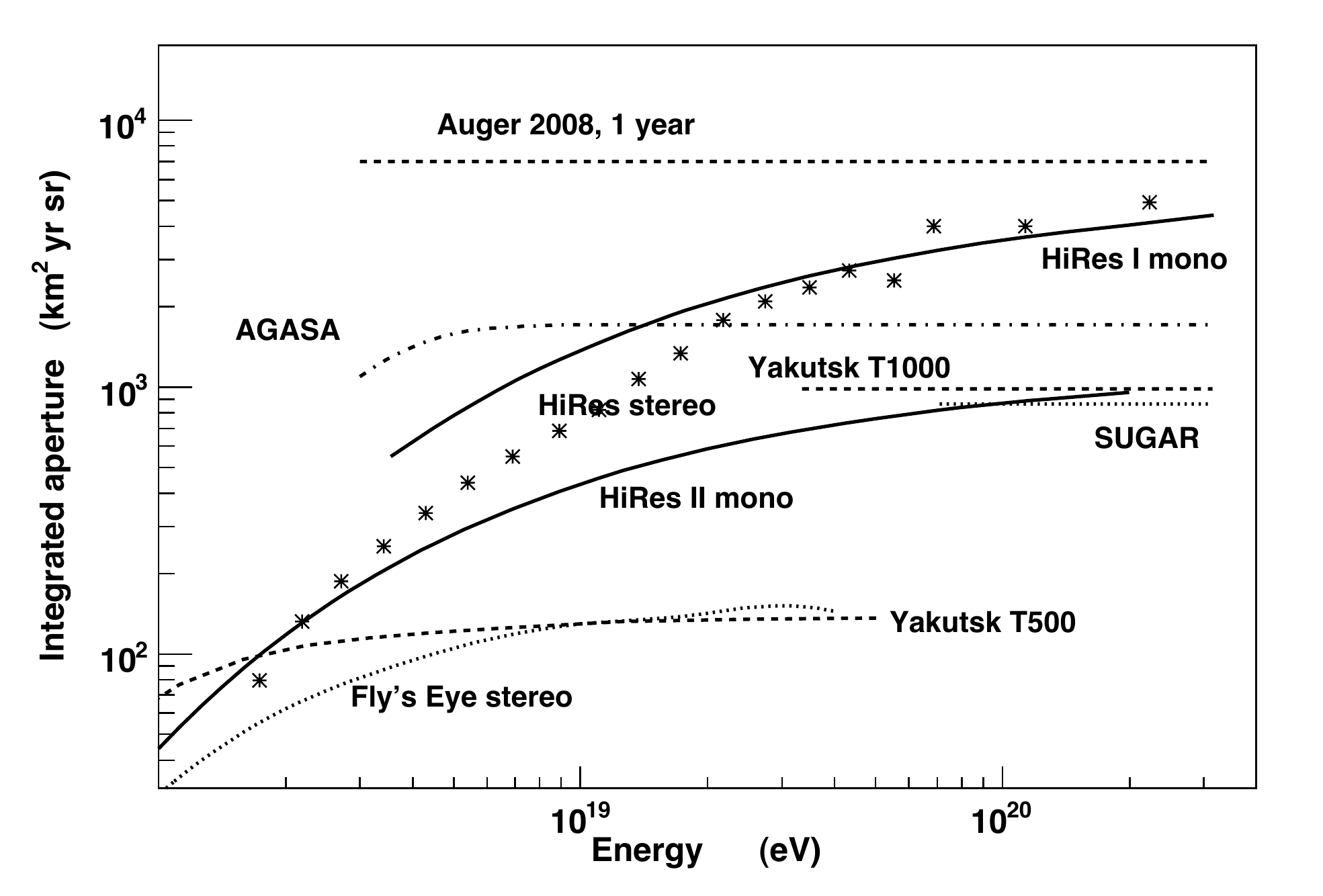}
\Caption{\label{fig:exposure}
Integrated aperture of different high-energy detectors corresponding to
the data shown in Fig.~\protect\ref{fig:flux3-all}. 
The AGASA
aperture refers to all air shower data with $\theta < 45^\circ$ up to May 2003.
The two HiRes detectors have different data taking periods: HiRes I from June
1997 to June 2005 and HiRes II from
December 1999 to August 2004 \protect\cite{Abbasi:2007sv}. The HiRes stereo
exposure is that used for the recent anisotropy study \protect\cite{Abbasi:2008md}.
The Auger exposure of 7000 km$^2$~sr~yr refers to data taking
during the construction from Jan 2004 until August 2007, excluding
events close to the array boundary \protect\cite{Abraham:2008ru}.
The integrated aperture of the
Yakutsk array includes data taken from September 1974 to 
June 2001 for T1000, and September 1979 till June 2001 for T500
\protect\cite{Pravdin:2004aa}. The exposure shown for SUGAR is based on the 
re-analysis of the 5 highest energy events reported in
\protect\cite{Anchordoqui:2003gm} and corresponds to 11 years of operation.
The Fly's Eye exposure in stereo mode is taken from \protect\cite{Bird:1994wp}.
The integrated aperture of the data set used in \protect\cite{Ave:2001hq} 
for calculating the Haverah Park flux  
is $7.39\times 10^{12}$\,m$^2$\,s\,sr.
}
\end{figure}

A compilation of the integrated aperture (i.e.\ total exposure)
reported by experiments with data above $10^{19}$\,eV is shown in
Fig.~\ref{fig:exposure}.  At ultra high-energy, about two times more events
are expected in the Auger data set than HiRes has
collected in monocular or stereo mode. The Yakutsk experiment is expected to have slightly more
than a fourth of the statistics of HiRes I and about one half of that
of AGASA.

\begin{figure}[t]\centering
\includegraphics[width=\breite]{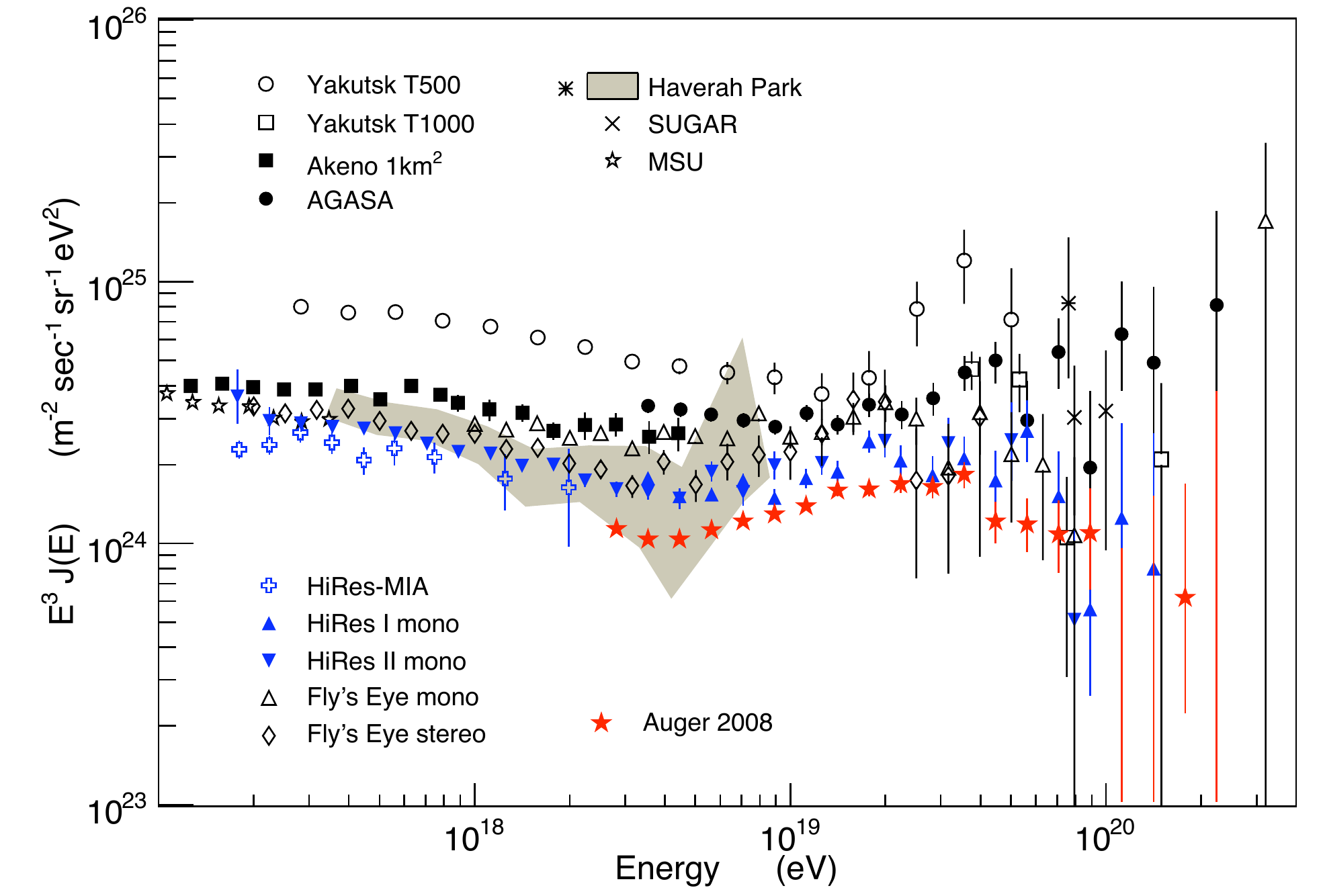}
\Caption{\label{fig:flux3-all} Comparison of flux measurements scaled
  by $E^3$. Only statistical errors are shown. 
Shown are the data of
AGASA~\protect\cite{Takeda:2002at,Takeda:2003aa}, 
Akeno~\protect\cite{Nagano:1992jz,Nagano:1984db}, 
Auger~\protect\cite{Abraham:2008ru},
Fly's Eye~\protect\cite{Bird:1994wp,Bird:1993yi},
Haverah Park~\protect\cite{Ave:2001hq},
HiRes-MIA~\protect\cite{AbuZayyad:2000ay,AbuZayyad:1999xa},
HiRes Fly's Eye~\protect\cite{Abbasi:2007sv}, 
MSU~\protect\cite{Fomin:2003aa},
SUGAR~\protect\cite{Anchordoqui:2003gm},
and
Yakutsk~\protect\cite{Glushkov:2003aa}. Yakutsk T500 (trigger 500)
refers to the smaller
sub-array of the experiment with 500m detector spacing and T1000 (trigger
1000) to the array with 1000m detector distance.
The data of the MSU array are included to show the
connection of the high-energy measurements to lower energy data covering
the knee of the cosmic-ray spectrum. }
\end{figure}

The measurements of the cosmic-ray fluxes obtained with these exposures are
shown in Fig.~\ref{fig:flux3-all} (error bars indicate statistical
uncertainties only). In case several analyses of the same data set are
available, only the most recent results are included in the plot.  The shaded
area, depicting the results of the analysis of the Haverah Park data
\cite{Ave:2001hq}, accounts for some systematic uncertainties by assuming
extreme elemental compositions, either fully iron or proton dominated.  The
highest energy point (Fly's Eye monocular observation) corresponds to the
highest energy event \cite{Bird:1995uy}. For sake of clarity upper limits are
not shown.

\begin{figure}[t]\centering
\includegraphics[width=0.85\textwidth]{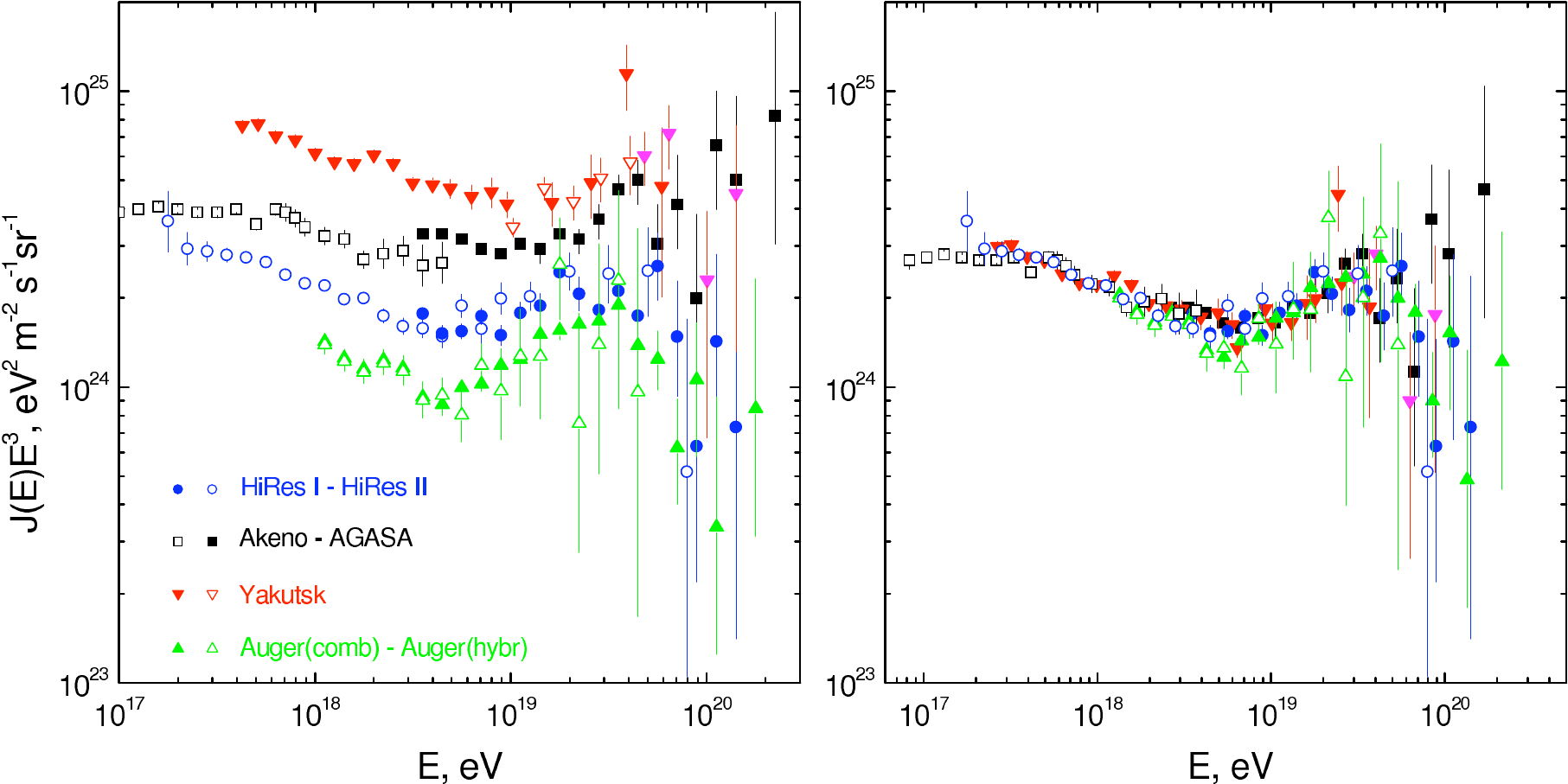}
\Caption{Flux of UHECRs as measured with the four
  detectors that have the largest exposures, namely
Yakutsk~\protect\cite{Glushkov:2003aa}
  AGASA~\protect\cite{Takeda:2002at,Takeda:2003aa},
  Auger~\protect\cite{Abraham:2008ru}, and
  HiRes~\protect\cite{Abbasi:2007sv}. 
\LLeft: Cosmic-ray spectra as derived by the Collaborations using the
calibration of the detectors.
\RRight: Cosmic-ray spectra after re-scaling of the energy scale
 of the experiments to obtain a common position of the dip, from
 \protect\cite{Aloisio:2006wv,Berezinsky:2008qh}.
The nominal energy scales of the experiments have been multiplied by
1.2, 1.0, 0.75, 0.625 for Auger, HiRes, AGASA, and Yakutsk, respectively.
\label{fig:UHECR-flux3-selected}
}
\end{figure}

It is common to present the data multiplied by $E^3$ to enhance
deviations from a $dN/dE \sim E^{-3}$ power law.  In this
representation, characteristic features such as the second knee at
about $10^{17.5}$\,eV and the ankle at $10^{18.5}$\,eV are better
visible.  However, it should be emphasized that 
scaling the flux with energy to some power
(e.g.\ $E^3 J(E)$ or $E^{2.5} J(E)$)
is misleading and does not reflect correctly the
uncertainties of the measurements. In this presentation, the
statistical uncertainties cannot be separated from the systematic
energy calibration uncertainties. The importance of the systematic
uncertainty of the experimental energy scale is demonstrated in
Fig.~\ref{fig:UHECR-flux3-selected} in which fluxes with the nominal
energy scale of the experiments are compared with that after a
model-motivated energy shift has been applied
\cite{Bahcall:2002wi,Aloisio:2006wv}. After shifting the energy scales
of the experiments, a very good overall agreement of the different
measurements is obtained. In particular, the ankle in the cosmic-ray
spectrum is clearly seen. The good agreement is a non-trivial
observation as the position of the ankle and the overall flux change
in a correlated way in the $E^3$-representation.

The data sets of the HiRes and Auger measurements provide evidence for
a flux suppression at ultra high-energy. The statistical significance
of the flux suppression is difficult to specify unambiguously. If one
compares to a power-law flux a significance of more than 5 $\sigma$ is
found in each of the data sets
\cite{Abraham:2008ru,Abbasi:2007sv}. Both spectra can be well
described by models with uniformly distributed sources and a  
GZK suppression \cite{Yamamoto:2007xj,Abbasi:2005ni}. A similarly
good agreement between GZK model predictions and the Yakutsk data was
shown by several authors (for example,
\cite{Bahcall:2002wi,Berezinsky:2003mq}).

Only the AGASA data seem to disfavor a flux suppression. Assuming uniformly
distributed sources of UHE protons and treating the normalization of the
expected energy spectrum as free parameter AGASA expects to observe 1.8 events
above $10^{20}$\,eV with 11 actually detected.  This corresponds to a
$4.5\sigma$ deviation from the GZK cutoff spectrum \cite{Takeda:2003aa}. Other
assumptions on the shape of the GZK proton spectrum lead to the prediction of
2.4 expected events \cite{Berezinsky:2002nc,Berezinsky:2003mq}, corresponding
to a deviation of $3.9\sigma$ from the GZK cutoff hypothesis.  Due to
uncertainties in the absolute energy scale and partially also low statistics,
the differences between the spectra of the different experiments are of limited
statistical significance only \cite{DeMarco:2003ig,Teshima:2003aa}.  

It should be kept in mind that a sudden and drastic change of
composition from hadrons to photons could give observational results
similar to a suppression of the flux \cite{Chou:2006jj} observed with 
fluorescence detectors, but not with surface arrays. So far there
are no indications for very deeply penetrating showers that would be
expected in such a case.

Given the
importance of the absolute energy assignment to a reconstructed shower we will
summarize the current systematic uncertainties below.

The energy reconstruction of showers detected with the AGASA array is
based on the scintillator signal $S(600)$ at 600\,m from the shower
core, where shower-to-shower fluctuations are the smallest and the
relation between the signal and the primary energy is almost
composition independent \cite{Dai:1988bd,Nagano:1999xk}. The
systematic error of energy assignment is analyzed in
\cite{Takeda:2002at} in detail, see also discussion
\cite{Drescher:2005bg}. AGASA finds a total systematic uncertainty of
the energy assignment of about $18$\%. The main
sources of uncertainty are related to shower phenomenology and the
simulation of the relation of $S(600)$ to the primary particle energy.
In particular, the observed discrepancy between the surface detector signal at
1000\,m from the shower core and the fluorescence-based calorimetric
energy measurement reported by Auger \cite{Engel:2007cm,*Schmidt:2009ge} indicates
that currently available shower simulations do not allow to obtain an
absolute energy scale with an systematic uncertainty smaller than
$~$20\%. Therefore it is not surprising that discrepancies between the
experimentally observed attenuation length for $S(600)$ and that
expected from simulations hamper a re-analysis of the AGASA data
\cite{Shinozaki2006private}.

The shower energy determination applied in HiRes is based on the track length
integral $ E_{\rm cal} = \alpha_{\rm eff} \int N(X) \ dX, $ where
$N(X)$ is a fit to the shower profile using the Gaisser-Hillas
function (\ref{eq:GaisserHillas}) and $\alpha_{\rm eff}$ denotes the
mean ionization energy deposit \cite{Song:1999wq}.
With HiRes being a
fluorescence detector, the energy reconstruction is closely related to
properties of the atmosphere which is serving as calorimeter. At the
same time, atmospheric properties also determine the aperture of the
detector. The HiRes flux measurements (HiRes I and HiRes II mono) are
found to have similar systematic uncertainties
\cite{AbuZayyad:2002sf,Abbasi:2002ta,Abbasi:2007sv}.  The main contributions to the
systematic uncertainty of the energy scale are the absolute
calibration of the PMTs ($10$\%), the limited knowledge of the
air fluorescence yield ($6$\%), and
atmospheric conditions ($9$\%).  About $10$\% uncertainty
results from the rescaling of the measured calorimetric energy to
obtain the total shower energy \cite{Song:1999wq}, see also
\sref{sec:ShowerProperties}.  Adding the individual contributions
in quadrature, the overall systematic uncertainty of the energy
reconstruction amounts to $17$\% \cite{Abbasi:2002ta}.

In contrast to surface arrays, the aperture of fluorescence detectors
has to be determined by simulations. Sources of uncertainty are here
varying atmospheric conditions, simulation of shower profiles and
detector trigger thresholds, and the primary cosmic ray composition.
The uncertainty due to varying atmospheric conditions, mainly that of
the vertical aerosol optical depth (VAOD), has been estimated to
contribute to the aperture uncertainty $15$\%
\cite{AbuZayyad:2002sf,Abbasi:2002ta}. In a recent study the other,
simulation-related sources of uncertainty were found not to contribute
significantly to the overall flux uncertainty of $30$\%
\cite{Abbasi:2006mb}.

The technique employed in the Auger measurement of the flux combines the
advantages of surface detector arrays with that of fluorescence detectors
\cite{Abraham:2008ru}. The surface array operates with almost
100\% duty cycle and the aperture can be calculated in a rather
straight-forward way for energies well above the trigger threshold.
Fluorescence telescopes allow the direct measurement of the calorimetric shower
energy, however, their duty cycle is only about 13\%. Using a set of
well-reconstructed hybrid events,\footnote{Events detected with both, the
fluorescence telescopes and the surface detectors are called hybrid events.}
one can calibrate the energy estimator for surface detector data in an almost
model-independent way. This is done in a two-step process. First the shower
signal at 1000\,m, $S(1000)$, is corrected for attenuation to that of an
equivalent shower of $38^\circ$ zenith angle, $S_{38}$. To avoid any possible
bias from simulations this is done with the constant intensity method by
requiring the same number of showers per unit of $\sin^2\theta$. In the second
step $S_{38}$ is converted to total shower energy.
 
The aperture of the Auger detector increased continuously during construction
and has an uncertainty of less than $3$\%.  The systematic uncertainty of the
energy scale coming from the fluorescence energy measurement is estimated to be
$22$\%.  The main contributions to this uncertainty are coming from the
uncertainty of the fluorescence yield ($14$\%), the calibration of the
fluorescence telescopes ($10$\%), and the reconstruction method ($10$\%)
\cite{Dawson:2007di,Abraham:2008ru}. The described calibration procedure 
for relating S(1000) to the primary particle energy leads
to an uncertainty of $7$\% at $10^{19}$\,eV increasing to $15$\% at
$10^{20}$\,eV.

\Section{Composition} \label{compsec}
\subsubsection*{Mean Logarithmic Mass}

At energies below $10^{14}$~eV the abundance of individual elements has been
measured with detectors above the atmosphere.  At higher energies this is
presently not possible due to the low flux values and the large fluctuations in
the development of extensive air showers. Thus, in the past, mostly the mean
mass has been investigated.  An often-used quantity to characterize the
composition is the mean logarithmic mass, defined as $\lnA= \sum_i
r_i \ln A_i$, $r_i$ being the relative fraction of nuclei of mass $A_i$.  
Experimentally, $\lnA$ is obtained applying two methods: (i) the quantity is
proportional to the ratio of the number of electrons and muons registered at
ground level $\lnA\propto\log_{10}(N_e/N_\mu)$, see \eref{nenmeq} and (ii)
$\lnA$ is proportional to the observed depth of the shower maximum, according
to the relation $X_{\rm max}^A=X_{\rm max}^p-X_R\ln A$, see \eref{xmaxeq}.  Hence, the
maximum of an iron induced shower should be about 150~\gcm2 higher up in the
atmosphere as compared to a proton induced shower ($X_{\rm max}^p$).  

\begin{figure}[t] \centering
 \includegraphics[width=\breite]{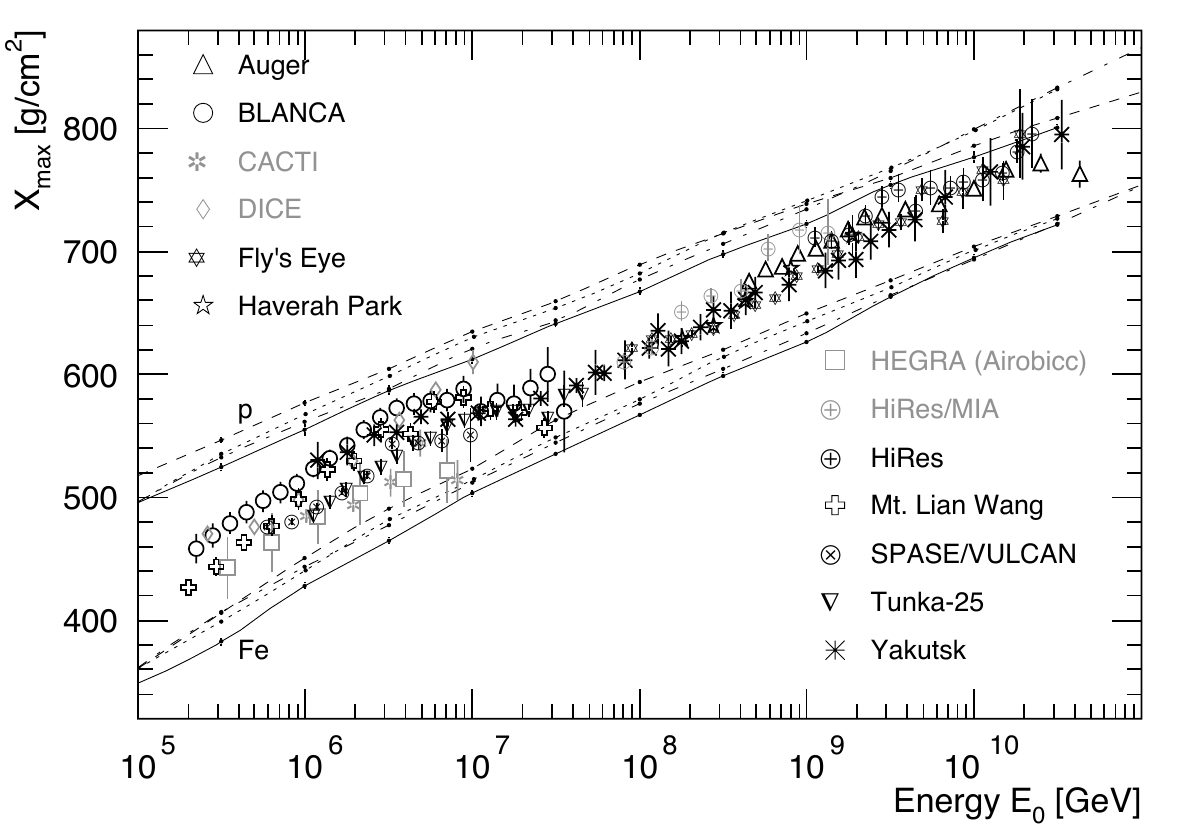}
 \Caption{Average depth of the shower maximum \Xmax as function of primary
           energy as obtained by 
           Auger \protect\cite{Unger:2007mc},
           BLANCA \protect\cite{blanca},
           CACTI \protect\cite{cacti},
           DICE \protect\cite{dice},
           Fly's Eye \protect\cite{flyseye},
           Haverah Park \protect\cite{haverahpark00},
           HEGRA \protect\cite{hegraairobic},
           HiRes/MIA \protect\cite{AbuZayyad:1999xa},
	   HiRes \protect\cite{Abbasi:2004nz},
           Mt. Lian Wang \protect\cite{mtlianwang},
           SPASE/VULCAN \protect\cite{spase99},
           Tunka-25 \protect\cite{tunka04},
           Yakutsk \protect\cite{Knurenko:2001ab}. 
           The lines indicate simulations for proton and iron induced showers
           using the CORSIKA code with the hadronic interaction model QGSJET~01
           (\line), QGSJET~II-3 (\dashed), SIBYLL~2.1
           (\dotted), and EPOS~1.6 (\dashdot). \label{xmax} }
\end{figure}

Recent measurements of the average depth of the shower maximum are compiled in
\fref{xmax}, covering energies from $10^5$ to almost $10^{11}$~GeV.  The
experimental results are compared to predictions of the average depth of the
shower maximum from simulations for primary protons and iron nuclei. The
CORSIKA code \cite{Heck98a} has been used with the hadronic interaction model
QGSJET~01~\cite{Kalmykov89a-e,*Kalmykov:1993qe,*Kalmykov:1997te},
QGSJET~II-3~\cite{Ostapchenko:2005nj,*Ostapchenko:2006vr}, 
SIBYLL~2.1~\cite{Engel:1999db}, and EPOS~1.6 \cite{Pierog:2007x1}.  There are significant
differences between the predictions of the different models concerning the
absolute values of \Xmax.  The differences become important when the model
predictions are compared to the experimental data to derive information on the
elemental composition of cosmic rays.

Below $4\times10^6$~GeV the values obtained by different experiments exhibit a
common trend, they seem to increase faster as function of energy than the
simulations, which implies that the average composition would become lighter as
function of energy.  Above the knee ($E\ga4\times10^6$~GeV) the measured values
flatten up to about $4\times10^7$~GeV, indicating an increase of the average
mass in this energy range, as expected from sequential breaks in the energy
spectra for individual elements, seen already in \fref{elementspek}.  Finally,
above $4\times10^7$~GeV the measured data exhibit about a constant slope for
\Xmax as function of energy. The slope is slightly steeper than the predicted
slope for iron nuclei for all models shown. But, on the other hand, a
comparison to the predicted proton values is not conclusive, while models like
QGSJET~01 favor an extremely light composition at the highest energies, models
like DPMJET~2.5 give a hint to an intermediate average mass.

\begin{figure}[t] \centering
 \includegraphics[width=\breite]{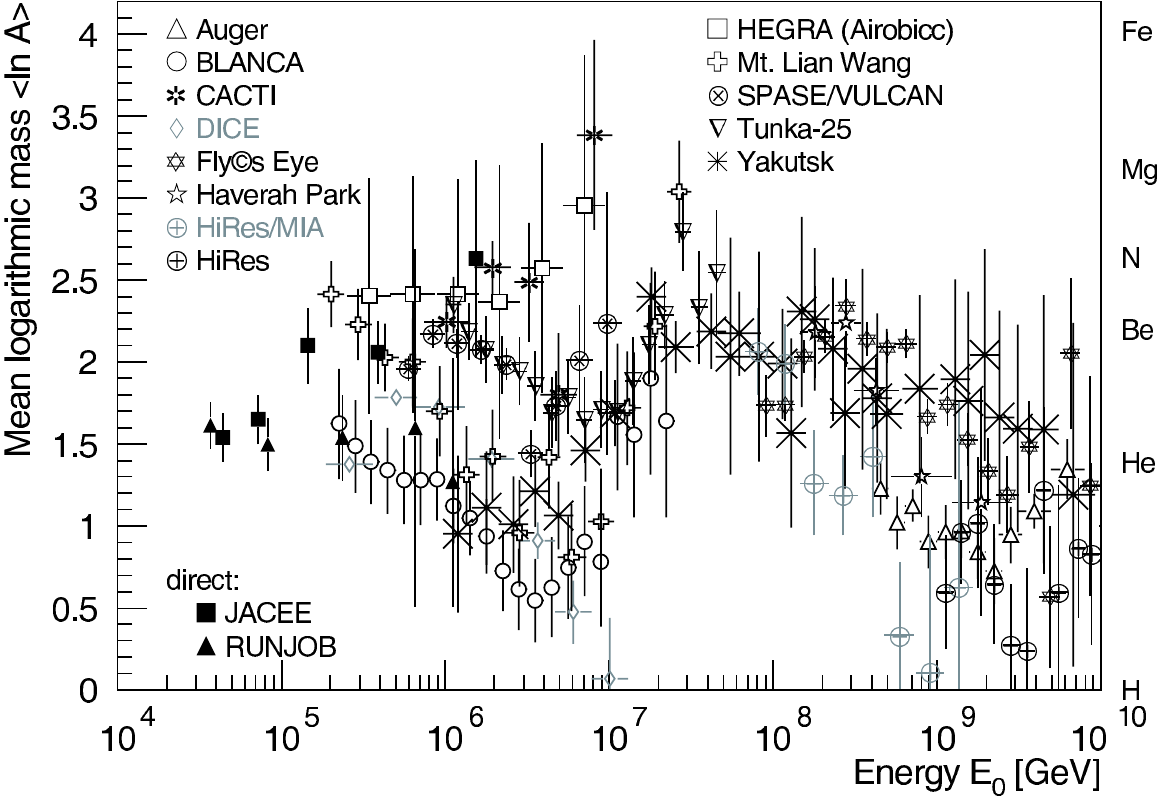}
 \includegraphics[width=\breite]{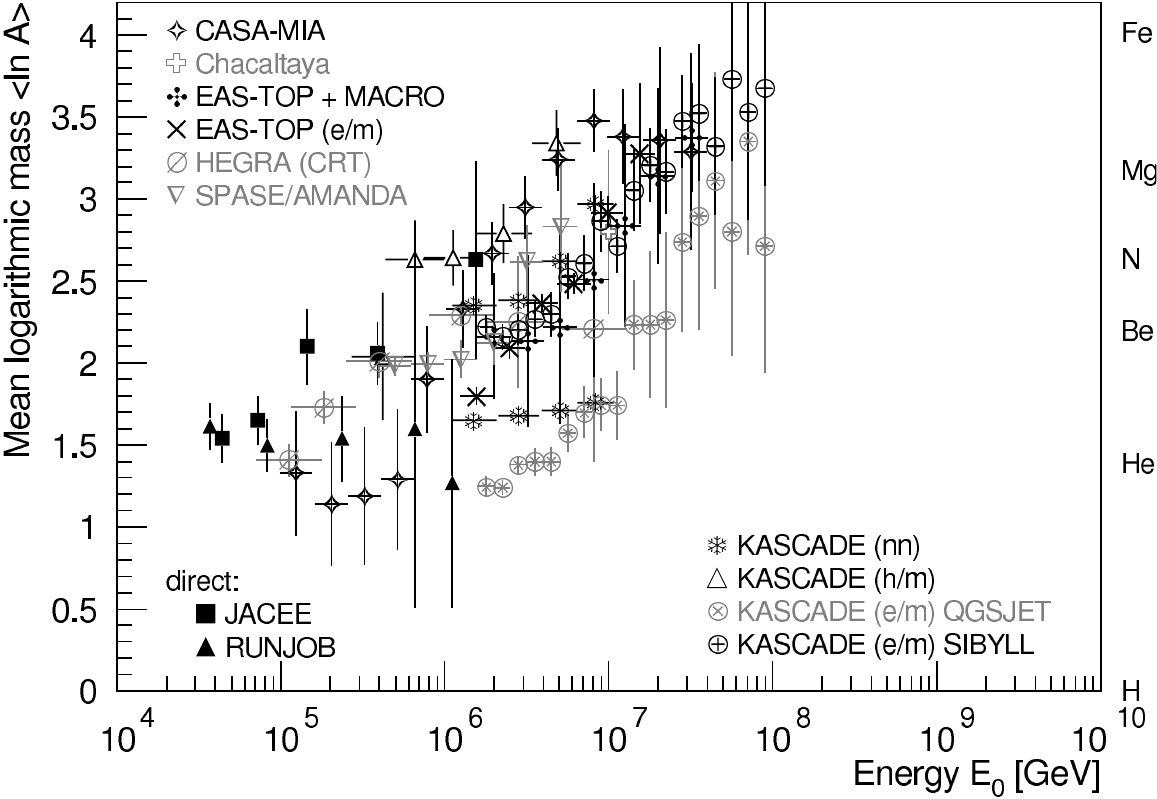}
  \Caption{\TTop: Mean logarithmic mass of cosmic rays derived from the average
           depth of the shower maximum, see \fref{xmax}.
	   The hadronic interaction model QGSJET~01 is used to interpret the
	   measurements.  For comparison, results from direct measurements are
	   shown as well from the JACEE \protect\cite{jaceemasse} and RUNJOB
	   \protect\cite{runjob05} experiments. 
           \BBottom: Mean logarithmic mass of cosmic rays derived from the
	   measurements of electrons, muons, and hadrons at ground level.
	   Results are shown from
	   CASA-MIA \protect\cite{casam},
	   Chacaltaya \protect\cite{chacaltaya},
	   EAS-TOP electrons and GeV muons \protect\cite{eastop-knee},
	   EAS-TOP/MACRO (TeV muons) \protect\cite{eastop-macro-lna},
	   HEGRA CRT \protect\cite{hegracrt},
	   KASCADE electrons and muons interpreted with two hadronic
	   interaction models \protect\cite{Antoni:2005wq},
	   hadrons and muons \protect\cite{kascadehm}, 
	   as well as an analysis combining different observables with a 
	   neural network \protect\cite{rothnn}, and
	   SPASE/AMANDA \protect\cite{spaseamandalna}.
  \label{masse}}
\end{figure}

Knowing the average depth of the shower maximum for protons $X_{\rm max}^p$ and
iron nuclei $X_{\rm max}^{Fe}$ from simulations, the mean logarithmic mass is
derived in the superposition model of air showers from the measured
$X_{\rm max}^{\rm meas}$ values using
$\lnA=(X_{\rm max}^{\rm meas}-X_{\rm max}^{p})/(X_{\rm max}^{Fe}-X_{\rm max}^{p}) \cdot \ln
A_{Fe}$.  The corresponding $\lnA$ values, obtained from the results shown in
\fref{xmax}, are plotted in \fref{masse} (\ttop) as function primary energy
using the hadronic interaction model QGSJET~01 to interpret the observed data.  
For comparison, also results of direct measurements are shown (JACEE and
RUNJOB).

In the figure three energy ranges may be distinguished for the indirect
measurements.  Below about $4\times10^6$~GeV the individual experiments seem to
indicate a decrease of $\lnA$ with energy, while above this energy up to about
$4\times10^7$~GeV an increase with energy is exhibited. At the highest energies
$E\ga4\times10^7$~GeV, again a decrease with energy can be stated.

Results of measurements of electrons, muons, and hadrons at ground level
interpreted with the hadronic interaction code QGSJET~01 are compiled in
\fref{masse} (\bbottom).  They yield a clear increase of the mean logarithmic
mass as function of energy.  There seems to be some tension between the results
obtained through the observation of the average depth of the shower maximum
shown in the top panel and values derived from particle ratios measured at
ground level depicted in the bottom panel \cite{Hoerandel:2002yg,Hoerandel:2003vu}.  In
particular, at energies below the knee ($E\ga4\times10^6$~GeV) the decrease of
$\lnA$ as derived from some \Xmax measurements is not visible in the particle
ratio results.

Using the latest version of QGSJET~II does not change the situation
qualitatively. The threefold
structure of the results obtained is about the same as using QGSJET~01.
The main difference are the absolute $\lnA$ values which
are shifted upwards by about 0.8 units for QGSJET~II-3 with respect to
QGSJET~01. Using the hadronic interaction model SIBYLL~2.1 yields about the same
values as for QGSJET~II.
Using lower inelastic hadronic cross sections in the QGSJET~01 code and larger
values for the elasticity of hadronic interactions the discrepancies between
\Xmax measurements and particle ratios at ground can be reduced
\cite{Hoerandel:2003vu,isvhecri04wq}.

\subsubsection*{Spectra for Elemental Groups}
In addition to the mean mass as discussed above, it is interesting to
investigate the energy spectra for individual elements or at least groups of
elements.

Information on the flux of primary protons can be inferred from the measurement
of the flux of unaccompanied hadrons at ground level.  With the KASCADE hadron
calorimeter the energy spectrum of single hadrons close to sea level has been
measured in the energy range from 100~GeV up to 50~TeV \cite{kascadesh}.  Based
on simulations using the CORSIKA code with the hadronic interaction model
QGSJET~01~\cite{Kalmykov89a-e,*Kalmykov:1993qe,*Kalmykov:1997te}
the energy spectrum of primary hadrons in the
energy range from 100~GeV to 1~PeV has been derived.  Over the whole four
decades in energy it can be described by a single power law.

\begin{figure}[t] \centering
 \includegraphics[width=0.49\textwidth]{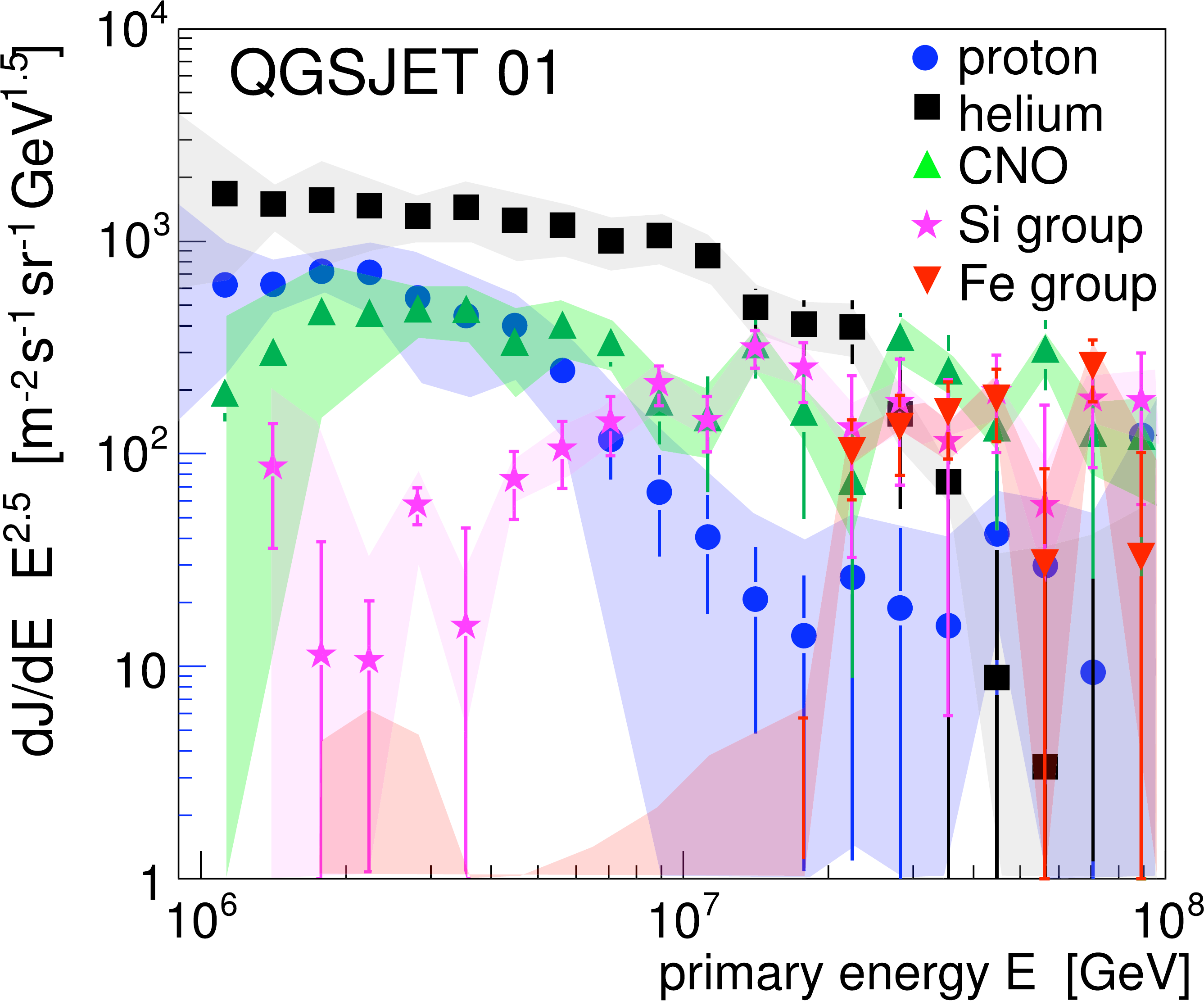}\hspace*{\fill}
 \includegraphics[width=0.49\textwidth]{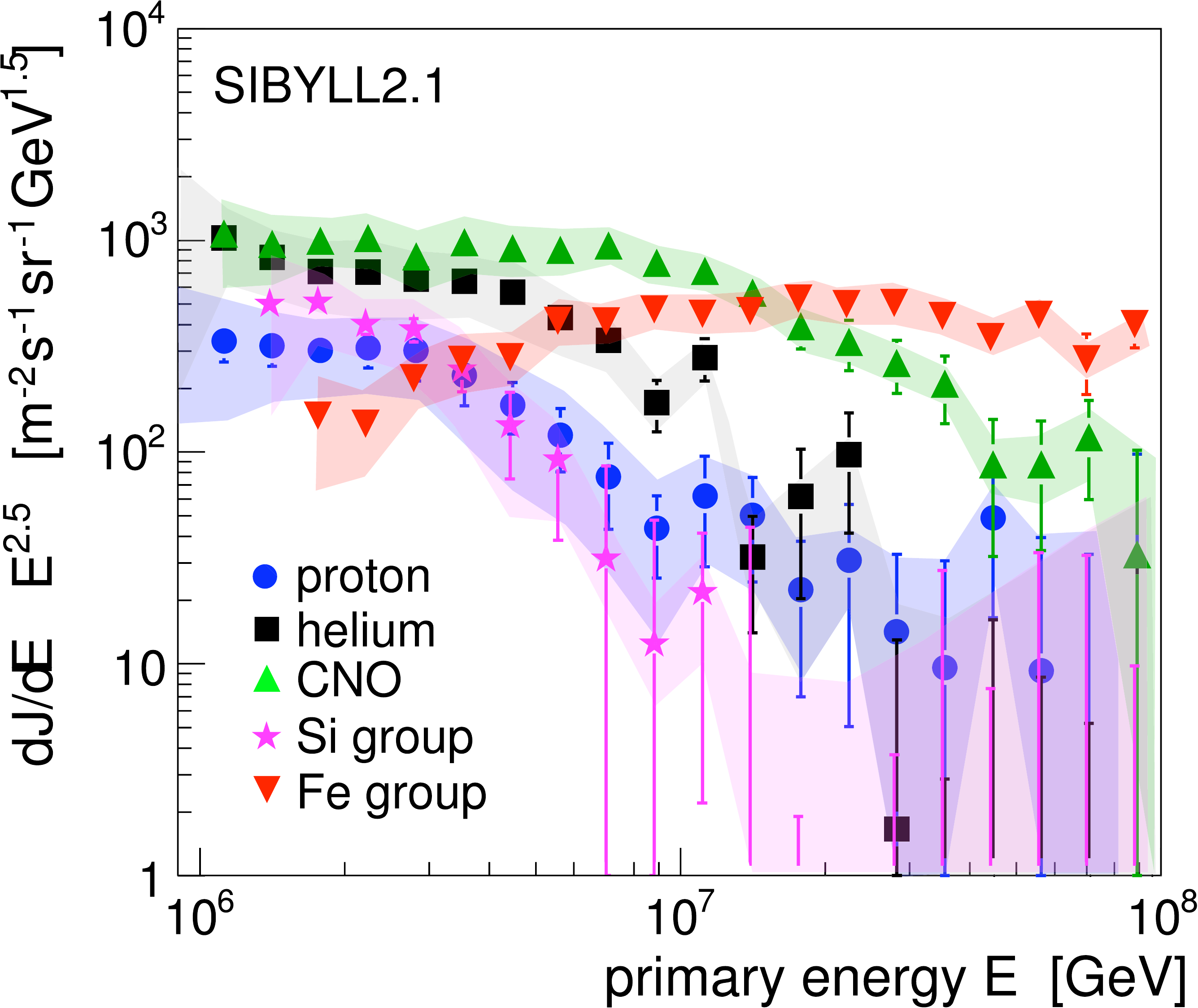}
 \Caption{Cosmic-ray energy spectrum for five groups of elements as
	 reconstructed by the KASCADE experiment using the hadronic interaction
	 models QGSJET~01 (\lleft) and SIBYLL~2.1 (\rright) to interpret the
	 measured data \protect\cite{Antoni:2005wq}.  \label{ulrichspek}}
\end{figure}

The KASCADE experiment used the number of electrons and muons ($E_\mu>230$~MeV)
measured in the scintillator array to reconstruct energy spectra for five primary
elemental groups \cite{Antoni:2005wq}.  Starting point of the analysis is the
correlated frequency distribution of the number of electrons $N_e$ and the
number of muons $N_\mu$. Unfolding algorithms were applied to derive energy
spectra for elemental groups.  For the analysis the primary particles H, He, C,
Si, and Fe were chosen as representatives for five mass groups.  Details of the
analysis and the used unfolding methods can be found in Ref.~\cite{Antoni:2005wq}.

The spectra obtained are presented in \fref{ulrichspek}. To check the influence
of the description of hadronic interactions in the atmosphere on the result,
the same experimental data were unfolded using two interaction models, namely
QGSJET and SIBYLL. The corresponding results are displayed in the figure.  The
resulting all-particle spectra for both models show a knee at about 4 PeV and
coincide within their statistical errors.  The decrease of light elements
across the knee, i.e.\ the occurrence of knee-like features in the light element
spectra is revealed independently of the used simulation code, as can bee seen
in \fref{ulrichspek}.  In contrast, the spectra of silicon and iron groups
differ significantly and look quite unexpected. This can be understood by
judging the ability of the simulations to describe the data.  It turns out that
both interaction models fail to reproduce the overall correlation between $\lg
N_e$ and $\lg N_\mu$ as observed in the data.  In the case of QGSJET
simulations the predictions are incompatible with the data in the low energy
regime (simulations look too heavy), for SIBYLL incompatibility occurs at
higher energies (simulations look too light).  Summarizing this analysis, for
the first time energy spectra for groups of elements were reconstructed from
air shower data.  The spectra indicate that the knee in the all-particle
spectrum is due to fall-offs in the light element spectra resulting in a
heavier composition above the knee. 

\begin{figure}[t] \centering
 \includegraphics[width=0.6\textwidth]{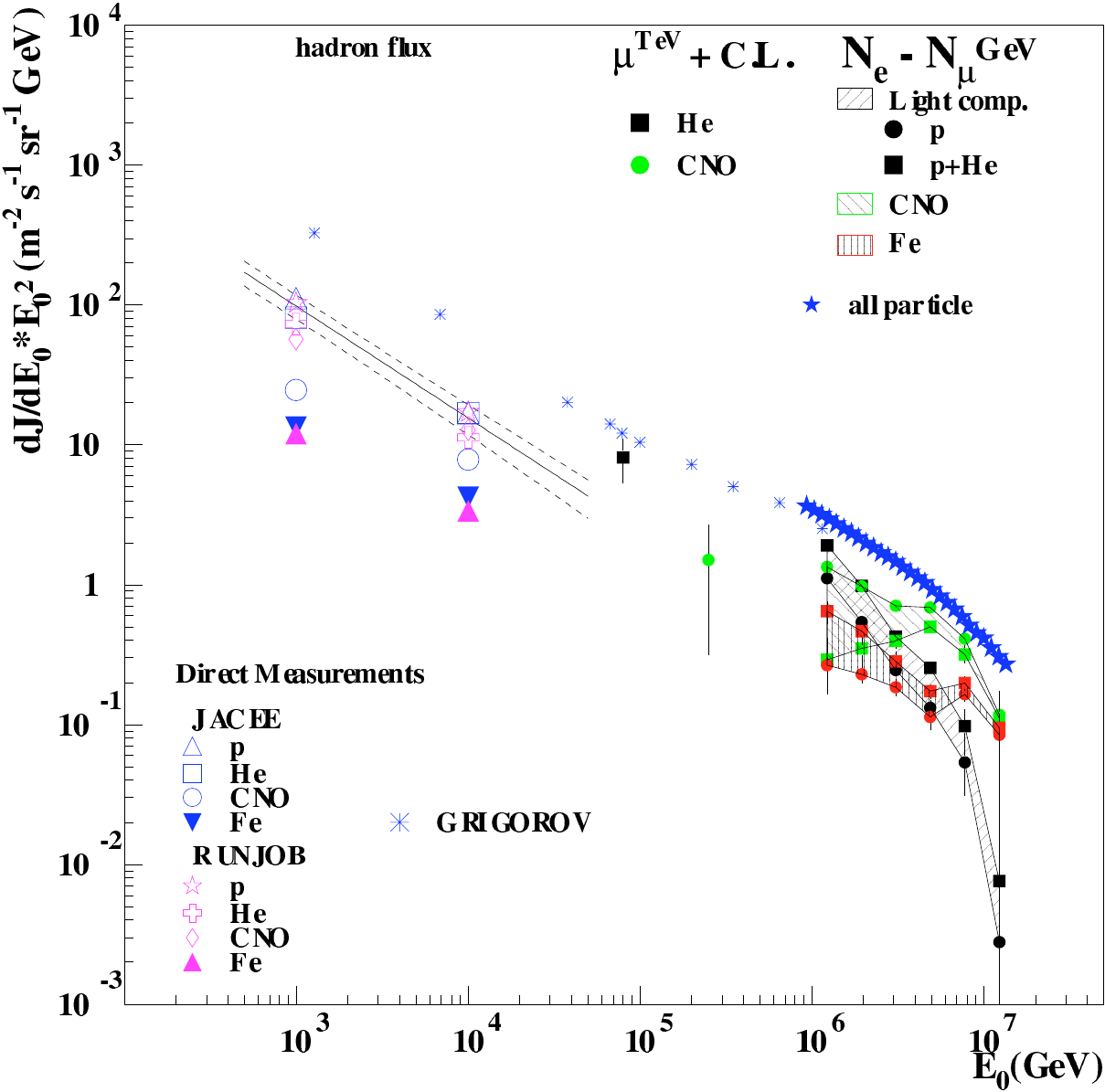}
 \Caption{Energy spectrum of cosmic rays as obtained by several detection
	 methods applied in the EAS-TOP experiment \protect\cite{eastopspec} compared
	 to results of direct measurements.\label{eastopsum}}
\end{figure}

The EAS-TOP collaboration combined several detection systems to obtain
information on the energy spectra of primary cosmic rays. The results are
summarized in \fref{eastopsum}.

The proton spectrum in the energy range 0.5 - 500 TeV \cite{eastopsh} has been
obtained from the measurement of unaccompanied hadrons with a calorimeter,
taking into account the contribution of helium nuclei as obtained by direct
measurements \cite{jaceephe,runjobapp01}.  The proton spectrum is described
over the whole energy range by a single power law.

The helium and CNO fluxes in the energy region from 80 to 200~TeV have been
studied from measurements of the \Cerenkov light and TeV muons registered with
the underground MACRO experiment \cite{eastopcermacro}.  Primaries are selected
through their energy/nucleon by means of the TeV muon information.  The shower
energy is inferred from the measurement of the \Cerenkov light yield at
distances from 125 to 185~m from the shower core.  The flux for p+He at 80~TeV
and for p+He+CNO at 250~TeV was obtained.  By subtracting the measured proton
flux the following values were calculated:
$\Phi_{He}(80~\mbox{TeV}) = (12.7 \pm 4.4) \times
10^{-7}$~m$^{-2}$sr$^{-1}$s$^{-1}$GeV$^{-1}$ and
$\Phi_{CNO}(250~\mbox{TeV}) = (0.24 \pm 0.19) \times
10^{-7}$~m$^{-2}$sr$^{-1}$s$^{-1}$GeV$^{-1}$.

The all-particle energy spectrum is obtained from the measured shower size
spectra in the knee region \cite{eastope}, showing the angular (i.e.\ depth)
dependence of the knee position. The knee is observed at $N_e = 10^{6.1}$ in
the vertical direction, corresponding to a primary energy $E_0 = (2-4) \times
10^{15}$~eV, and intensity of $10^{-7}$~m$^{-2}$s$^{-1}$sr$^{-1}$ with about
20\% uncertainty. The obtained power law indices of the energy spectrum are:
$\gamma_1= 2.76 \pm 0.03$ and $\gamma_2=3.19\pm0.06$, below and above the knee,
respectively.

A composition analysis at knee energies was performed for vertical showers
($1.00<\sec(\Theta)<1.05$) through measurements of the number of electrons
$N_e$ and the number of muons with energies above 1~GeV recorded in the muon
detector at core distances $r=180-210$~m ($N_{\mu180}$). The experimental
$N_{\mu180}$ distributions, measured in different shower size intervals are
fitted with simulated data to obtain energy spectra for groups of elements.
Intrinsic fluctuations and measurement accuracies allow a three component
analysis: light (constructed either with protons, and a mixture of 50\% proton
and 50\% helium), intermediate (CNO), and heavy (Fe).  The shaded areas in
\fref{eastopsum} indicate the energy dependence of the flux thus obtained for
the three components.  

The Tibet air shower array has the advantage to be located at high altitude
(4300~m, 606~\gcm2). It comprises a scintillation counter array as well as
emulsion chambers and burst detectors. The data were used to derive spectra for
primary protons and helium nuclei, see \fref{elementspek}
\cite{tibetbdp,tibetasg03}. However, one has to keep in mind that only a few
hundred events remain after quality cuts and are included in the analysis,
which may indicate that the results are limited by their statistical
significance.

Similar to KASCADE and EAS-TOP, the GRAPES-3 experiment uses the correlation
between the registered number of electrons and muons to derive the flux for
individual elemental groups. Spectra for protons, helium, as well as the CNO,
silicon, and iron groups have been obtained \cite{grapes05}, see \fref{elementspek}.

\subsubsection*{Highest Energies}

In 1993 observations with Fly's Eye \cite{Bird:1993yi} gave the first
indication of a systematic change of the cosmic-ray mass composition
at very high energy. Analyzing the mean depth of shower maximum, $\langle
X_{\rm max}\rangle$, a change from an iron dominated composition at
$10^{17}$\,eV to a proton dominated composition at $10^{19.3}$\,eV was
found. However, an analysis entirely based on the mean $X_{\rm max}$
is strongly model dependent 
(e.g.~\cite{Gaisser:1993ix,Knapp96a,Ding:1997bv,Heck:2002yf,Hoerandel:2003vu}).

\begin{figure}[t]
\includegraphics[width=0.5\textwidth]{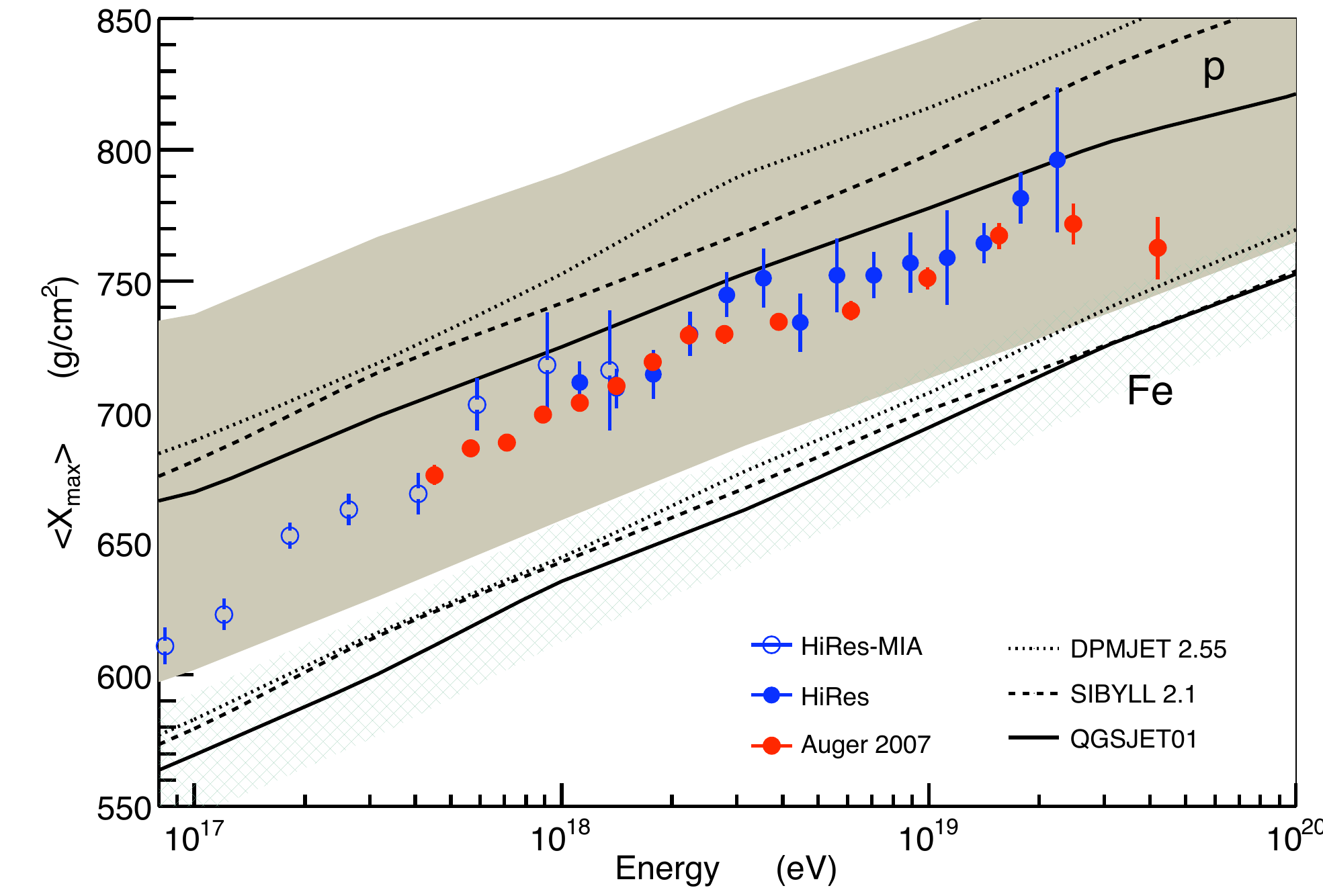}\hspace*{\fill}
\includegraphics[width=0.44\textwidth]{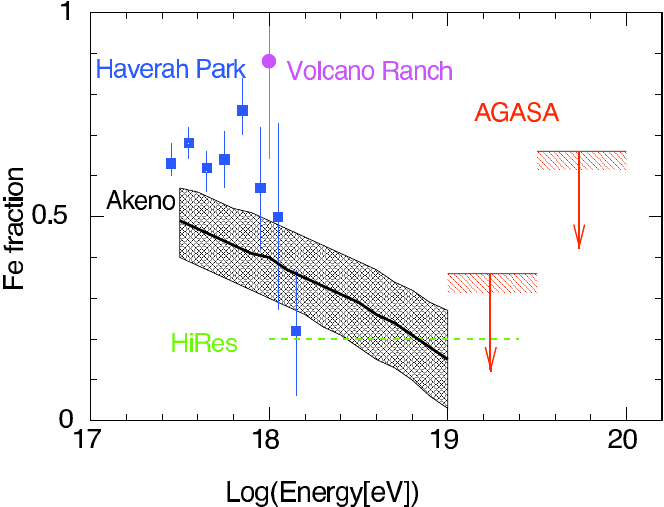}
\Caption{\label{fig:mean-Xmax-2008} \LLeft: Compilation of
  fluorescence-based measurements of the mean $X_{\rm max}$ of very
  high-energy air showers. The data are from Auger~\protect\cite{Unger:2007mc},
  HiRes-MIA~\protect\cite{AbuZayyad:2000ay}, and HiRes
  stereo~\protect\cite{Abbasi:2004nz}.  The model predictions are calculated
  with CORSIKA~\protect\cite{Heck98a} and are taken from
  \protect\cite{Heck:2002yf,Heck2006private}.  The QGSJET predictions on the
  shower-to-shower fluctuations of the depth of maximum are indicated
  by the shaded (cross-hatched) area for proton (iron) primaries.
  \RRight: Fraction of iron, if data are interpreted with a
  hypothetical bi-modal composition of proton and iron primaries
  only. Shown are Akeno data \protect\cite{Hayashida:1995tu}, as well as results from
  HiRes \protect\cite{Abbasi:2004nz}, Haverah Park \protect\cite{Ave:2002gc}, and Vulcano Ranch
  \protect\cite{Dova:2002yt,Dova:2003an}.  Upper limits to the iron fraction are
  obtained from AGASA data \protect\cite{Shinozaki:2006kk}.
         \label{fig:CompositionMuons}
       }
\end{figure}

In Fig.~\ref{fig:mean-Xmax-2008} (\lleft) a compilation of measurements of
$\langle X_{\rm max}\rangle$ is shown together with model predictions. 
Adopting the QGSJET~01 model \cite{Kalmykov89a-e,*Kalmykov:1993qe,*Kalmykov:1997te} the
conclusions of \cite{Bird:1993yi} still hold, though a mixed composition is
expected at $10^{17}$\,eV. On the other hand, on the basis of models like
SIBYLL 2.1 \cite{Engel:1992vf,Fletcher:1994bd,Engel:1999db} or DPMJET 2.55
\cite{Ranft:1994fd} a much more moderate change of the composition is derived.
The model ambiguity of the interpretation of $\langle X_{\rm max}\rangle$ can
be resolved to some degree by studying the measured distribution of $X_{\rm
max}$ \cite{Gaisser:1993ix,Abbasi:2006vp}.

In contrast to the old measurements of Fly's Eye \cite{Bird:1993yi}
and Yakutsk \cite{Dyakonov:1993ab} the HiRes data indicate a change
from an iron-like to a proton dominated composition already at
$10^{18}$\,eV. The two independent measurements are consistent in the
overlap region.  The large elongation rate of the low-energy data of
$\sim93$\,g/cm$^2$ \cite{AbuZayyad:1999xa} can only be understood in
terms of a change of composition (see
\sref{sec:ShowerProperties}). Any model with scaling violations
will predict a change to a lighter composition
\cite{Linsley:1981gh,Ding:1997bv,Alvarez-Muniz:2002ne}. Also the muon
density measured with the MIA detector \cite{Borione:1994iy} in the
HiRes-MIA setup indicates a change from a heavy to light
composition. The observed muon densities, however, are higher or
similar than the expectation for iron primaries and not compatible with
medium or light nuclei \cite{AbuZayyad:1999xa,AbuZayyad:2000ay}.

The new Auger data on the average depth of shower maximum
\cite{Unger:2007mc} are, within the systematic uncertainties of about
$~ 15$\,g/cm$^2$ in good agreement with the published HiRes data. On
the other hand, if one would just analyze Auger data, a break in the
elongation rate is found at a higher energy, at about
$10^{18.35}$\,eV. The elongation rate is $71\pm 5$\,g/cm$^2$ below
this break and $40 \pm 4$\,g/cm$^2$ above \cite{Unger:2007mc}. The
highest energy point of the Auger data is smaller than expected
from a constant elongation rate. It is calculated from 13 observed showers and
could indicate a transition to heavier elements as one would expect in
case of a rigidity-dependent maximum energy of sources.

The right panel of Fig.~\ref{fig:CompositionMuons} shows a compilation of 
composition measurements based on particle detectors, as published in
\rref{Shinozaki:2006kk}. The Akeno result given here is re-scaled according to
the predictions of the QGSJET model \cite{Shinozaki2006private}.\footnote{The
original composition analysis of Akeno data \cite{Hayashida:1995tu} in the
energy range $10^{16.5} - 10^{19.5}$\,eV appeared to be in contradiction to the
Fly's Eye composition interpretation \cite{Bird:1993yi}, see also the analysis
in \cite{Dawson:1998kk}.} 

The muon density of ultra high-energy showers measured with AGASA was analyzed
in \rref{Shinozaki:2003ab}.  Comparing the data with model predictions from
AIRES \cite{Sciutto:1999jh} and QGSJET~01 the following limits on a two
component composition were derived: less than 35\% iron in the energy range
$10^{19} - 10^{19.5}$\,eV and less than 76\% at higher energy (90\% c.l.)
\cite{Shinozaki:2003ab,Shinozaki:2006kk}.  

A re-analysis of Haverah Park data was done in \rref{Ave:2002gc}. The
authors use the sensitivity of the steepness of the lateral particle
distribution to the shower development height, which in turn depends
on the depth of shower maximum as composition sensitive observable. It
was found that the predictions of CORSIKA with the QGSJET~98 model give
a good description of the data if a two-component composition with
about ($66\pm2$)\% iron is used in the energy range from $2\times
10^{17}$ to $10^{18}$\,eV. At higher energy (from $10^{18}$ to
$2\times 10^{18}$\,eV), indications are seen for a transition to a
lighter composition (see Fig.~\ref{fig:CompositionMuons}).  This is
supported by the number of inclined showers with $E>10^{19}$\,eV that
have triggered the Haverah Park array \cite{Ave:2001xn}.  The data
analyzed in \cite{Ave:2001xn} agree well with simulations assuming all
primary particles are protons, though no mass composition study was
done.

On the other hand, a first study of the time structure of Haverah Park
showers with zenith angles less than 45$^\circ$ finds 
a more iron-dominated composition in the same energy range
\cite{Ave:2003ab}.  Also a re-analysis of Volcano Ranch data similar
to the Haverah Park favors a large fraction of iron
\cite{Dova:2004nq}.  Using $\sim 370$ showers in the energy range from
$5\times 10^{17}$ to $10^{19}$\,eV a fraction of $\sim 90$\% ($75$\%)
iron is found for a two component composition and QGSJET~98 (QGSJET
01).

The discrepancy between muon density-based composition measurements
and others based on features of the longitudinal profile underlines
the shortcomings of the hadronic interaction models currently
available. There seems to be a systematic deficit of muons predicted
in simulations in comparison with data
\cite{AbuZayyad:1999xa,Engel:2007cm,*Schmidt:2009ge}. First progress in addressing
this problem has been made in \cite{Pierog:2006qv} by increasing the 
number of pair-produced baryons in the simulation.

\subsubsection*{Ultra High-Energy Photons}

\begin{figure}[t] 
 \includegraphics[width=0.43\textwidth]{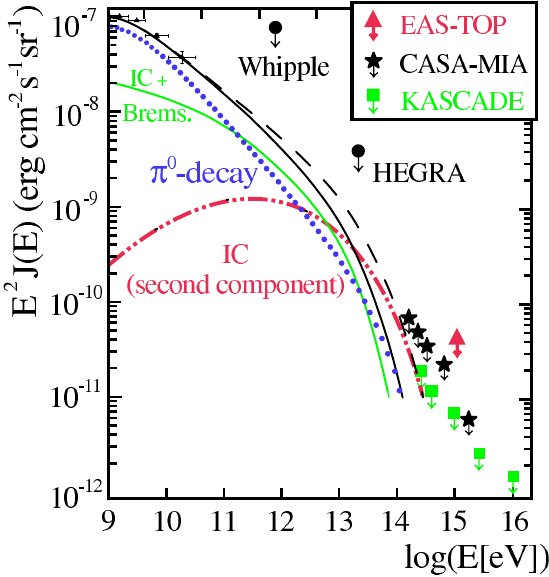}\hspace*{\fill}
 \includegraphics[width=0.55\textwidth]{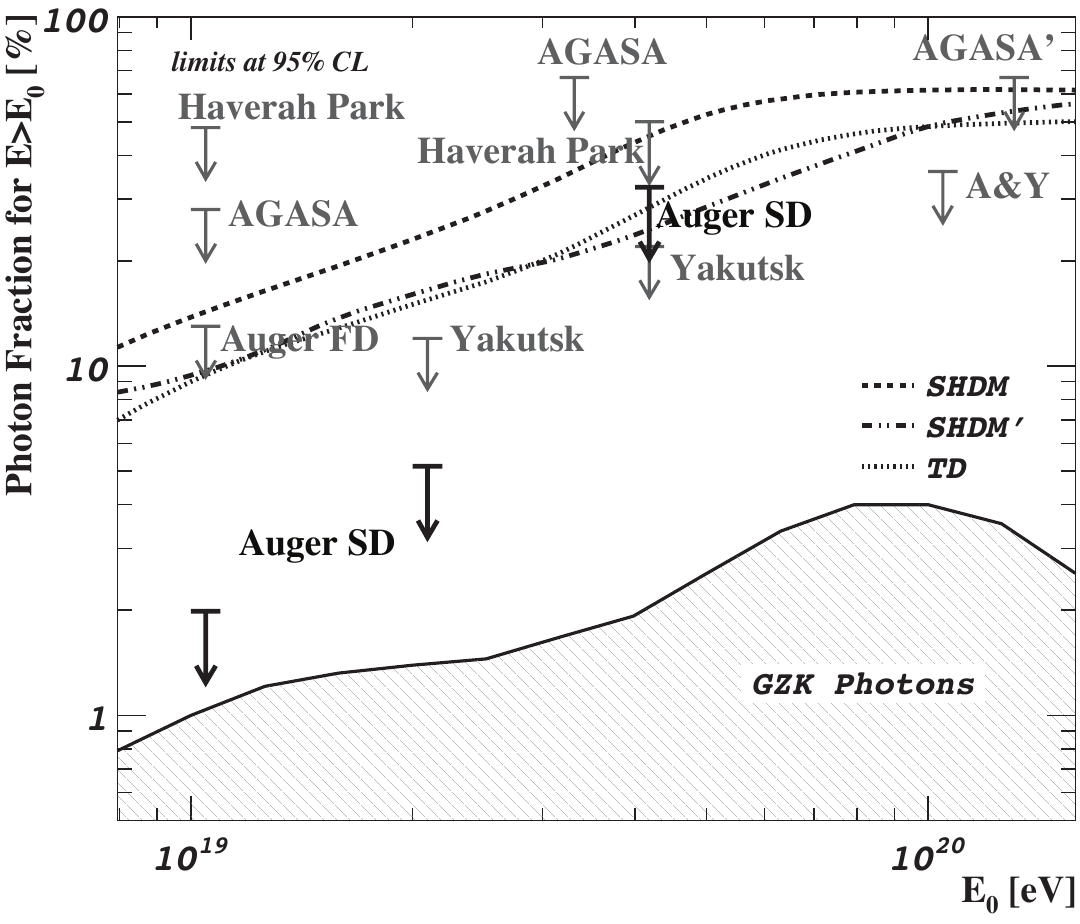}
 \Caption{\LLeft:
	  Upper limits for the photon flux derived from air shower observations
	  by the  CASA-MIA \protect\cite{casamiagamma}, EAS-Top \protect\cite{eastopgamma}, and
	  KASCADE \protect\cite{Schatz:2003x1} experiments compared to theoretical
          predictions \protect\cite{Aharonian:2000iz}.
          \RRight: 
          Upper limits on the fraction of photons in the integral cosmic-ray
          flux compared to predictions for GZK photons and top-down scenarios 
          (SHDM, SHDM', TD)
          \protect\cite{Aglietta:2007yx}. Experimental data are from the Auger surface
          detectors (Auger SD) \protect\cite{Aglietta:2007yx} and 
          a hybrid analysis (Auger FD) \protect\cite{Abraham:2006ar},
          Haverah Park \protect\cite{haverahphoton},
          AGASA (marked AGASA and AGASA') \protect\cite{Shinozaki:2002ve,Risse:2005jr},
          AGASA and Yakutsk (A\&Y) \protect\cite{Rubtsov:2006tt}, as well as
          Yakutsk \protect\cite{yakutskphoton}.
          \label{gammaflux}}
\end{figure}

Using gamma-ray telescopes, such as MAGIC, H.E.S.S., or VERITAS, photons
with energies of about 50~TeV have been observed from several sources in the sky, e.g.\
\cite{Weekes:2008dr,Hinton:2008ka}.
At higher energies various air shower experiments searched for gamma rays.
Air showers induced by primary photons develop an almost pure electromagnetic
cascade. Experimentally they are identified by their relatively low muon
content or their relatively deep shower maximum. Since mostly electromagnetic
processes are involved in the shower development, the predictions are more
reliable and don't suffer from uncertainties in hadronic interaction models.  

Results up to energies of $10^{16}$~eV are summarized in \fref{gammaflux}
(\lleft) \cite{Schatz:2003x1}.
There are no indications of a substantial fraction of gamma-rays in the
high-energy cosmic-ray flux.

The longitudinal profile of the highest energy event observed by Fly's
Eye ($E \sim 3.2\times10^{20}$\,eV) \cite{Bird:1995uy} has been
studied by several groups.  Comparing the measured shower profile with
Monte Carlo simulations shows that this event is well described by a
hadronic primary particle
\cite{Bird:1995uy,Halzen:1995gy,Risse:2004mx}.  However, due to the
large reconstruction uncertainty of the atmospheric depth of the
shower profile, a photon cannot be excluded \cite{Risse:2004mx}.

The deeply penetrating muon component of inclined showers 
of hadronic origin is employed in an analysis of
Haverah Park data in \cite{Ave:2000nd,Ave:2001xn}. Using the primary
cosmic-ray flux parametrization of \cite{Nagano:2000ve} less than 48\%
of the observed events above $10^{19}$\,eV can be photons (95\% c.l.).
At energies above $4\times10^{19}$\,eV this limit is 50\%.

Based on the analysis of muons observed in high-energy showers at
AGASA the following upper limits were derived in
\cite{Shinozaki:2002ve,Shinozaki:2003ab}: 34\%, 59\% and 63\% for
primary energies above $10^{19}$, $10^{19.25}$, and $10^{19.5}$\,eV,
respectively (95\% c.l.). A separate analysis of the 6 highest energy
events of AGASA was performed in \cite{Risse:2005jr}. Using a new
method that accounts for the arrival direction of each individual
shower, a limit of 67\% at 95\% c.l.\ could be derived for $E > 1.25 \times 10^{20}$\,eV.

This new analysis method was also employed in a recent study of
Auger shower longitudinal profile data \cite{Abraham:2006ar}. No more
than 16\% photons are expected at 95\% c.l. It should be noted that
this limit as well as the one derived in \cite{Risse:2005jr} are
independent of assumptions on hadronic multiparticle production at
very high energy. It relies only on the simulation of photon showers
which is much better under control than that of hadron-induced
showers.

In \rref{Rubtsov:2006tt}, the scintillator signals of 10 showers above
$10^{20}$\,eV from AGASA and Yakutsk were analyzed to derive a photon
fraction limit. The authors report the strongest limit on the photon
fraction for showers above $10^{20}$\,eV available so far, less than
$36$\% photons at 95\% c.l. To obtain this limit, new energies are
assigned to the considered showers, partially being very much
different from those originally reconstructed. This shows the
importance of simulating and understanding detector effects which can
only be done in a limited way in such studies. A confirmation by the
AGASA and Yakutsk experiments would be very important to establish this
limit beyond doubt.

A compilation of recent upper limits on the contribution of photons to the
all-particle flux is shown in \fref{gammaflux} \cite{Aglietta:2007yx}.  The best
photon limits are the latest results of the Pierre Auger Observatory
\cite{Aglietta:2007yx} setting rather strong limits on the photon flux.  They are
based on measurements with the Auger surface detectors, taking into account
observables sensitive to the longitudinal shower development, the signal rise
time, and the curvature of the shower front.  The photon fraction is smaller
than 2\%, 5.1\%, and 31\% above energies of $10^{19}$, $2\cdot10^{19}$, and
$4\cdot10^{19}$~eV, respectively with 95\% confidence level.

In top-down scenarios for high-energy cosmic rays, the particles are decay
products of super-heavy objects. The decay process yields relatively high fluxes of
photons, a typical feature of such models \cite{Kachelriess:2004ax}.  Several
predictions are shown in the figure
\cite{Aloisio:2003xj,Ellis:2005jc}. These scenarios are strongly disfavored by
the recent Auger results. It should also be noted that the upper limits are 
already relatively close to the fluxes expected for
photons originating from the GZK effect \cite{Gelmini:2005wu}, shown in the figure as 
shaded area.

\subsubsection*{Ultra High-Energy Neutrinos}

\begin{figure}[t] \centering
 \includegraphics[width=0.6\textwidth]{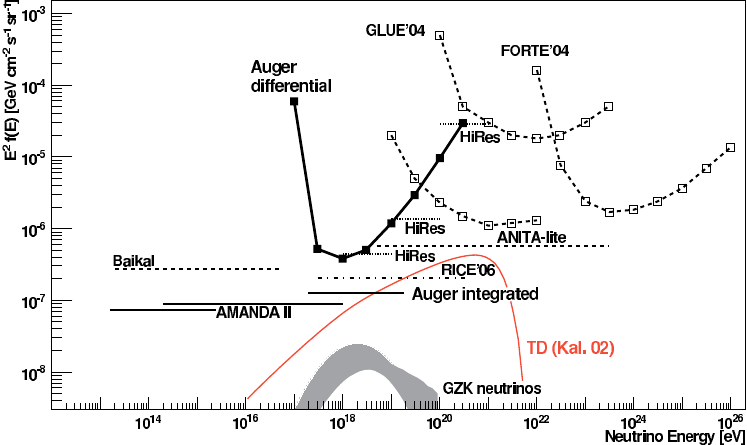}
 \Caption{Limits at 90\% confidence level for a diffuse flux of $\nu_\tau$
          assuming a 1:1:1 ratio of the three neutrino flavors at Earth
          \protect\cite{Abraham:2007rj,Kampert:2008pr}. The experimental results are
          compared to predictions for GZK neutrinos and a top-down model
          \protect\cite{kalashev}.
          \label{neutrinoflux}}
\end{figure}

The detection of ultra high-energy cosmic neutrinos is a long-standing
experimental challenge. Many experiments are searching for such neutrinos, and
there are several ongoing efforts to construct dedicated experiments to detect
them \cite{Halzen:2002pg,falckenar}. Their discovery would open a new window to the
Universe \cite{Becker:2007sv}.  However, so far no ultra high-energy neutrinos have
been detected. 
\footnote{Neutrinos produced in air showers (atmospheric neutrinos)
   \cite{superkatmospheric}, in the Sun \cite{superksolar,sno04}, and during
   supernova 1987A \cite{hirata,bionta} have been detected, but are at
   energies well below our focus. }

Due to interactions in the source region or during the propagation processes,
ultra high-energy cosmic rays are expected to be accompanied by ultra
high-energy neutrinos.  The neutrinos are produced with different abundances
for the individual flavors, e.g.\ pion decay leads to a ratio
$\nu_e:\nu_\mu=2:1$. However, due to neutrino oscillations the ratio expected
at Earth is $\nu_\tau:\nu_\mu:\nu_e=1:1:1$.

To discriminate against the huge hadronic background in air shower detectors,
neutrino candidates are identified as nearly horizontal showers with a
significant electromagnetic component 
\cite{Bertou:2001vm,*Feng:2001ue,*Zas:2005zz,*Aramo:2004pr,*Fargion:2005fa,*Miele:2005bt,*Gora:2007nh}.
The Pierre Auger Observatory is sensitive to Earth-skimming tau neutrinos that
interact in the Earth's crust. Tau leptons from $\nu_\tau$ charged-current
interactions can emerge and decay in the atmosphere to produce a nearly
horizontal shower with a significant electromagnetic component.  Recent results
from the Pierre Auger Observatory together with upper limits from other
experiments are presented in \fref{neutrinoflux}.  Assuming an $E_\nu^{-2}$
differential energy spectrum, the Auger Collaboration derived a limit at 90\% confidence level of
$E_\nu^2 \, \mbox{d}N_{\nu_\tau}/\mbox{d}E_\nu < 1.3 \cdot 10^{-7}$~GeV
cm$^{-2}$ s$^{-1}$ sr$^{-1}$ in the energy range between $2\cdot10^{17}$ and
$2\cdot 10^{19}$~eV \cite{Abraham:2007rj}.

According to top-down models for ultra high-energy cosmic rays a large flux of
ultra high-energy neutrinos is expected. As an example, the predictions of a
model \cite{kalashev} are shown in the figure as well. This model is
disfavored by the recent upper limits.  It should also be noted that the
current experiments are only about one order of magnitude away from predicted
fluxes of GZK neutrinos.

\Section{Anisotropy} \label{anisosec}

The search for anisotropies in the arrival direction of cosmic rays on
different angular scales can contribute to the understanding of the cosmic-ray
origin, in particular the identification of source regions or individual sources.

\subsubsection*{Galactic Cosmic Rays}

Large-scale anisotropies are connected to the propagation process of cosmic
rays in the Galaxy, while small-scale anisotropies would be a hint towards 
cosmic-ray sources.  However, it has to be considered that the Larmor
radius \eref{larmorradeq}
of protons with an energy around 1~PeV in the galactic magnetic
field ($B=3~\mu$G) is of the order of 0.4~pc.  Hence, it is not expected to
find any point sources for galactic cosmic rays.  The situation changes for the
highest energies, see below.

An excess of charged particles with energies above $10^{15}$~eV from the
direction of a SNR (Monogem ring, $d\approx300$~pc) has been reported
\cite{maketani-points} and later withdrawn \cite{Chilingarian:2006cn}.  This
supernova remnant with an age of about $10^5$~yr has been suggested as possible
single source of galactic cosmic rays \cite{ewmonogem}.  A signal could not be
confirmed neither by the KASCADE experiment \cite{kascade-points} investigating
cosmic rays with energies above 0.3~PeV nor by the Tibet experiment, looking
for PeV $\gamma$-ray emission \cite{tibet-monog}.

The Tibet air shower experiment has performed a northern sky survey, looking
for TeV $\gamma$-ray point sources in a declination range from 0\deg to 60\deg
\cite{tibet-aniso}.  A small excess is found, most likely caused by well known
$\gamma$-ray sources such as in the Crab Nebula and Mrk 421. 

At higher energies, the KASCADE experiment has performed a detailed search for
point sources, covering the whole visible sky at energies $E_0>0.3$~PeV
\cite{kascade-points}.  The visible sky has been divided in cells with a size
of 0.5\deg.  Two distributions have been investigated, all events and a
selection of muon poor showers.  The latter have been investigated in order to
look for potential gamma rays.  They would manifest themselves in air showers
with no or little muons only. Both investigations indicate an isotropic
distribution of the arrival direction of cosmic rays.  Special attention has
been given to the region of the galactic plane, as well as to the vicinity of
known SNRs and TeV-$\gamma$-ray sources.  No significant excess could be found
in either sample.

\begin{figure}[t] 
 \includegraphics[width=0.52\textwidth]{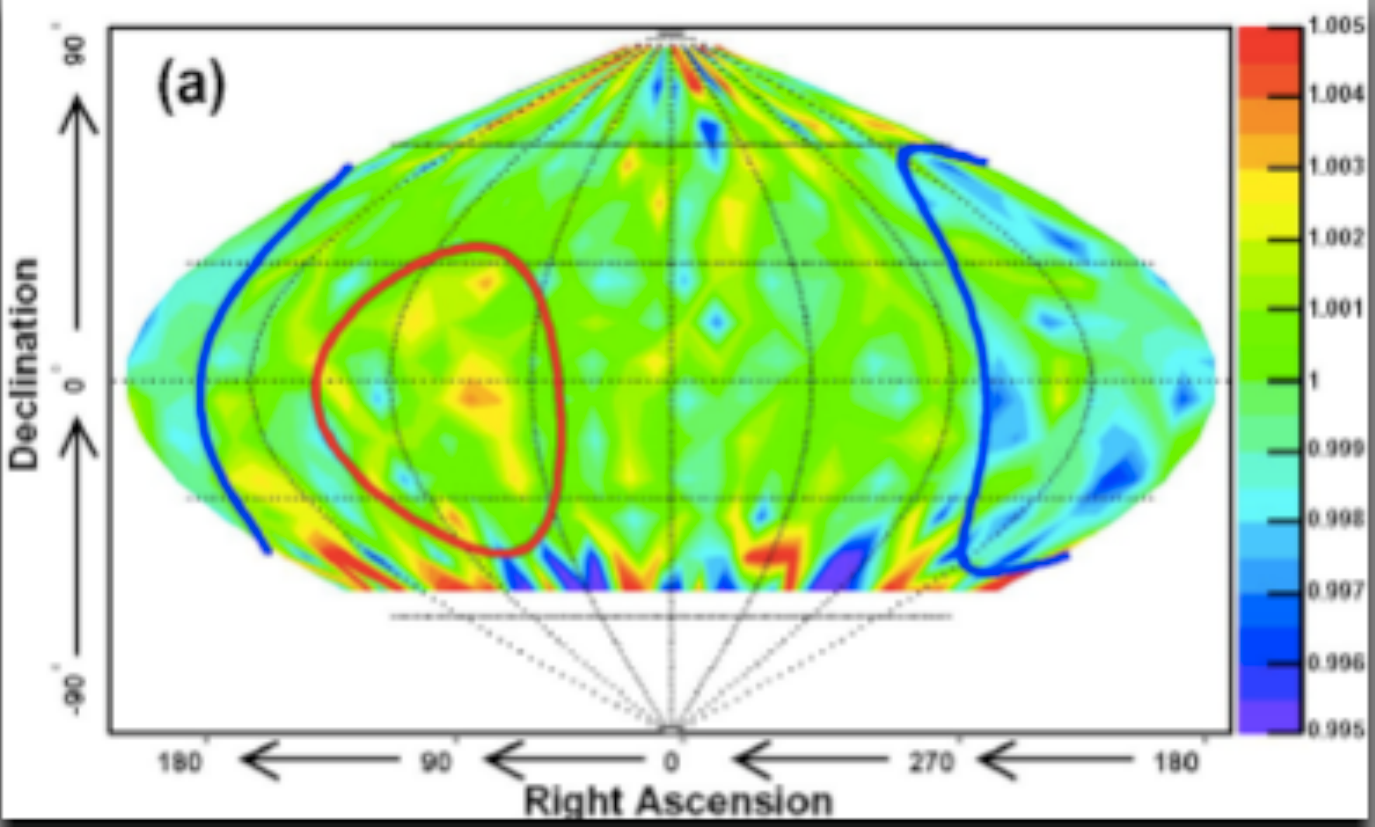}\hspace*{\fill}
 \includegraphics[width=0.47\textwidth]{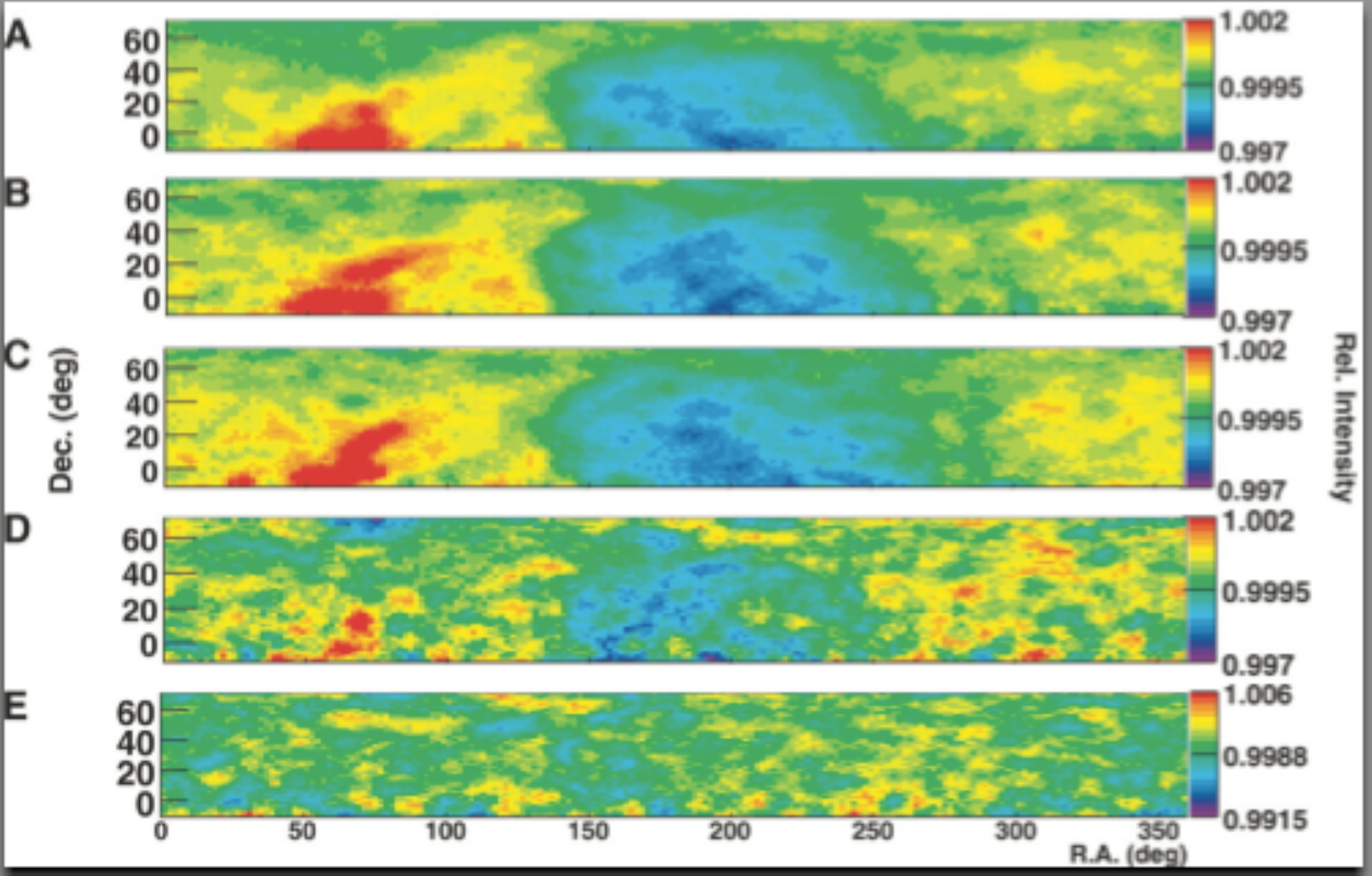}
 \Caption{\LLeft: arrival direction of cosmic rays in celestial coordinates
	  observed by the Super-Kamiokande experiment. Deviations from the
	  average value for the same declination are shown (amplitude $\pm5\%$)
          \protect\cite{kamioka-anisomoriond,*kamioka-aniso}.
	 \RRight: Cosmic-ray intensity map as observed by the Tibet experiment
	 for various energy thresholds, from top to bottom: 4~TeV, 6.2~TeV,
         12~TeV, 50~TeV, 300~TeV \protect\cite{tibetscience}.
         \label{kamiokaaniso}}
\end{figure}

The Super-Kamiokande experiment investigated large-scale anisotropies for cosmic
rays with energies around 10~TeV \cite{kamioka-anisomoriond,*kamioka-aniso}.
The experiment registers muons from air showers with an energy threshold of
0.8~TeV.  An anisotropy map in celestial coordinates is obtained by binning the
data in pixels of $10^\circ \times 10^\circ$. The result after applying a
smoothing algorithm is shown in \fref{kamiokaaniso} (\lleft).  Each cell
represents the relative deviation from an isotropic distribution.  A $3\sigma$
excess (``Taurus excess'') is found with an amplitude of $1.04\pm0.20 \times
10^{-3}$ at right ascension $\alpha=75^\circ\pm7^\circ$ and declination
$\delta=-5^\circ\pm9^\circ$.  On the other hand, a deficit (``Virgo deficit'')
is found with an amplitude of $-(0.94\pm0.14)\times10^{-3}$ at
$\alpha=205^\circ\pm7^\circ$ and $\delta=5^\circ\pm10^\circ$.  The angular
difference between the Taurus excess and the Virgo deficit amounts to about
130\deg.  A large-scale anisotropy would be expected if the Earth moves
relative to a cosmic-ray rest system (Compton Getting effect
\cite{comptongetting}).  For such a scenario the angular difference between
maximum and minimum flux should be 180\deg. Within the relatively large angular
uncertainties the anisotropy observed by the Super-Kamiokande experiment would
be compatible with a Compton Getting effect caused by a velocity of about
50~km/s relative to the rest frame. This velocity is smaller than the rotation
speed of the solar system around the galactic center ($\approx200$~km/s).  This
implies that the rest frame of cosmic rays (presumably the galactic magnetic
fields) is co-rotating with the Galaxy.

The Tibet experiment reported anisotropies in the same regions on the sky
\cite{tibetscience}. The observed intensity for different energy thresholds is
displayed in \fref{kamiokaaniso} (\rright). For energies below 12~TeV the
anisotropies show little dependence on energy, whereas above this energy
anisotropies fade away, consistent with measurements of the KASCADE experiment
in the energy range from 0.7 to 6~PeV \cite{kascade-aniso}.  A Compton Getting
effect caused by the orbital motion of the solar system around the galactic
center would cause an excess at $\alpha\approx315^\circ$, $\delta=40^\circ$ and
a minimum at $\alpha=135^\circ$, $\delta=-49^\circ$ with an amplitude of
0.35\%. However, the measurements at 300~TeV yield an anisotropy amplitude of
$0.03\%\pm0.03\%$, consistent with an isotropic cosmic-ray intensity. Hence, a
galactic Compton Getting effect can be excluded with a confidence level of
about $5\sigma$. This implies, similar to the result of the Super-Kamiokande
experiment, that galactic cosmic rays co-rotate with the local galactic magnetic
field environment.

The Tibet experiment finds also an excess with a 0.1\% increase of the cosmic
ray intensity in the direction of the Cygnus region ($\alpha\approx309^\circ$,
$\delta\approx38^\circ$N) \cite{tibetscience}.  In the same region an excess of
$\gamma$-rays with energies around 10~TeV is seen by the Milagro experiment
\cite{goodmanerice04,milagrocygnus}.
Recently, the Milagro experiment reported also an excess of (charged) cosmic
rays with a significance greater than 12$\sigma$ \cite{milagrocrexcess}.
Presently, different explanations are discussed in the literature: the excess
could be related to the Geminga pulsar as local cosmic-ray source
\cite{salvatisaccomilagro} or cosmic rays from a local source could reach the Earth
via a magnetic mirror \cite{druryaharonianmilagro}.

At energies around the knee the Rayleigh formalism has been applied by several
groups to characterize the large-scale anisotropy.  The two-dimensional
distribution of the arrival directions of cosmic rays is reduced to one
coordinate because of the limited field of view (typical experiments cover only
a fraction of the whole sky) and the small amplitudes expected from theory. A
first order approximation of the multipole expansion of the arrival directions
of cosmic rays is a harmonic analysis of the right ascension values of
extensive air showers. 

Applying the Rayleigh formalism to the right ascension distribution of
extensive air showers measured by KASCADE yields no hints of anisotropy  in
the energy range from 0.7 to 6~PeV \cite{kascade-aniso}. This accounts for
all showers, as well as for subsets containing showers induced by
predominantly light or heavy primary particles.  Also other experiments, like
Mt.  Norikura \cite{norikura-aniso}, EAS-TOP
\cite{eastop-aniso96,*eastop-aniso03}, Akeno \cite{akeno-aniso}, and Adelaide
\cite{adelaide-aniso} have derived Rayleigh amplitudes. Some experiments find
anisotropies, however the phases do not agree between the different results.
Hence, it seems to be more likely that all amplitudes derived should be
considered as upper limits.

\subsubsection*{Extragalactic Cosmic Rays}

Different groups have searched for large-scale anisotropies but no
confirmed deviation from an isotropic arrival direction distribution
has been found. 
A dipole amplitude that is compatible with full isotropy was found in an
analysis of more than 135000 showers with energies from $3\times 10^{16}$ to
$10^{17.5}$\,eV  of the Yakutsk array \cite{Pravdin:2001jk}. Also no
significant anisotropy is found in the AGASA data \cite{Hayashida:1998qb} in
the energy range between $10^{17}$ to $10^{17.5}$\,eV.  

In the energy
range around $10^{18}$\,eV, however, AGASA found an excess of showers coming from
directions near the galactic center and the Cygnus region
\cite{Hayashida:1998qb}.  
Furthermore, analysis of SUGAR
data \cite{Bellido:2000tr} indicates an excess of cosmic rays coming from a
similar region.  
The Haverah
Park and Yakutsk arrays are located too far north to be able to see the excess
regions of AGASA and SUGAR. 

With the galactic center being in the field of view of the southern Auger
Observatory, it is ideally suited to search for these possible source
regions, though the energy threshold for reaching 100\% acceptance of
the Auger surface array is just below $10^{18.5}$\,eV.  Recently, the
Auger Collaboration has performed an analysis of their low-energy showers
from the direction of the galactic center.  In the energy range from
$10^{17.9}$\,eV to $10^{18.5}$\,eV, no abnormally over-dense regions
were found in the neighborhood of the galactic center
\cite{Aglietta:2006ur}. There are 506 (413.6) events found (expected)
in the AGASA data set for a circle of $20^\circ$ around $280^\circ$
right ascension and $-17^\circ$ declination, corresponding to an excess
ratio of 1.2. In the same region, the Auger Collaboration observed 2116
events while 2170 were expected.

At energies above $10^{18.5}$\,eV, the large-scale structure of the arrival
direction distribution appears, within the limited statistics of the AGASA
array, isotropic \cite{Hayashida:1998qb}.  This finding agrees with that of the
HiRes Collaboration, performing a global anisotropy search based on $\sim1500$
events observed by HiRes~I in monocular mode \cite{Abbasi:2003tk}.  
By combining data from arrays of the northern and southern hemispheres a full
sky anisotropy study is done in \cite{Swain:2004na}.  Considering in total 99
showers from AGASA and SUGAR with $E>10^{19.6}$ no large-scale anisotropy is
found. The data set of the Auger Observatory analyzed in \cite{Armengaud:2007hc}
contains more than 7000 events with $E > 10^{18.3}$\,eV. No large scale 
anisotropy is found. The upper limit to the amplitude of a dipole-like anisotropy 
in right ascension is 3\% with 95\% CL \cite{Armengaud:2007hc}.


Small angle clustering could be an indication for point sources. At
energies above $4\times 10^{19}$\,eV, clustering at an
angular scale of $2.5^\circ$ has been
reported by the AGASA Collaboration
\cite{Hayashida:1996bc,*Takeda:1999sg,*Teshima:2003ab}.
This result could not be confirmed. An analysis of a HiRes I data set
corresponding to an exposure similar to that of AGASA did not reveal
any evidence for small scale clustering \cite{Abbasi:2004dx}.  Also no
significant clustering is seen in a data set from HiRes stereo
observations of more than 270 showers with $E>10^{19}$\,eV
\cite{Abbasi:2004ib} and in the
combined AGASA-HiRes stereo data set with $E>4\times 10^{19}$\,eV
\cite{Abbasi:2004vu}. Clustering at the angular scale of the AGASA
signal is also not found in the Auger data \cite{Mollerach:2007vb}.

Uchihori et al.\ included showers above $4\times 10^{19}$\,eV from all four
surface arrays of the northern hemisphere in their small scale correlation
analysis  \cite{Uchihori:1999gu}.  The combined data set is found to contain
many clusters, however, the statistical significance is low ($\sim 10$\%). On
the other hand, there are indications for a correlation with the super galactic
plane. Restricting the considered arrival directions to the range of
$\pm10^\circ$ off the super galactic plane the chance probability for finding
doublets and triplets decreases to the order of 1\%.

Studies of small angle correlations with the Yakutsk array are difficult
because of the angular resolution of the shower axis reconstruction,
which is about $4^\circ$ \cite{Antonov:1999aa}. Nevertheless,
at much lower energy, clusters of the arrival directions of showers 
in the energy range $(1.3 - 4) \times 10^{17}$\,eV were reported
in \cite{Glushkov:2003ny}. The direction of these clusters seem to
support a correlation with the super galactic plane. Dividing the
observed cosmic-ray showers into isotropic and cluster components this
correlation can be enhanced significantly \cite{Glushkov:2003gj}.

A recent search for correlations at a medium angular scale has been carried out
in \cite{Kachelriess:2005uf}. Publicly available data sets from AGASA, HiRes,
SUGAR, and Yakutsk with $E > 4\times 10^{19}$\,eV were combined by adjusting
their energy scales to shift the ankle in all data sets to the same energy. In
addition events with $E>10^{20}$\,eV from Haverah Park, Volcano Ranch, and
Fly's Eye were considered. In this set of 107 events in total, a signal at a
scale of $25^\circ$ is found in the autocorrelation function. Results from 
the Pierre Auger Observatory also indicate some clustering on
intermediate scales with an angular separation of order of 15\deg to 25\deg
\cite{Mollerach:2007vb}. The medium angle correlation reported in
\cite{Mollerach:2007vb} has a 2\% chance probability to originate from an
isotropic distribution.


There is a long history of searches for correlations with astrophysical point
sources such as colliding galaxies and powerful radio galaxies.  It appears
almost impossible to assess unambiguously the chance probability of such
correlations. First of all, highly incomplete catalogs of astrophysical objects
necessarily have to be used in these analyses. Secondly, the penalty factor for
scans of several catalogs, selecting different classes of objects, distance and
angular ranges, and other parameters cannot be calculated reliably. Part of
these problems can be avoided by defining a prescription before analyzing a
data set. This has been done by the Auger Collaboration \cite{Revenu:2005qb}.

A correlation with BL Lacertae, at a distance exceeding the GZK energy loss
length, has been claimed for the AGASA and Yakutsk high-energy data ($E > 4\times
10^{19}$\,eV) \cite{Tinyakov:2001ic,*Tinyakov:2001nr,*Gorbunov:2002hk}. The
significance of this correlation is highly debated \cite{Evans:2002ry,*Tinyakov:2003bi,*Evans:2004ba,*Stern:2005fh} as there might
be ``hidden'' trials involved that cannot be corrected for with a Monte Carlo
simulation. These correlations were tested with the independent data
set of HiRes stereoscopic observations \cite{Abbasi:2005qy}. None of the
previous claims could be confirmed.
Recently, a $\sim 0.8^\circ$ correlation between BL Lacertae objects of the
Veron-Cetty \& Veron catalog \cite{VeronCetty:2003xx} with HiRes data ($E > 10^{19}$\,eV) was
pointed out \cite{Gorbunov:2004bs}.  This correlation was confirmed at a
nominal significance of about $0.5$\% not only for the high-energy part but
also for the entire set of HiRes stereo data \cite{Abbasi:2005qy}. 
An analysis of data from the Pierre Auger Observatory did not confirm a
correlation of the arrival direction of cosmic rays with the positions of BL
Lacertae objects in the southern hemisphere \cite{Harari:2007up}.

\begin{figure}[t] \centering
  \includegraphics[width=\breite]{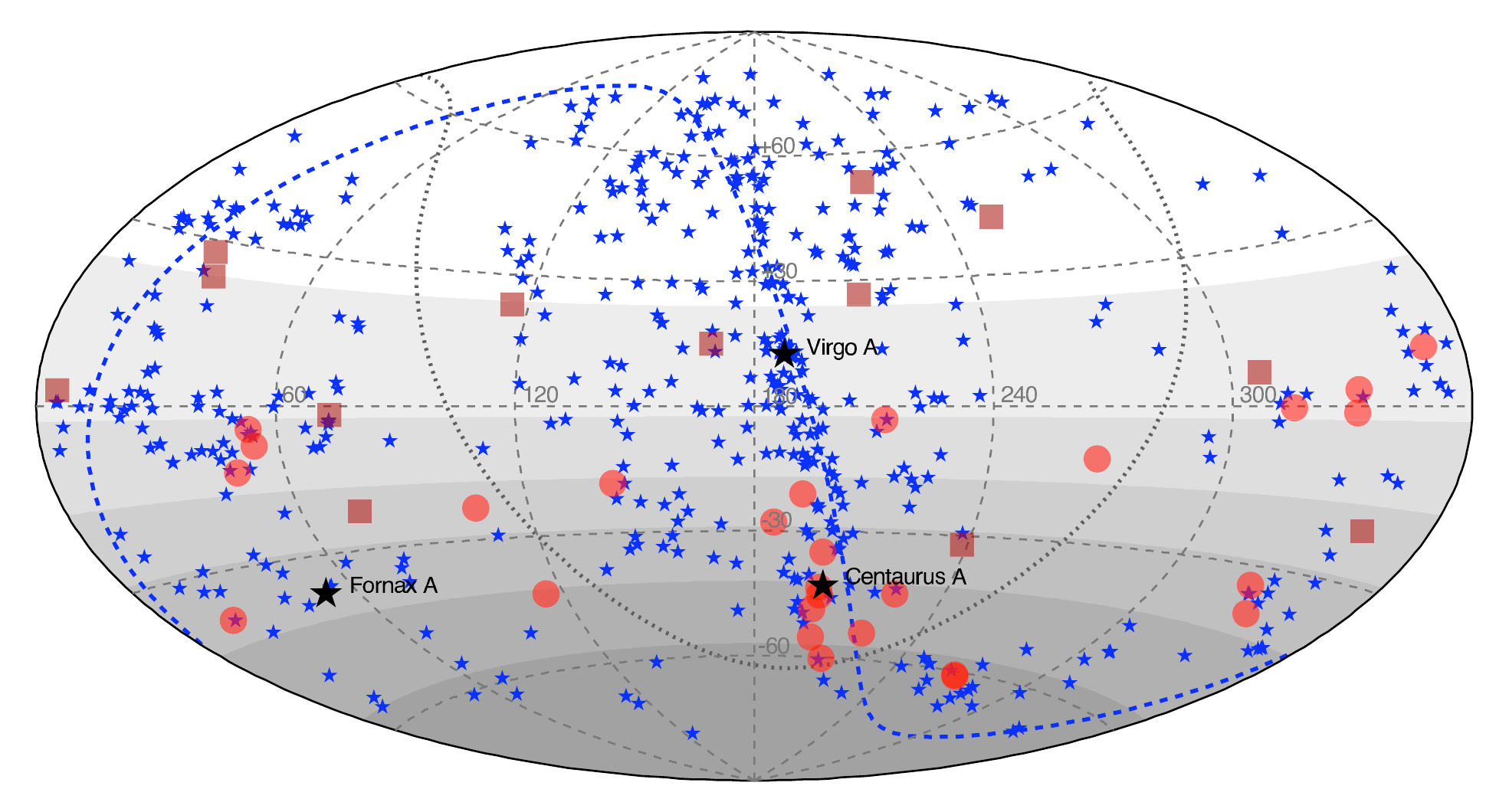}
  \Caption{Arrival directions (equatorial coordinates) of the highest
    energy cosmic rays observed with the Pierre Auger Observatory
    \protect\cite{Abraham:2007bb} (circles) and the HiRes telescopes
    \protect\cite{Abbasi:2008md} (squares). The asterisks indicate the position of
    active galactic nuclei (AGN) from the Veron-Cetty Veron catalog
    \protect\cite{VeronCetty:2003xx} up to a distance of 75 Mpc. The shaded
    area indicates the relative exposure of the Auger data set. The
    dotted line marks the galactic disk and the dashed curve is the
    super galactic plane.
\label{fig:agn}  }
\end{figure}

A breakthrough in the anisotropy searches is the correlation discovered by the
Auger Collaboration \cite{Abraham:2007bb,Abraham:2007si}. The arrival
directions of the very highest energy events ($E > 5.7\times10^{19}$\,eV) are
found to be correlated with the positions of active galactic nuclei (AGNs) from
the Veron-Cetty \& Veron catalog \cite{VeronCetty:2003xx}. Out of 27 events 
observed with an integrated
aperture of 9000 km$^2$ sr yr, 20 are correlated with AGNs within an angular
distance of $3.1^\circ$. The correlation was initially found in an exploratory
search with different catalogs (12 out of 15 events were correlated). A
prescription was set up to verify or reject the correlation hypothesis using an
independent data set. Of the next 8 events that were detected, 6 were
correlated with AGNs within the prescribed phase space, corresponding to a
chance probability of incorrectly accepting the hypothesis of correlation of
less than 1\%. After accepting the correlation hypothesis, the Auger group made
a scan to refine the correlation parameters and found an energy threshold of
$E_{\rm th} = 5.7\times 10^{19}$\,eV, an ``source'' distance of less than $z=
0.017$ ($D \approx 75$\,Mpc) and a maximum angular difference of $3.2^\circ$ as
optimal parameters. If one does not account for the penalty factor due to different
trials and the parameter scans, the nominal chance probability of being
compatible with isotropy of $\sim 10^{-10}$ would be obtained.

A sky map of the measured arrival direction distributions is shown in
Fig.~\ref{fig:agn} using equatorial coordinates. The relative exposure
of the Auger Observatory is indicated by the shaded regions. The Veron
catalog of AGNs is not a complete catalog. As expected, in the
direction of the galactic plane the density of detected AGNs is lower
than in other directions. 

The nearby AGNs are very good tracers of
the local matter distribution. In particular, the super galactic plane
is clearly visible. A recent update of the super galactic plane
position even improves the correlation between AGNs and this matter
over-density \cite{Stanev:2008sd}.  The correlation of the arrival
direction distribution of UHECRs with nearby AGNs is also reflected by
the autocorrelation function which shows some indications of
anisotropy in the 15\deg to 25\deg range, as one would expect from
the typical thickness of the super galactic plane.

Given the limited statistics one cannot conclude from the found correlation
that AGNs are sources of UHECRs. Subsequent studies of the published
highest energy events of the Auger Observatory revealed correlations
with the large scale structure of galaxies 
\cite{Kashti:2008bw,*Ghisellini:2008gb,*Takami:2008ri}. A correlation
of UHECRs with the large scale structure in the cosmological neighborhood
is also found in the Yakutsk data set \cite{Ivanov:2008it}.

The HiRes Collaboration has used the correlation parameters published by
Auger to perform a search in their stereo data set. To obtain the same energy
threshold as used in the Auger analysis, they scaled their
reconstructed energies down by $10$\%~\cite{Abbasi:2008md}. Using all
stereo data an exposure of roughly 4000 km$^2$\,sr\,yr is obtained at the
highest energy (see \fref{fig:exposure}). For a total of 13 events two
associations with AGNs were found, while 3.2 such correlations are
expected for an isotropic arrival direction distribution. No
correlation signal is found. Also the autocorrelation function of the
highest energy events from HiRes is perfectly in agreement with the
expectations for isotropy. The arrival directions of the selected 13
events are shown in Fig.~\ref{fig:agn} as well. The exposure distribution
of the HiRes data set is very similar to that of the Auger Observatory, but North
exchanged with South.

The discrepancy between the Auger and HiRes results are currently not
understood, but it is clear that a possible difference of the energy scale of
the two experiments could lead to such effects. In addition the reconstruction
resolution has to be very good to reproduce the very sharp threshold of the
correlation found in Auger data. An independent data set of similar size as the
published one will allow to test the anisotropy signal. 

\Section{Astrophysical Interpretation} \label{astrosec}

\subsubsection*{Galactic Cosmic Rays and the Knee}

The measurements indicate that the knee in the all-particle energy spectrum is
caused by a break in the spectra for the light elements, yielding an increase
of the mean mass of cosmic rays in this energy region.  Several scenarios are
discussed in the literature as possible origin for the knee, see e.g.\
\cite{Hoerandel:2004gv}. In the following, a current astrophysical picture of
the origin of high-energy cosmic rays is sketched, based on recent
observations.

\begin{figure}[t] \centering
 \includegraphics[width=0.49\textwidth]{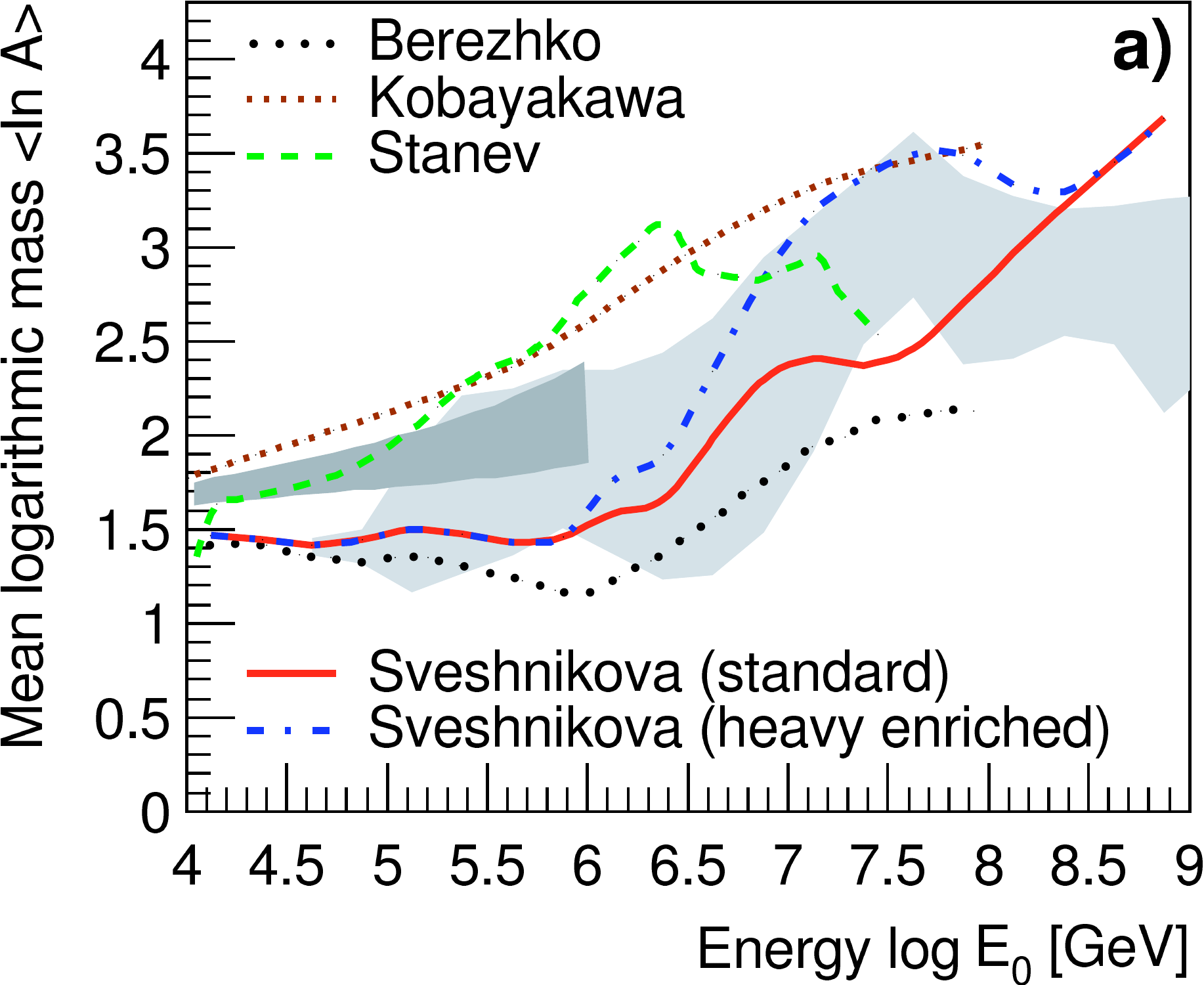}
 \includegraphics[width=0.49\textwidth]{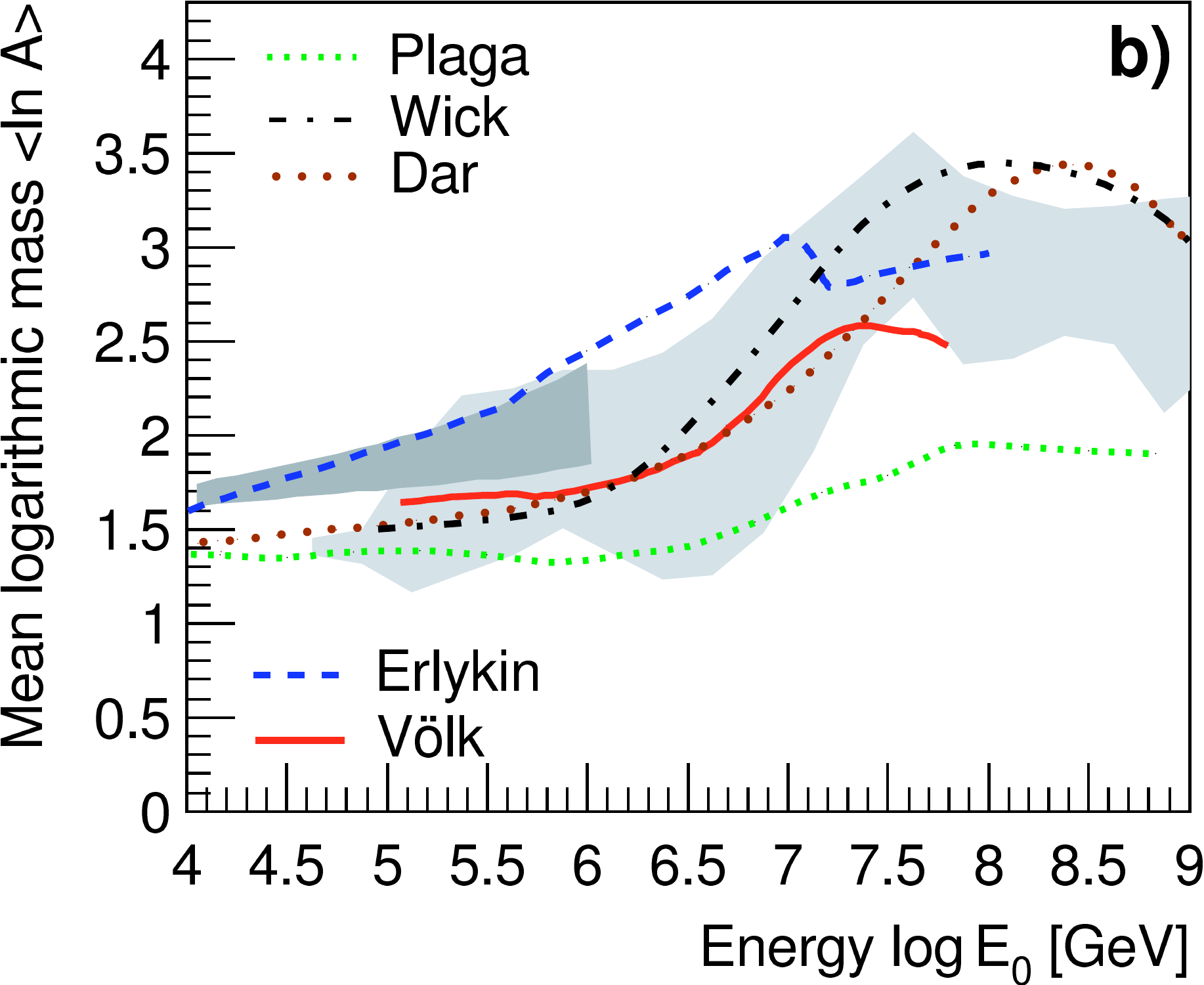}
 \includegraphics[width=0.49\textwidth]{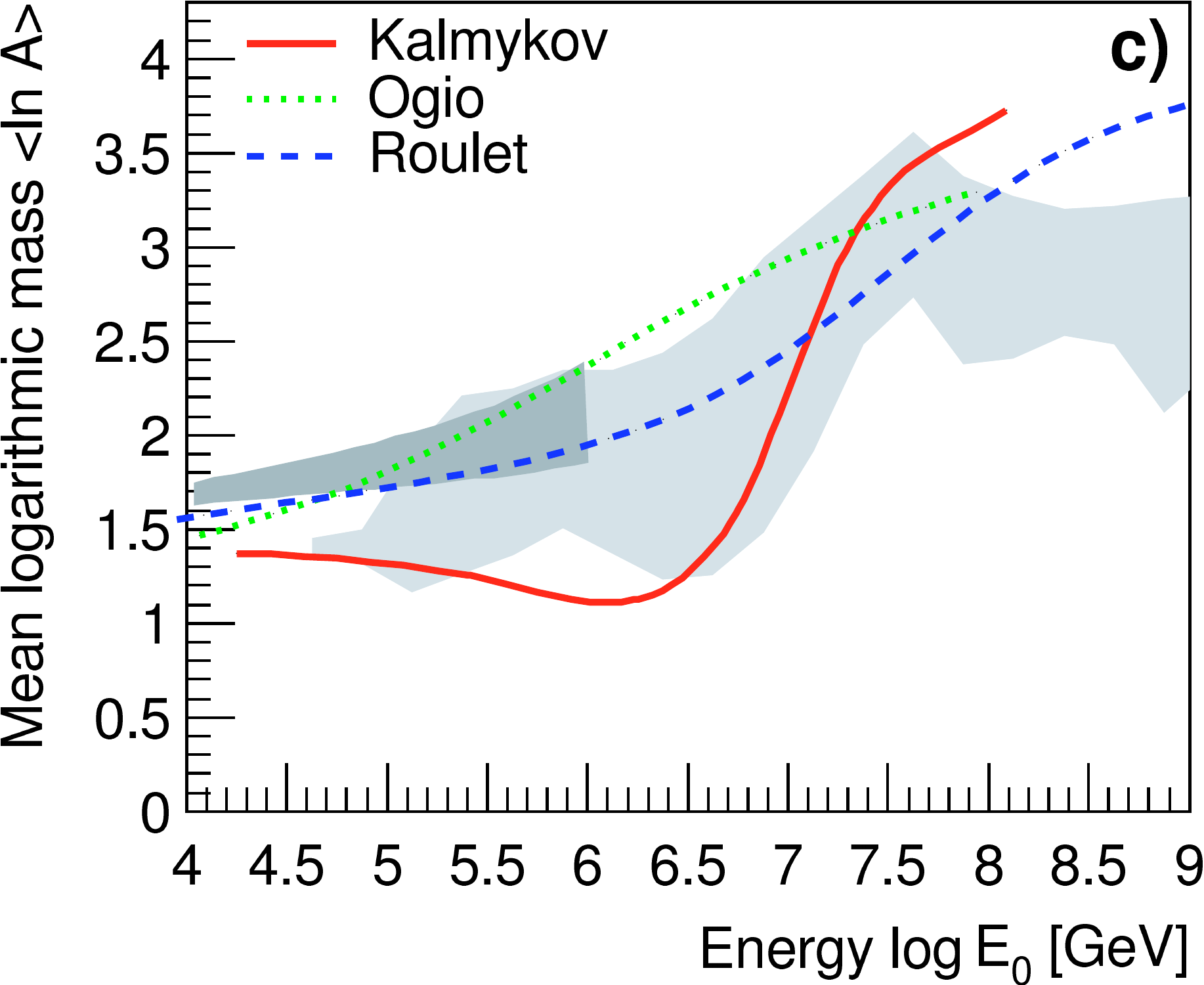}
 \includegraphics[width=0.49\textwidth]{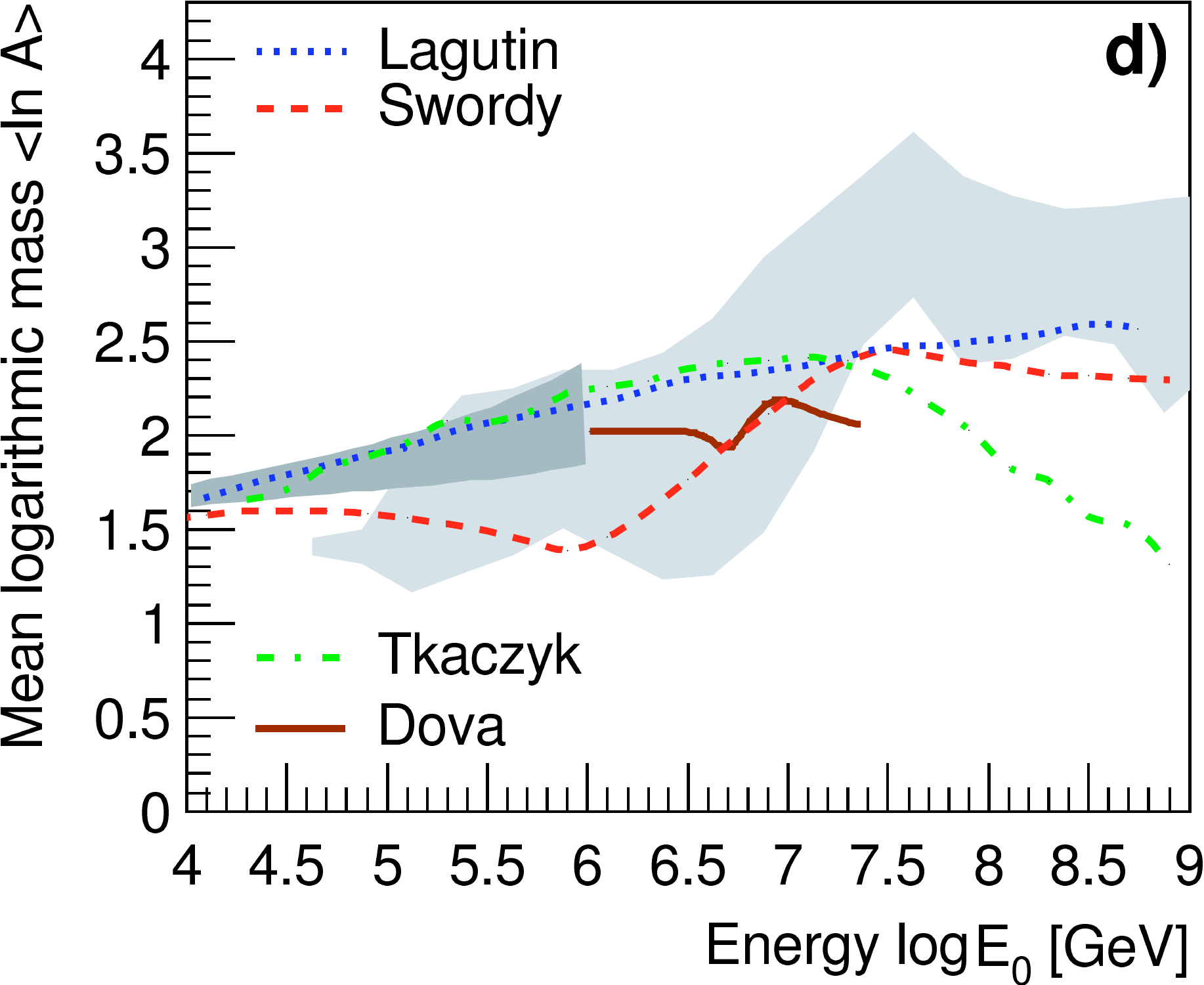}
 \Caption{Mean logarithmic mass as function of energy obtained by
  direct observations (dark gray area) and air shower experiments (light gray
  area, obtained as weighted average of the results of many experiments, see
  \protect\cite{Hoerandel:2004gv}) compared with different models (lines).
  {\bf a)} Acceleration in SNRs 
           (Berezhko \etal \protect\cite{berezhko},
           Kobayakawa \etal \protect\cite{kobayakawa},
           Stanev \etal \protect\cite{Stanev:1993tx,Biermann:2001m1},
           Sveshnikova \etal \protect\cite{sveshnikova});
  {\bf b)} acceleration in GRBs 
           (Plaga \protect\cite{plaga},
           Wick \etal, \protect\cite{wick}
           Dar \protect\cite{Dar:2004jy}),
           single source model (Erlykin \& Wolfendale \protect\cite{wolfendale}),
           reacceleration in the galactic wind (V\"olk \etal \protect\cite{voelk});
  {\bf c)} diffusion in the Galaxy 
           (Kalmykov \etal \protect\cite{kalmykov},
            Ogio \etal \protect\cite{ogio},
            Roulet \etal \protect\cite{roulet});
  {\bf d)} propagation in the Galaxy 
           (Lagutin \etal \protect\cite{lagutin}, Swordy \protect\cite{swordy}), as well as
           interaction with background photons 
	   (Tkaczyk \protect\cite{tkaczyk}) and
           neutrinos (Dova \etal \protect\cite{dova}).
  For details see \protect\cite{Hoerandel:2004gv}.\label{lnamod}}
\end{figure}

One of the most popular explanations for the origin of the knee is that the
spectra at the source exhibit a break.
The bulk of cosmic rays is assumed to be accelerated in strong shock fronts of
SNRs \cite{axford,*krymsky,*bell,*Blandford:1978ky,*Blandford:1987pw}. 
The finite lifetime of a shock front ($\sim 10^5$~a)
limits the maximum energy attainable for particles with charge $Z$ to
$E_{\rm max}\sim Z\cdot (0.1 - 5)$~PeV.  Many versions of this scenario have been
discussed \cite{berezhko,Stanev:1993tx,Biermann:2001m1,kobayakawa,sveshnikova}. The models differ in
assumptions of properties of the SNRs like magnetic field strength, available
energy, ambient medium, etc.\ The differences of the predicted $\lnA$ can be inferred from
\fref{lnamod}a.  While older models \cite{Stanev:1993tx} limit the maximum energy to
about 0.1~PeV, recent ideas \cite{sveshnikova}, taking into account latest
observations of SNRs, predict maximum energies above 1~PeV.  In such a model
sufficient energy is released from SNRs to explain the observed spectra.  A
special case of SNR acceleration is the single source model  \cite{wolfendale},
which predicts in the knee region pronounced structures in the all-particle
energy spectrum, caused by a single SNR. Such structures can not be seen in the
compilation of \fref{knee-zoom}.

In the literature also other acceleration mechanisms, like the acceleration of
particles in $\gamma$-ray bursts, are discussed \cite{plaga,wick,Dar:2004jy}.  They
differ in their interpretation of the origin for the knee.  The approach by
Plaga, assuming Fermi acceleration in a ``cannon ball'' is not compatible with
the measured $\lnA$ values, see \fref{lnamod}b.  A different interpretation of
acceleration in the cannon ball model yields -- at the source -- a cut-off for
individual elements proportional to their mass due to effects of relativistic
beaming in jets.  The predictions of the actual model are compatible with
recent data \cite{Dar:2004jy,Dar:2006dy}. However, it remains to be clarified how a detailed
consideration of the propagation processes, e.g., in a diffusion model, effects
the cut-off behavior observed at Earth.  Gamma-ray bursts as a special case of
supernova explosions are proposed \cite{wick} to accelerate cosmic rays from
0.1~PeV up to the highest energies ($>10^{20}$~eV). In this approach the
propagation of cosmic rays is taken into account and the knee is caused by
leakage from the Galaxy leading to a rigidity dependent cut-off behavior.

Also frequently discussed is the idea that the knee is a propagation effect.
The propagation is accompanied by leakage of particles from the Galaxy. With
increasing energy it becomes more and more difficult to confine the nuclei to
the Galaxy.  As mentioned above, the path length decreases as $\Lambda\propto
E^{-\delta}$.  Such a decrease will ultimately lead to a complete loss of the
particles, with a rigidity dependent cut-off of the flux for individual
elements.  Many approaches have been undertaken to describe the propagation
process \cite{ptuskin,ogio,roulet,swordy,lagutin}.  The Leaky Box model
\cite{swordy} and the anomalous diffusion model \cite{lagutin} yield cut-offs
significantly weaker than the data shown in \fref{elementspek}  and
\cite{Hoerandel:2004gv}.

The propagation as described in diffusion models \cite{kalmykov,ogio,roulet}
yields $\lnA$-values which are presented in \fref{lnamod}c.  The models are
based on the same principal idea \cite{ptuskin}, but take into account different
assumptions on details of the propagation process, like the structure of
galactic magnetic fields etc.\ This results in a more or less strong cut-off for
the flux at the individual knees and, accordingly, in a more or less strong
increase of $\lnA$. 
The observed break of the spectra is relatively sharp. It seems to be difficult
to generate such a behavior by a cut-off at the source or due to propagation
only. More likely seems to be a combined approach assuming a break of the
spectra at the source and leakage from the Galaxy, as e.g. discussed in
\cite{Hoerandel:2004gv}.
During the propagation phase, reacceleration of particles has been suggested at
shock fronts in the galactic wind \cite{voelk}. Also this mechanism yields a
rigidity dependent cut-off.

Another hypothetical explanation for the knee are interactions of cosmic rays
with background particles like massive neutrinos \cite{dova,wigmans} or 
photodisintegration in dense photon fields \cite{tkaczyk,candia}.  Such models
appear to be excluded with a high level of confidence.  The interactions would
produce a large amount of secondary protons, which results in a light mass
composition at high energies, not observed by the experiments, see
\fref{lnamod}d.  Furthermore, a massive neutrino, proposed in
\cite{dova,wigmans} can be excluded by measurements of the WMAP and 2dFGRS
experiments \cite{hannestad}.

\begin{figure}[t] \centering
 \includegraphics[width=0.6\textwidth]{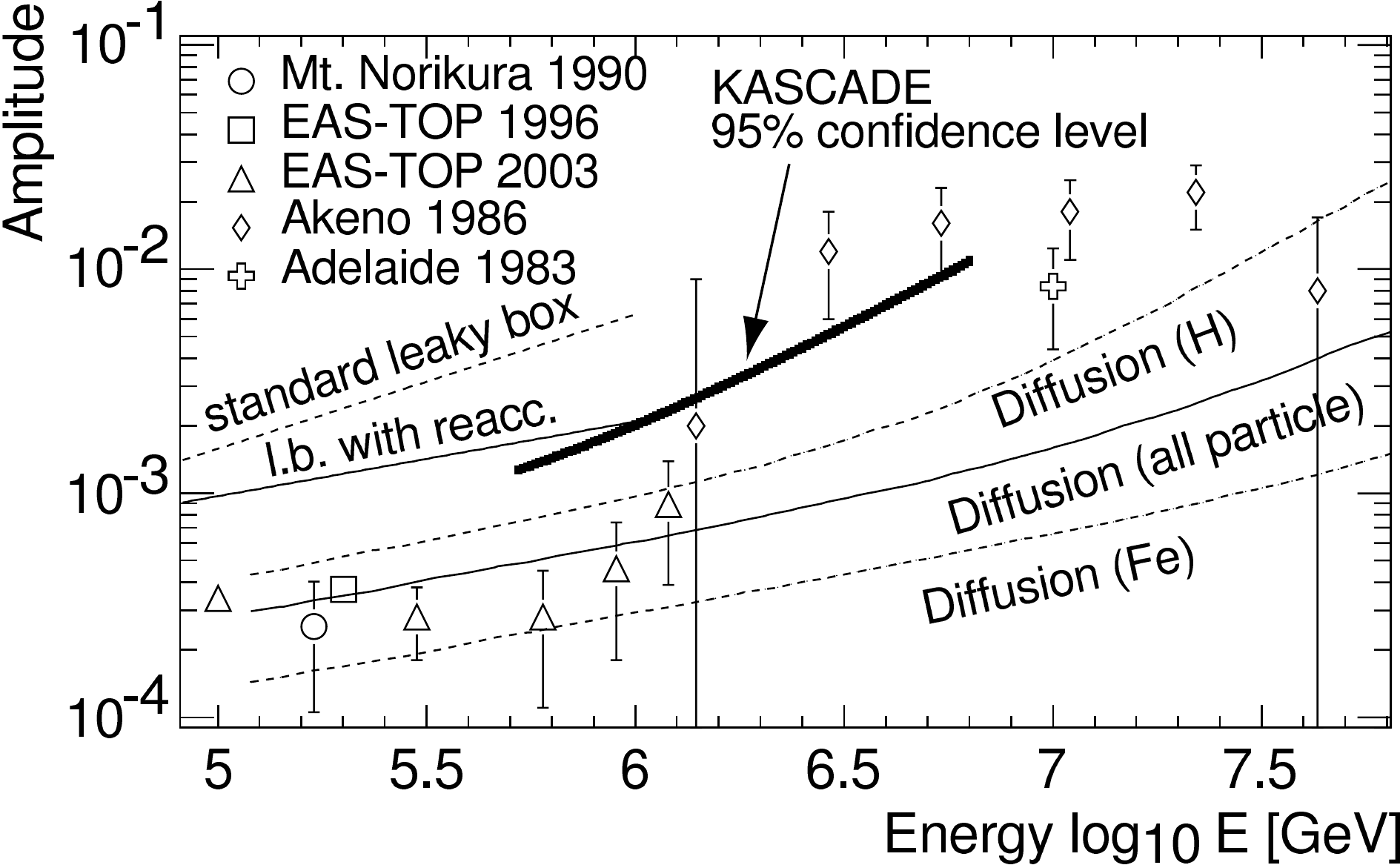}
 \Caption{Rayleigh amplitudes as function of energy for various experiments,
	  taken from \protect\cite{kascade-aniso}.  The results obtained by Mt.
	  Norikura \protect\cite{norikura-aniso}, EAS-TOP
	  \protect\cite{eastop-aniso96,*eastop-aniso03}, Akeno \protect\cite{akeno-aniso},
	  Adelaide \protect\cite{adelaide-aniso}, and KASCADE \protect\cite{kascade-aniso} are
	  compared to predictions of leaky-box models \protect\cite{ptuskinaniso} and a
	  diffusion model \protect\cite{Candia:2003dk}.  For the diffusion model,
	  predictions for primary protons, iron nuclei, and all particles are
	  displayed. 
  \label{kascadeaniso}}
\end{figure}

A completely different reason for the knee is the idea to transfer energy in
nucleon-nucleon interactions into particles, like gravitons \cite{kazanas} or
extremely high-energy muons \cite{petrukhin}, which are not observable (or not
yet observed) in air shower experiments. The latter proposal seems to be
excluded by measurements of the Baikal experiment \cite{baikal-mu}
setting upper limits for the flux of muons above $10^5$~GeV.

No point sources of charged cosmic rays were found in the knee region.  The
observed (large-scale) anisotropy amplitudes in the energy region of the knee
are compared to model predictions in \fref{kascadeaniso}.  Two versions of a
leaky-box model \cite{ptuskinaniso}, with and without reacceleration are shown.
Leaky-box models, with their extremely steep decrease of the path length as
function of energy ($\lambda\propto E^{-0.6}$), yield relatively large
anisotropies even at modest energies below $10^6$~GeV, which seem to be ruled
out by the measurements, see also \cite{ptuskinaniso,Hoerandel:2004gv}.  The measured
values are almost an order of magnitude smaller.  On the other hand, a
diffusion model \cite{Candia:2003dk}, which is based on the idea of
Ref.~\cite{ptuskin}, predicts relatively small values at low energies and a modest
rise only. In the figure, predictions for pure protons and iron nuclei are
shown together with calculations for a mixed composition. The predicted
Rayleigh amplitudes are compatible with the measured values. This may indicate
that diffusion models are a realistic description of cosmic-ray propagation in
the Galaxy at PeV energies.

\begin{figure}[t] 
 \includegraphics[width=0.40\textwidth]{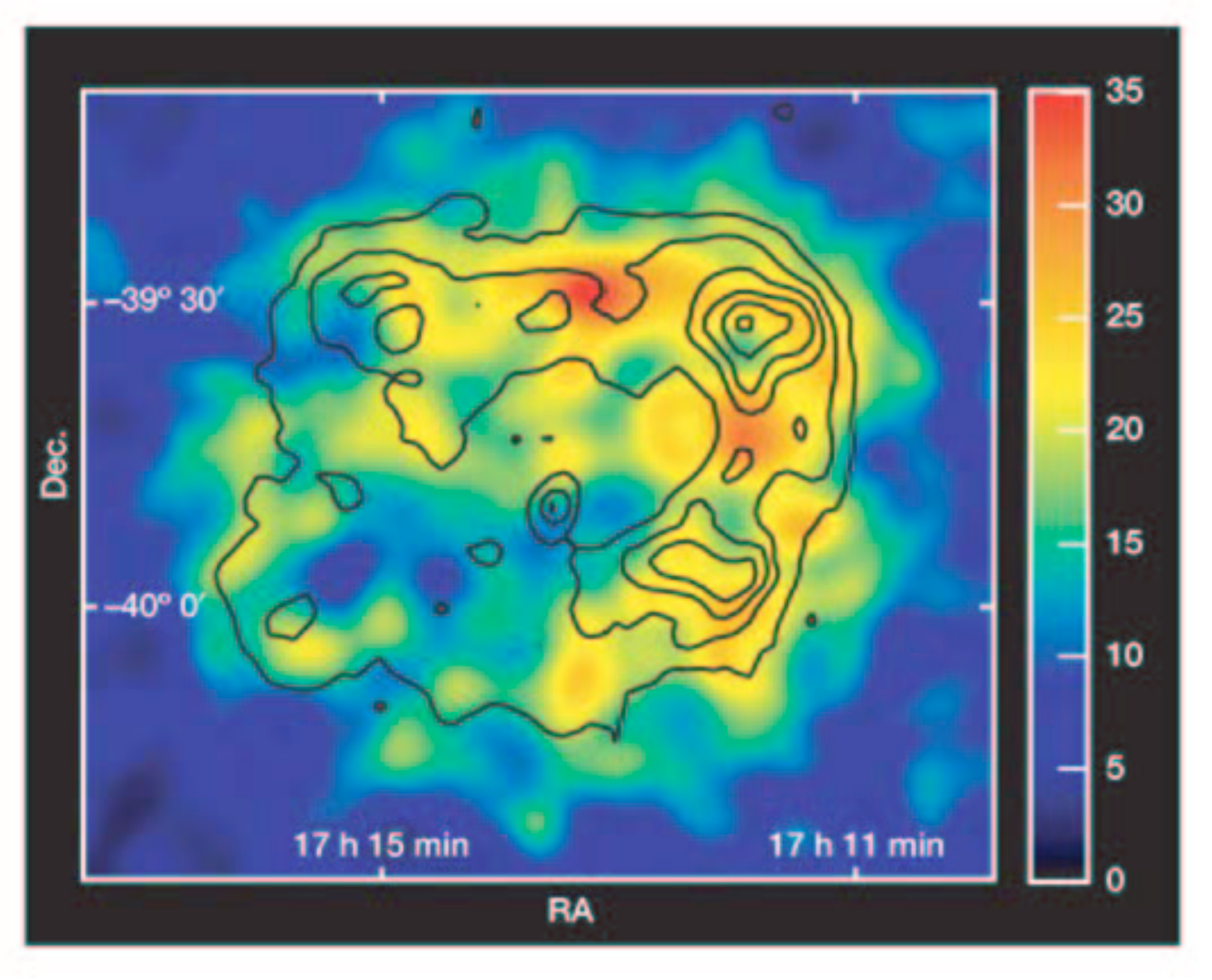}\hspace*{\fill}%
 \includegraphics[width=0.59\textwidth]{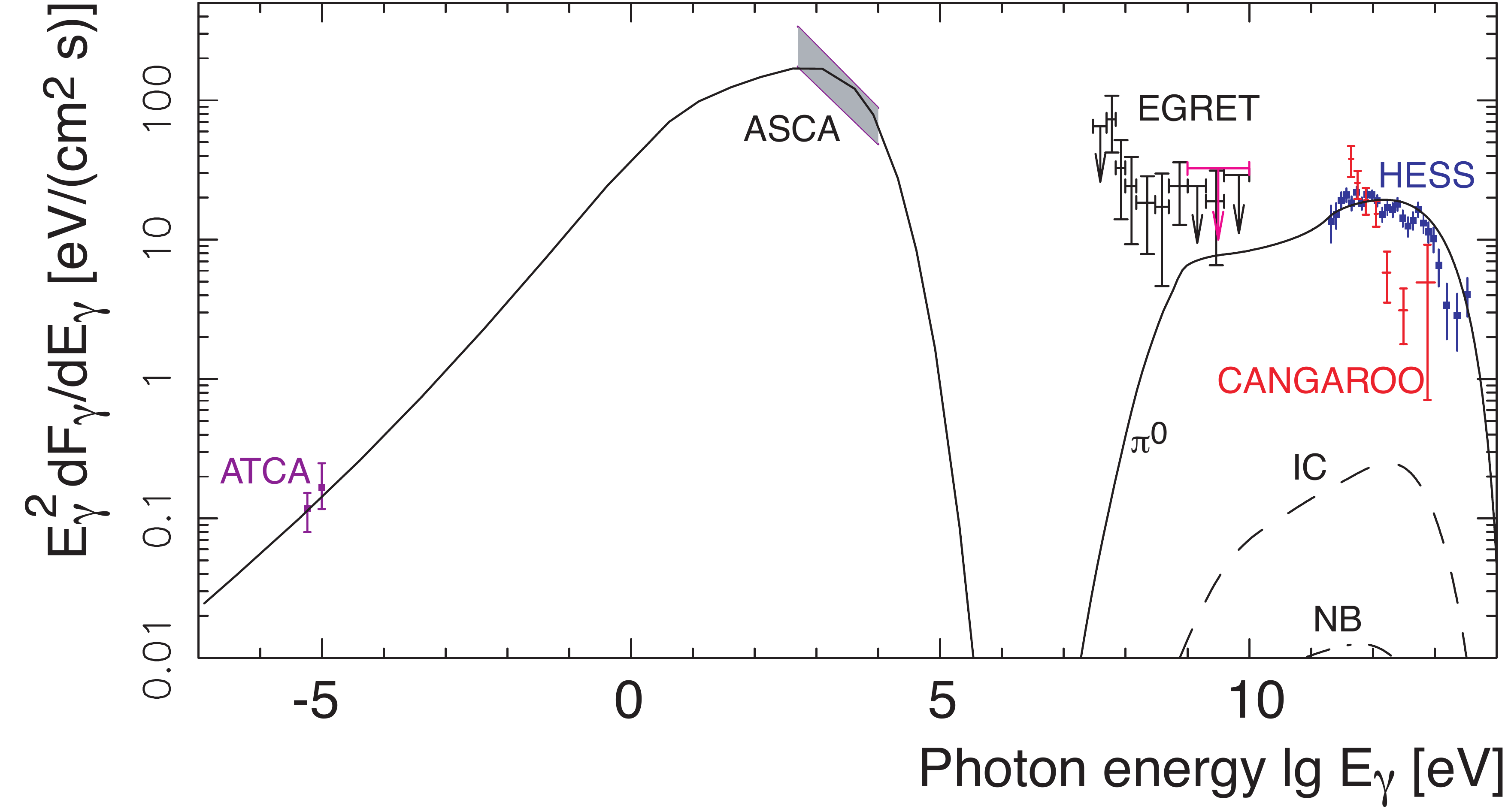}
 \Caption{\LLeft: $\gamma$-ray image of the supernova remnant RX J1713.7-3946
	  obtained with the H.E.S.S. telescope. The superimposed contours show
	  the x-ray surface brightness as seen by the ASCA satellite in the
          1-3~keV range \protect\cite{hesssnr}.\label{rxj1713}
	  \RRight: Spatially integrated spectral energy distribution of the
	  supernova remnant RX J1713. The solid line above $10^7$~eV
	  corresponds to $\gamma$-ray emission from $\pi^0$-decay, whereas the
	  dashed and dash-dotted curves indicate the inverse Compton and
	  non-thermal brems\-strahlung emissions, respectively
	  \protect\cite{voelkrxj1713}.}
\end{figure}

In addition to information extracted from measurements of charged particles,
important hints towards the origin of (hadronic) cosmic rays may be derived
from observations of high-energy $\gamma$-rays.  Recent observation of the
H.E.S.S.  experiment improve significantly the knowledge about galactic
cosmic-ray sources.  For the first time a spatially resolved image of a
supernova remnant has been obtained with multi-TeV gamma rays as shown in
\fref{rxj1713} (\lleft) \cite{hesssnr}. The shell type supernova remnant RX
J1713.7-3946 has been studied in detail \cite{hessrxj1713}.  The image of the
remnant has been divided into 14 regions to study the energy spectrum of gamma
rays up to eight TeV. The indices of the power law spectra vary between
$\gamma=1.95\pm0.08$ and $2.24\pm0.06$ in the different fields with an overall
index of the remnant $\gamma=2.13\pm0.03$.  The spectral indices are very close
to the spectral steepness expected from Fermi acceleration at strong shocks.
The morphology of the TeV $\gamma$-ray image agrees well with emissions in the
1-3~keV  x-ray regime as measured by the ASCA satellite. 

The photon energy spectrum of the supernova remnant RX J1713 is presented in
\fref{rxj1713} (\rright).  Measurements in various energy ranges (ATCA at radio
wavelengths, ASCA x-ray, EGRET GeV $\gamma$-ray, CANGAROO and H.E.S.S. TeV
$\gamma$-ray) are compared to predictions of a model by Berezhko \& V\"olk
\cite{voelkrxj1713}.  The solid line below $10^6$~eV indicates synchrotron
emission from electrons ranging from radio frequencies to the x-ray regime.
The observed synchrotron flux is used to adjust parameters in the model, which
in turn, is used to predict the flux of TeV $\gamma$-rays.  The solid line
above $10^6$~eV reflects the spectra of decaying neutral pions, generated in
interactions of accelerated hadrons with material in the vicinity of the source
(hadron + ISM $\rightarrow \pi^0 \rightarrow \gamma\gamma$).  This process is
clearly dominant over electromagnetic emission generated by inverse Compton
effect and non-thermal brems\-strahlung, as can be inferred from the figure.  The
results are compatible with a nonlinear kinetic theory of cosmic-ray
acceleration in supernova remnants and imply that this supernova remnant is an
effective source of nuclear cosmic rays, where about 10\% of the mechanical
explosion energy are converted into nuclear cosmic rays
\cite{voelkrxj1713,berezhkovoelk2008}.
Further quantitative evidence for the acceleration of hadrons in supernova
remnants is provided by measurements of the HEGRA experiment \cite{hegra-casa}
of TeV $\gamma$-rays from the SNR Cassiopeia~A \cite{berezhko-casa} and by
measurements of the H.E.S.S. experiment from the SNR ``Vela Junior''
\cite{Volk:2006nm}.

In conclusion, it may be stated that a standard picture of the origin of
galactic cosmic rays seems to emerge from the data. The measurements seem to be
compatible with the assumption that (hadronic) cosmic rays are accelerated at
strong shock fronts of supernova remnants. The particles propagate in a
diffusive process through the Galaxy. As origin for the knee a combination of
the maximum energy attained in the acceleration process and leakage from the
Galaxy seems to be favored. 

\subsubsection*{Transition Region}

Different scenarios are discussed in the literature for the transition from
galactic to extragalactic cosmic rays.  The transition most likely occurs at
energies between $10^{17}$ and $10^{18}$~eV.

\begin{figure}[t] \centering
 \includegraphics[width=\breite]{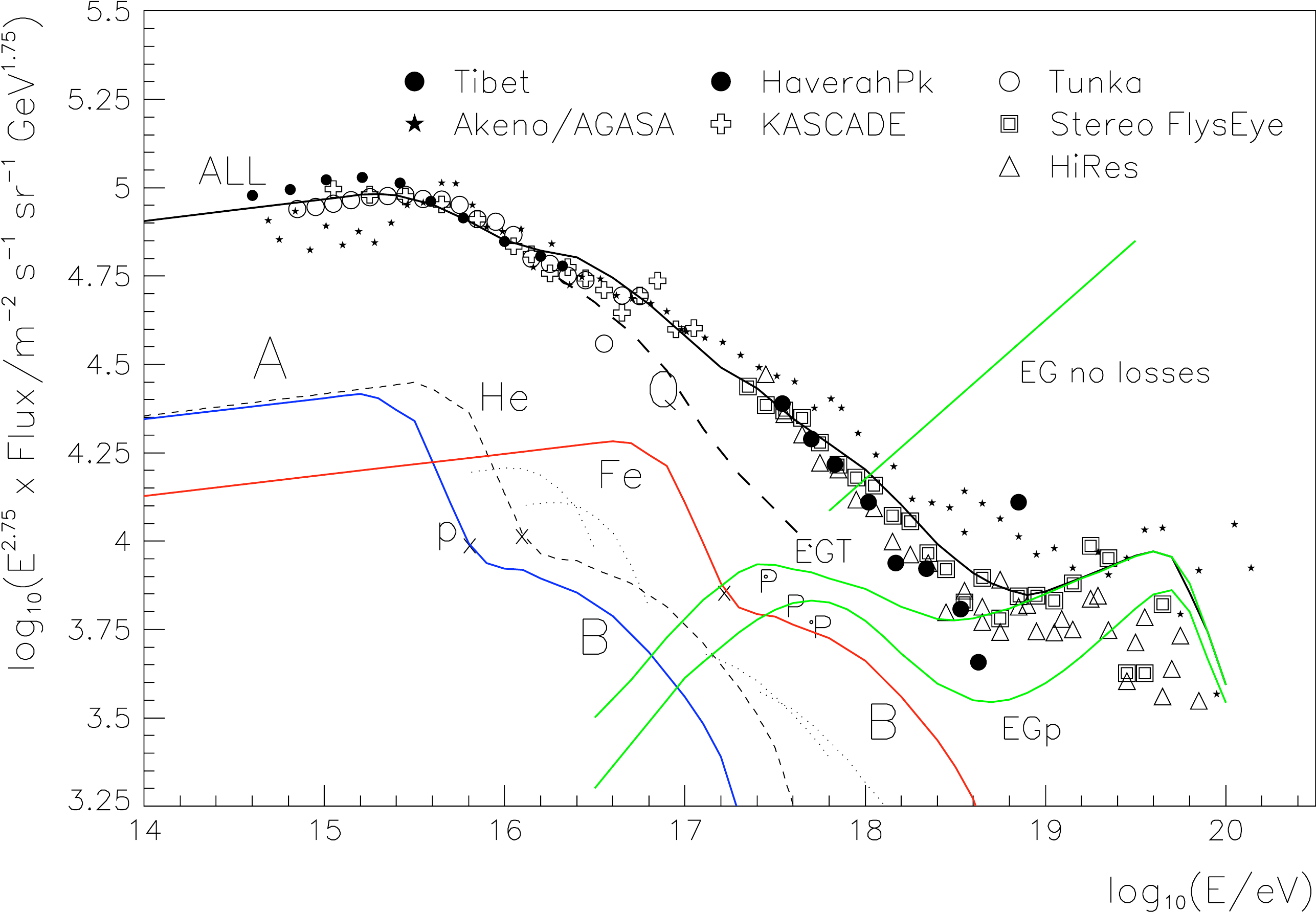}
 \Caption{Breakdown of the cosmic-ray spectrum according to a model of
   Hillas \protect\cite{Hillas:2005cs} as the sum of galactic H, He, CNO,
   Ne-S, and Fe components with the same rigidity dependence, and
   extragalactic H + He having a spectrum $\propto E^{-2.3}$ before
   suffering losses by cosmic microwave background and starlight
   interactions. The galactic components were given a turn-down shape
   based on a KASCADE knee shape as far as the point marked $x$. The
   dashed line $Q$ is the total galactic SNR flux if the extended tail
   (component $B$) of the galactic flux is omitted.
   \protect\cite{Hillas:2005cs}.\label{hillas}}
\end{figure}

The flux for elemental groups of the model of Hillas is shown in \fref{hillas}
\cite{Hillas:2005cs}. The spectra are constructed with rigidity-dependent knee
features at high energies.  Reviewing the properties of cosmic rays accelerated
in SNRs, and using the fluxes as derived by the KASCADE experiment (marked as
component $A$ in \fref{hillas}) Hillas finds that the obtained all-particle
flux (dashed line, marked with $Q$) is not sufficient to explain the observed
all-particle flux, see \fref{hillas} \cite{Hillas:2005cs}.  Hillas proposes a
second (galactic) component to explain the observed flux at energies above
$10^{16}$~eV, marked as component ``B'' in the figure.  An extragalactic
component, marked as $EGT$, dominates the all-particle spectrum above
$10^{19}$~eV, for details see \cite{Hillas:2005cs}.  Very similar is the model
proposed by Wibig and Wolfendale with a transition at higher energies between
$10^{18}$ and $10^{19}$~eV \cite{wibigwolfendale}. In this model the galactic
cosmic-ray flux extends to higher energies.  Thus, a significant contribution
of the extragalactic component is required beyond $10^{18}$~eV only.

\begin{figure}[t] 
 \includegraphics[width=0.51\textwidth]{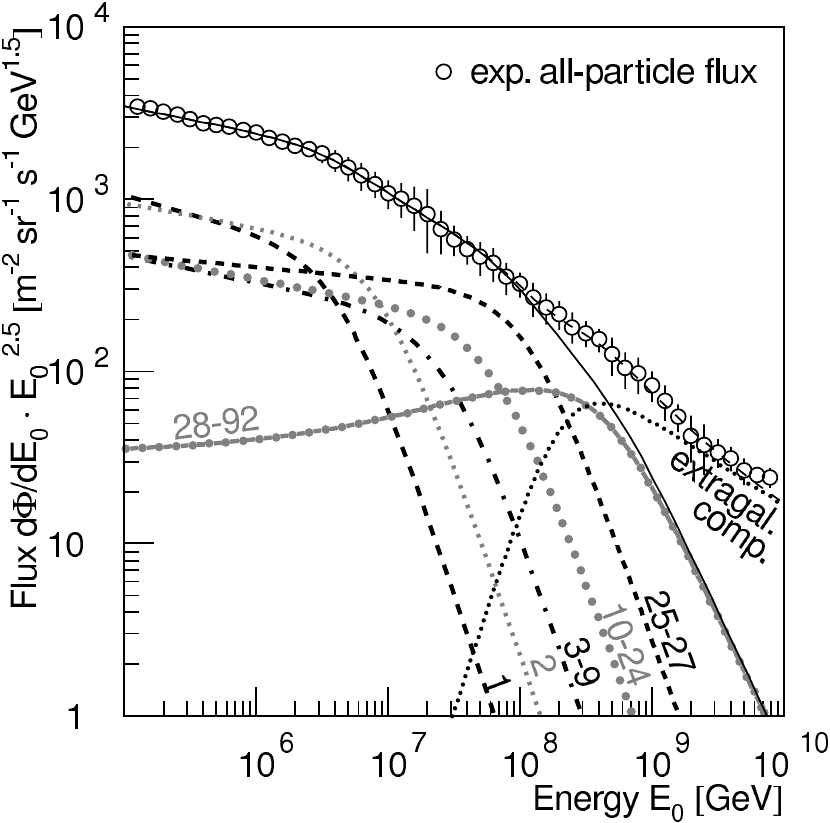}\hspace*{\fill}%
 \includegraphics[width=0.48\textwidth]{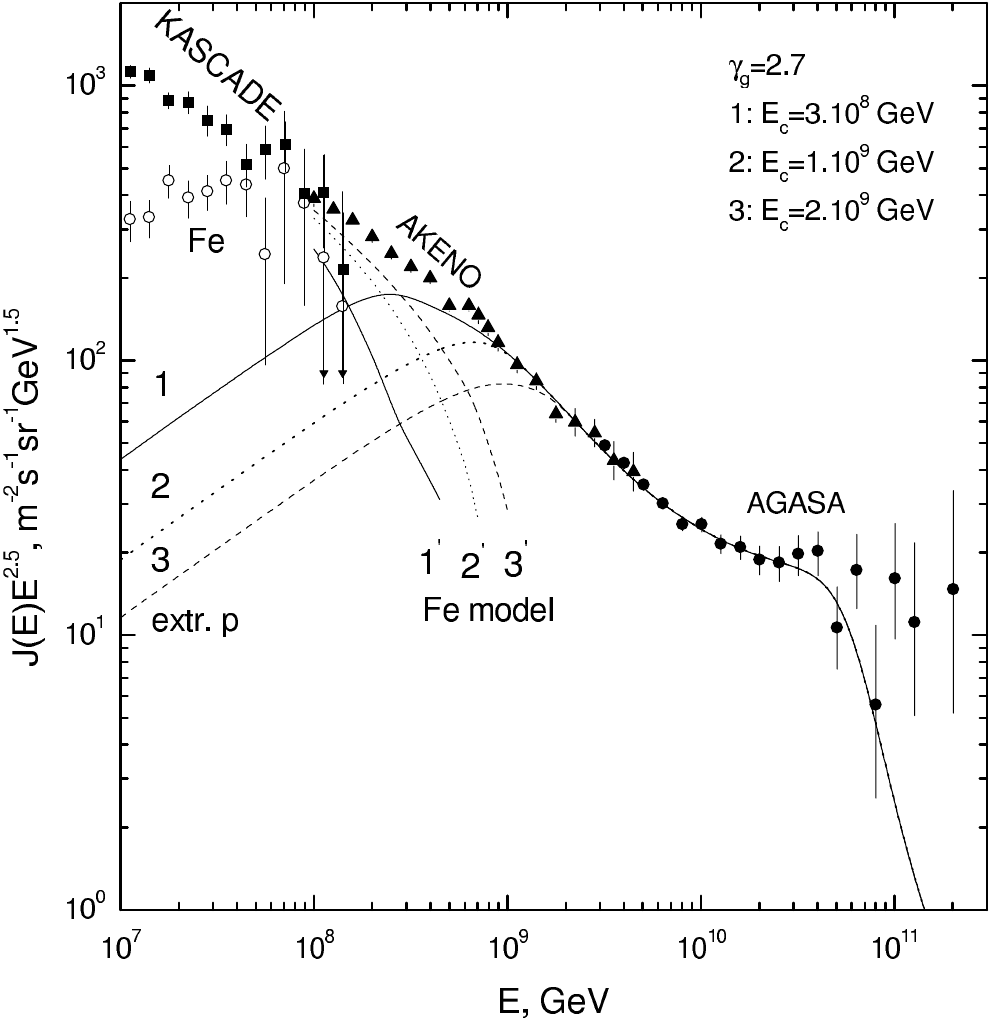}
 \Caption{\LLeft: Cosmic-ray energy spectra according to the poly-gonato model
	  \protect\cite{Hoerandel:2002yg}. The spectra for groups of elements are
          labeled by their
	  respective nuclear charge numbers. The sum of all elements yields the
	  galactic all-particle spectrum (\line) which is compared to the
	  average measured flux. In addition, a hypothetical extragalactic
	  component is shown to account for the observed all-particle
	  flux (\dashed).\label{polygonato}\newline
	  \RRight: Transition from galactic to extragalactic cosmic rays
	  according to Berezinsky \etal \protect\cite{Berezinsky:2004wx}. Calculated
	  spectra of extragalactic protons (curves 1, 2, 3) and of galactic
	  iron nuclei (curves 1', 2', 3') are compared with the all-particle
	  spectrum from the Akeno and AGASA experiments.  KASCADE data are
	  shown as filled squares for the all-particle flux and as open
	  circles for the flux of iron nuclei.}
\end{figure}

Another possibility to match the measured all-particle flux is a significant
contribution of ultra-heavy elements (heavier than iron) to the all-particle
spectrum at energies around $4\times 10^{17}$\,eV \cite{Hoerandel:2002yg,Hoerandel:2004gv},
as illustrated in \fref{polygonato} (\lleft).  The figure shows spectra for
elemental groups with nuclear charge numbers as indicated, derived from direct
and indirect measurements according to the \modell \cite{Hoerandel:2002yg}.
The sum of all elements is shown as solid line and is compared to the average
experimental all-particle flux in the figure.  In this approach the second knee
is caused by the fall-off of the heaviest elements with $Z$ up to 92. It is
remarkable that the second knee occurs at $E_{2nd}\approx92\cdot E_k$, the
latter being the energy of the first knee.  In this scenario a significant
extragalactic contribution is required at energies $E\ga4\times10^{17}$~eV.

In the model of Berezinsky and collaborators
\cite{Berezinsky:2004wx,Berezinsky:2005cq}, the dip in the
all-particle spectrum between $10^{18}$ and $10^{19}$~eV, see
\fref{polygonato} (\rright), is interpreted as a structure caused by
electron-positron pair production on cosmic microwave background
photons $p+\gamma_{3K}\rightarrow p+e^++e^-$.  Assuming a power law
injection spectrum with a spectral index between $\gamma=-2.7$
(without cosmological source evolution) and $-2.4$ (with cosmological
source evolution), the spectrum can be described for $E>
10^{17.5}$\,eV with a proton-dominated composition
\cite{Berezinsky:2004wx}. The shape of the dip is confirmed by data of
the Akeno, AGASA, HiRes, Yakutsk, and Fly's Eye detectors after
energy-rescaling \cite{Berezinsky:2005cq}.  Below a characteristic
energy $E_c\approx1\times10^{18}$~eV the spectrum flattens and the
steeper galactic spectrum becomes dominant at $E<E_c$. The transition
energy $E_{tr}<E_c$ approximately coincides with the position of the
second knee $E_{2nd}$ observed in the all-particle spectrum. The
critical energy $E_c$ is determined by the energy $E_{eq} = 2.3
\times10^{18}$~eV, where adiabatic and pair-production energy losses
are equal. Thus, the position of the second knee is explained in this
scenario by proton energy losses on cosmic microwave background
photons.  The extragalactic component required in the poly-gonato
model is somewhere between scenarios 1 and 2 shown in
Fig.~\ref{polygonato} (\rright).  It should be emphasized that the
pair production mechanism requires the primary particles to be
dominated ($\ga80\%$) by protons \cite{Aloisio:2006wv,Allard:2005cx}.

Traditionally, the ankle is interpreted as the characteristic
signature for the transition between galactic and extragalactic
cosmic rays \cite{Hillas:2005cs,Wibig:2004ye}. In such a scenario,
extragalactic cosmic rays dominate the flux above about
$10^{19}$\,eV. This picture of the transition to extragalactic cosmic
rays is supported by the pioneering observations of the Fly's Eye
experiment that the composition changes at about $10^{18.5}$\,eV
\cite{Bird:1993yi,Gaisser:1993ix}. New observations by HiRes-MIA and
HiRes find a rather sharp transition from a heavy to a light
composition at much lower energy, $E \sim 10^{17.5}$\,eV. It is clear
that the HiRes data are difficult to understand within a model in
which naturally heavy elements should dominate the end of the spectrum
of galactic cosmic rays just below $10^{19}$\,eV.  

If one assumes that extragalactic cosmic rays are accelerated in
processes qualitatively similar to those in our Galaxy then, at
injection, the composition of extragalactic cosmic rays should be
similar to that of cosmic rays at lower energy. Indeed, model
calculations show that a mixed or even predominantly heavy source
composition could, after taking propagation effects into account,
be compatible with existing data \cite{Allard:2005cx,Hooper:2006tn}.

On the other hand, the model by Berezinsky {\it et al.} predicts a
proton-dominated composition at energies as low as $10^{18}$\,eV. One
of the advantages of this model is the natural explanation of the
energy and the shape of the ankle. To obtain a good description of
the ankle, there should not be more than $\sim 20$\% He 
in the extragalactic cosmic-ray flux \cite{Allard:2005cx,Aloisio:2006wv}.
This could be interpreted as indication for either strong magnetic 
fields in the accelerating shock fronts or top-down source scenarios,
which predict proton-dominated fluxes at not too high an energy. 

Understanding the nature of the ankle in the
cosmic-ray spectrum has direct implications on the spectrum at
much higher energy. For example, if the $e^+e^-$ pair production model
is confirmed one can conclude that (i) extragalactic cosmic rays are
mainly protons, (ii) sources are cosmologically distributed, (iii)
there should be a GZK suppression of the flux, (iv) an arrival
direction correlation with local sources can be expected, and (v) the
injection spectrum of extragalactic cosmic rays is rather steep
($dN/dE \sim E^{-\gamma}; \gamma > 2.4$).

Finally it should be noted that the neutrino flux contains 
complementary information for distinguishing between 
different model scenarios for the ankle 
\cite{Seckel:2005cm,Allard:2006mv,Takami:2007pp,*Ahlers:2009rf}.

\subsubsection*{Extragalactic Cosmic Rays} \label{phys-extragal}

Many authors assume extragalactic particles to be nuclei of intermediate to
light mass.  The discrepancy of elemental compositions derived from
mean depth of shower maximum and electron/muon number measurements, however, makes it
impossible to use currently available composition measurements as reliable
criteria to disfavor models (see also discussion in \cite{Watson:2003ba}).
This is most strikingly seen in the prototype HiRes-MIA measurements
\cite{AbuZayyad:1999xa,AbuZayyad:2000ay}. Whereas the mean depth of maximum
data clearly showed the transition to a proton-dominated composition
(QGSJET-based interpretation), the muon density at 600\,m -- of the same
showers -- appeared to correspond to primaries as heavy or even heavier than
iron. A pure iron composition is obviously not compatible with data.  The
rather wide distribution of $X_{\rm max }$ cannot be described with iron
primaries only.

The measurements of the HiRes experiment and the Pierre Auger
Observatory have given evidence for a suppression of the flux at energies
exceeding $4\cdot10^{19}$~eV \cite{Abbasi:2007sv,Abraham:2008ru}.  The
question arises whether this steepening is due to the GZK effect, 
due to the maximum energy achieved during the acceleration processes,
or due to both, a GZK energy loss process and an upper energy limit in
the sources.

\begin{figure}[t]
\includegraphics[width=0.49\textwidth]{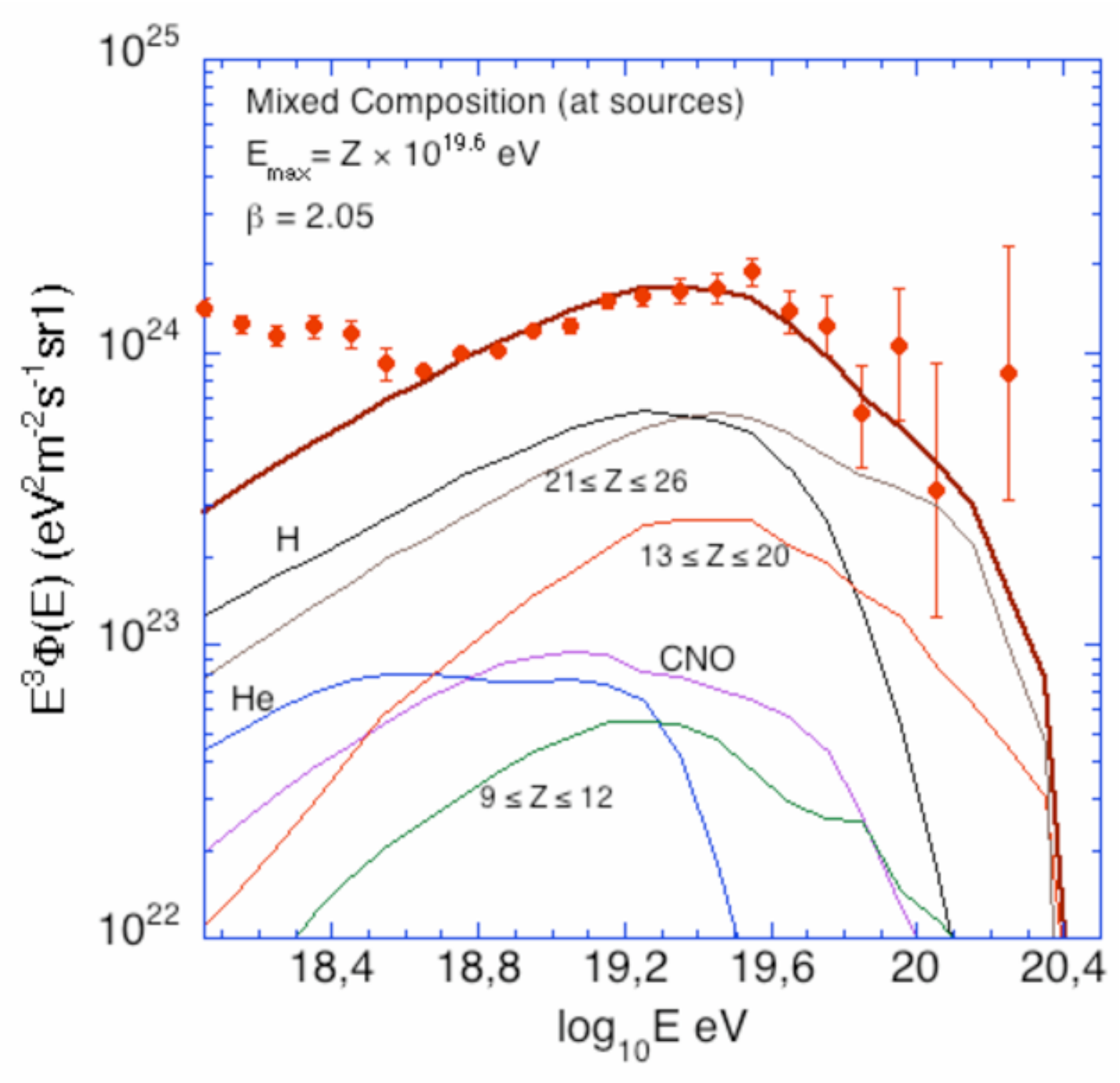}\hspace*{\fill}
\includegraphics[width=0.49\textwidth]{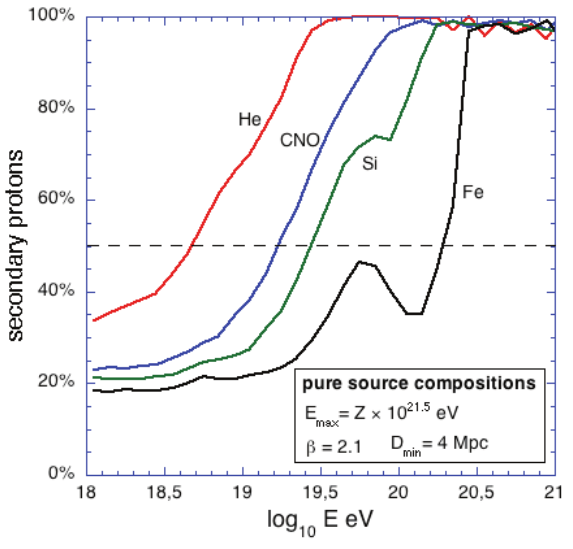}
\Caption{Results of a propagation model of highest energy cosmic rays by Allard
         et al.\ \protect\cite{Allard:2008gj}.
         \LLeft: Auger data compared to propagated spectra obtained, assuming a
	 mixed source composition. The contributions of spectra for groups of
         elements to the all-particle spectrum are shown.
	 \RRight: Expected relative abundance of secondary protons as function
	 of energy for different pure source composition hypotheses, assuming a
         spectral index of $-2.1$ at the sources.
         \label{allard}}
\end{figure}

The support for the existence of the GZK effect is provided by the correlation
of the arrival directions with AGN, which imply a strong anisotropy of the
arrival direction distribution. The anisotropy appears sharply at an energy of 57~EeV.
At this energy, the flux measured by the Pierre Auger Observatory is about 50\%
lower than expected from a power law extrapolation from lower energies. Thus,
there seems to be a connection between the steepening in the spectrum and the
AGN correlation. However, if the observed suppression is due to the GZK effect
one would expect either a light or rather heavy elemental composition above the
GZK threshold. Intermediate mass nuclei are expected to break up very rapidly
in interactions with the photons of the 3-K microwave background.  The relative
abundance of secondary protons produced during propagation according to a
recent model \cite{Allard:2008gj} is displayed in \fref{allard}~(\rright).
Observations of the average depth of the shower maximum at the highest energies
indicate a mixed composition, see e.g.\ \fref{xmax}.

The expected flux at Earth according to a propagation model of highest energy
cosmic rays by Allard et al.\ \cite{Allard:2008gj} is presented in
\fref{allard} (\lleft). In this article different scenarios for the properties
of the sources are discussed, like different elemental compositions and
different spectral indices for the energy spectra at the sources. \Fref{allard}
shows the result assuming a mixed elemental composition at the sources. The
contributions of individual elemental groups to the all-particle spectrum are
shown. The shape of the energy spectrum measured by the Pierre Auger
Observatory is well represented by the all-particle spectrum derived from the
model.

Another important source of information are fluxes of particles produced during
the propagation of cosmic rays in the Universe.  
\footnote{In the literature such particles are frequently called "secondary
particles". They should not be confused with secondary particles generated in
the Earth's atmosphere.}
Fluxes of neutrinos and
gamma-rays due to the interaction of ultra high-energy cosmic rays with the
background photon radiation provide complementary information for
discriminating models of UHECRs 
\cite{Semikoz:2003wv,Seckel:2005cm,Hooper:2004jc,Anchordoqui:2007fi}.  In
\fref{neutrinoflux} measured upper limits for the neutrino flux have been
compared to a top-down model \cite{kalashev}. The latest Auger data disfavor
this model.

Photons of ultra high energy are easier to detect but their energy loss
distance is very short. Nevertheless the measurement of the photon fraction in
the primary cosmic-ray flux is one of the most promising methods of
distinguishing between different source scenarios of extragalactic cosmic rays
\cite{Bhattacharjee:1998qc,Kachelriess:2004ax}.  Recent experimental upper
limits strongly disfavor predictions of top-down models
\cite{Aloisio:2003xj,Ellis:2005jc}, see \fref{gammaflux} (\rright).
Due to the down-cascading of photons in the extragalactic radiation background,
fluxes of GeV photons are also a complementary source of information.  For
example, neutrino and photon fluxes of a Z-burst model are discussed in
Ref.~\cite{Semikoz:2003wv}. Both, the neutrino flux limit by the FORTE satellite
\cite{Lehtinen:2003xv} and the EGRET diffuse extragalactic photon flux analysis
\cite{Strong:2004ry} severely constrain this model. Recently, also even more
stringent limits were set by the ANITA experiment \cite{Barwick:2005hn}, ruling out
most of the parameter space of the Z-burst model.
In summary, top-down models are disfavored by recent experimental results of
various experiments.


One of the key questions in the field of high-energy astroparticle
physics is the understanding of the observed
anisotropy above the GZK energy threshold. 

The angular scale of about $3^\circ$ would favour protons if the AGNs
within this correlation angle are indeed the sources of UHECRs
\cite{Abraham:2007si}. This would be at variance with the Auger data
on the mean depths of shower maximum. Most likely, the AGN correlation
has to be interpreted merely as a signature of anisotropy and
correlation with the nearby matter distribution
\cite{Stanev:1995my}. First of all, the rate of misidentification of a
potential source along the line of sight of the arrival direction of a
cosmic ray is very large. Secondly, the AGNs within the correlation
window are found to be often less powerful Seyfert 2 galaxies
\cite{Zaw:2008is}. Thirdly, the AGNs in our cosmological neighborhood
are strongly clustered, making it difficult to distinguish between a
true AGN correlation and a correlation with larger scale structures.

Instead of assuming a single source in each of the directions of the
measured UHECRs, one can assign most of the arrival directions to
about three sources or source regions
\cite{Wibig:2007pf}. This model scenario would require rather weak galactic
magnetic fields and particles in the mass range up to carbon.

In all anisotropy studies even at the highest energies, the galactic magnetic
fields play a central role \cite{Kachelriess:2006ip}. Knowing their
structure to sufficient detail would, for example, allow the
determination of the charge of UHECRs.  Other interesting applications
are the search for the cosmological Compton-Getting effect, a 0.6\%
dipole anisotropy that is expected for cosmological sources
\cite{Kachelriess:2006aq}.  

A much higher number of UHECRs has to
be collected for detailed anisotropy studies of the required
sensitivity. If UHECRs are confirmed to be protons, the data can be
used for proton astronomy including studies of energy spectra of
individual sources and magnetic field spectroscopy, e.g.~\cite{Stanev:1996qj,*Tinyakov:2001ir,*Alvarez-Muniz:2001vf ,Kachelriess:2005qm}. 
This will, of course, require very large-aperture
installations as the relevant energy range is just in the GZK
suppression region.

\Section{Importance of Modeling Hadronic Interactions\label{accelsec}}

There are strong indications for shortcomings in the shower simulations,
probably due to limitations of modeling hadronic interactions.

Detailed studies of the shower development in the atmosphere have been
performed with the KASCADE multi-detector set-up and interaction models have
been improved
\cite{wwtestjpg,kascadelateral,rissejpg,Haungs:2003tz,%
*kascadeabslength,*hadrisvhecricern,*annaprd,*Apel:2007zz,*Apel:2009sv}.
A particularly valuable tool to test high-energy interaction models are
correlations between different shower components \cite{jenskrakow,jenspune}.
Some years ago several models like SIBYLL\,1.6, DPMJET\,2.5, or NEXUS\,2 failed
to describe the measurements of particular correlations. On the other hand, for
modern models like QGSJET\,01, SIBYLL\,2.1, or DPMJET~2.55, the KASCADE
measurements are compatible with predictions for various correlations between
the electromagnetic, muonic, and hadronic components, i.e.\ the measurements
are bracketed by the extreme assumptions of primary protons and iron nuclei
\cite{jenskrakow,jenspune}.  In previous analyses  pure proton or iron
compositions have been assumed as extreme cases.  However, at present, more
detailed analyses are performed \cite{jenspune,jrhtorun}. They take into
account the spectra for elemental groups as obtained from investigations of the
electromagnetic and muonic components (as discussed above, see
\fref{ulrichspek}).  These investigations reveal deviations between
measurements and simulations for the hadronic component of the order of 10\% to
20\% \cite{jenspune}.

The situation is similar at very high energy. For example, the mean $X_{\rm
max}$ of the HiRes-MIA data is not consistent with the measured muon densities
of the same events \cite{AbuZayyad:1999xa}. The
conclusions on mass composition from Haverah Park data are different if the time structure of the
shower front is used instead of the muon yield that determines the rate of
inclined showers. The Auger data indicate that the energy scale derived from
surface detector simulation
seems to be of the order of 25\% higher than that obtained from fluorescence
measurements \cite{Engel:2007cm,*Schmidt:2009ge}.

\begin{figure}[t]
\includegraphics[width=0.49\textwidth]{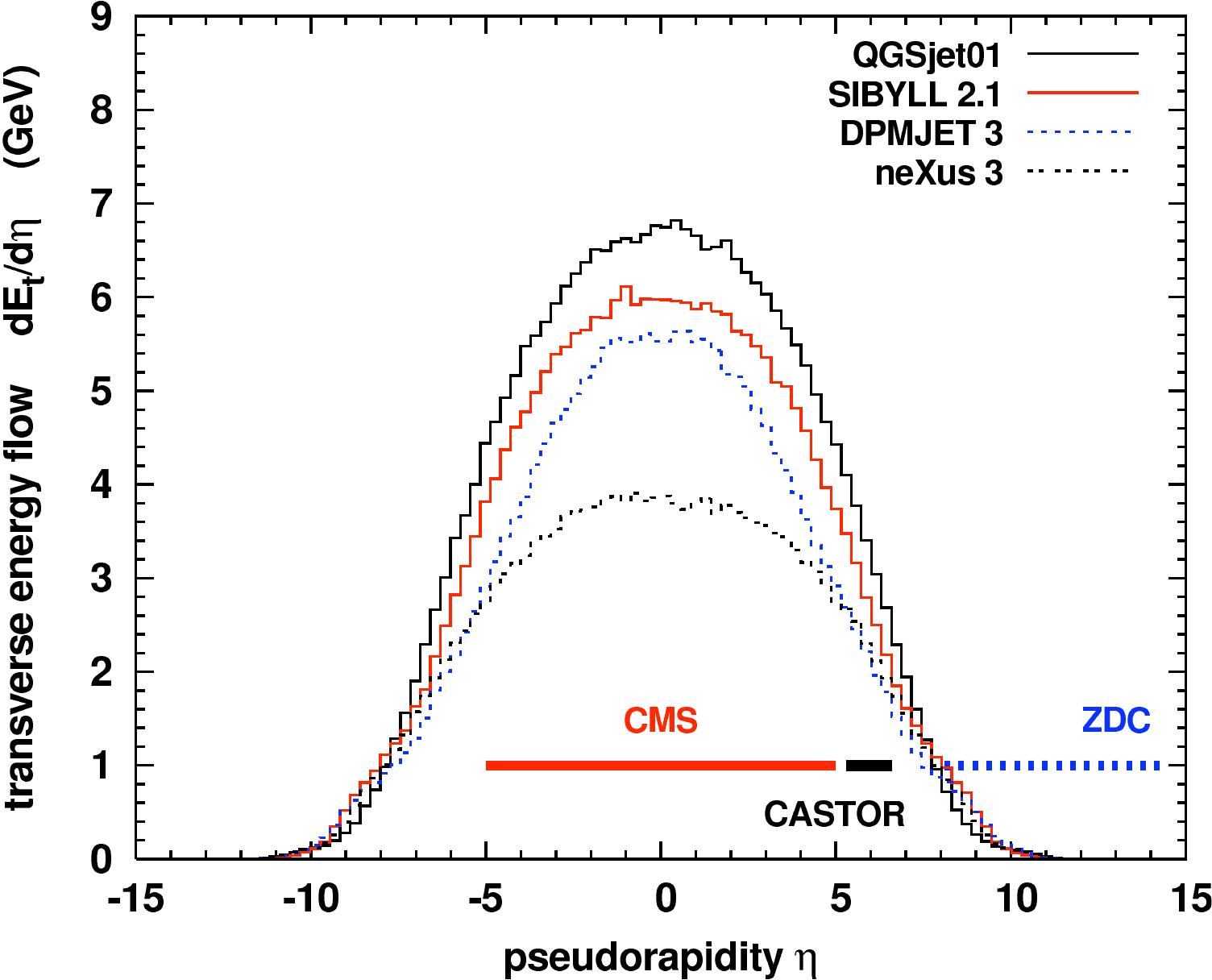}
  \hspace*{\fill}
\includegraphics[width=0.49\textwidth]{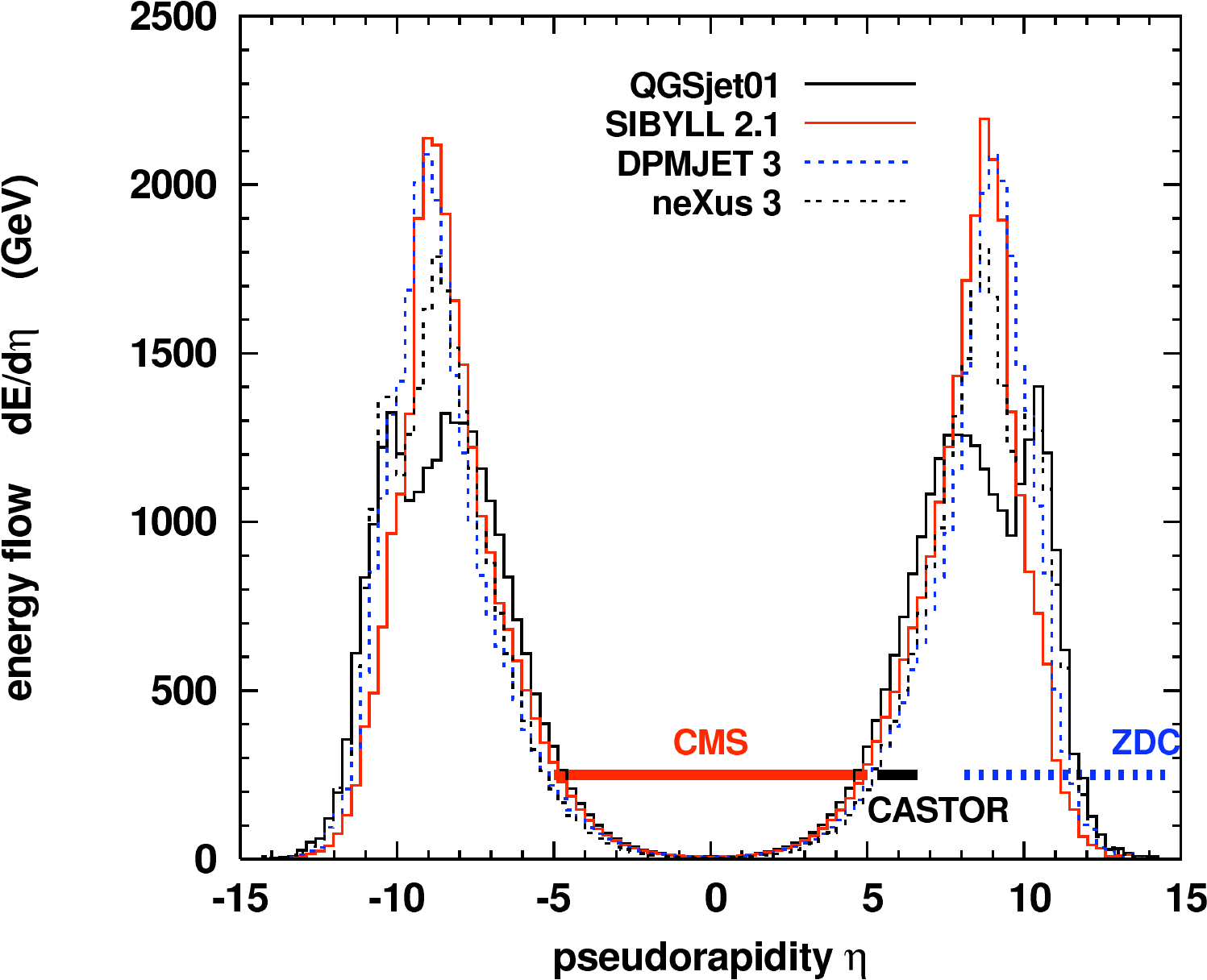}
\Caption{ Prediction of the transverse (\lleft) and total (\rright) energy
  flow produced in proton-proton collisions at LHC as obtained with
  cosmic-ray interaction models. Also shown are the acceptance ranges
  of the CMS central calorimeter, the CASTOR calorimeter and the zero
  degree calorimeter for neutral particles
  (ZDC).  \label{fig:ModelsLHC}
}
\end{figure}

In the foreseeable future soft multi-particle production will not be
calculable within QCD. Therefore the modeling of cosmic-ray
interactions will continue to depend strongly on the input from
accelerator experiments that is needed to tune phenomenological
particle production models. Measurements at both, fixed target and
collider experiments can substantially contribute to reducing the
uncertainties of the models and, hence, determine the composition of
cosmic rays. 

High-energy interactions are very important for the overall shower
profile but hadronic multiparticle production is least known in this
energy range. This is illustrated in Fig.~\ref{fig:ModelsLHC} in which
predictions of hadronic interaction models are shown for proton-proton
collisions at $\sqrt{s}=14$\,TeV. The acceptance ranges for different
detector components of the CERN CMS \cite{cms} detector for the 
LHC \cite{lhc} are also shown.

Every ultra high-energy air shower contains many sub-showers of lower
energy.  For example, the slope of the lateral
distribution of particles in a shower is a measure of the mass of the
primary particle.  Even for high-energy showers of $10^{19}$\,eV, this
slope is very sensitive to assumptions on hadronic multiparticle
production in the energy range of ten to a few hundred GeV \cite{Hillas:1997tf,*Engel:1999vy,*Heck:2003br,*Drescher:2002vp,*Drescher:2003gh}.
The energy distribution of
hadronic interactions, in which at least one meson was produced that
in turn decayed to a muon that reached sea level, has a maximum in the
range between 80 and 200 GeV.  Most of the interactions are induced by
pions (70\%) and nucleons (20\%). 

In addition to accelerator measurements of hadronic
multiparticle production \cite{NEEDS-web,*Engel:2002id}, measurements
and understanding of air shower data at lower energy are very
important to tune and validate the used hadronic interaction models
\cite{Engel:2003ac}.


\Section{Conclusions and Outlook\label{concsec}}

The all-particle flux of cosmic rays is reasonably well known up to the highest
energies.  Recent measurements by the HiRes and Auger Collaborations
established a GZK-like suppression of the flux at energies exceeding
$6\cdot10^{19}$~eV.

In the knee region the mean mass of cosmic rays is found to increase as
function of energy. The knee is caused by sequential breaks in the spectra of
individual elements, starting with the light elements.  At present, a rigidity
dependence of the cut-off energies for the individual elements is likely but
not yet clear beyond doubt.  Above $10^{17}$~eV the situation becomes very
uncertain.  Almost no data are available in the important energy range where
the galactic cosmic-ray component is expected to end ($10^{17} - 10^{19}$~eV),
at present there are only limited experimental efforts in this region.  At the
highest energies ($\ga10^{19}$~eV), several experiments indicate a light to
mixed composition with a strong dependence on the model used to describe
high-energy interactions in the atmosphere.

Large scale anisotropies have been found at low energies, being
compatible with the movement of the Earth around the Sun (Compton Getting
effect). In the knee region, the anisotropies disappear, indicating that the
rest frame of galactic cosmic rays co-rotates with the Galaxy. At ultra-high
energies anisotropy measurements provide independent information on the
composition due to deflections in the galactic magnetic fields.
The results of the Pierre Auger Observatory indicate a correlation between the
arrival direction of cosmic rays and the super galactic plane.

The energy of the transition from galactic to extragalactic particles is
discussed controversially.  The scenarios considered limit the transition energy
to the range between $10^{17}$ and $10^{19}$~eV.  Precise composition
measurements in this region will be decisive.

The knowledge about the cosmic-ray composition is presently limited by the
uncertainties in the hadronic interaction models used to describe the air
shower development. The ambiguities can not be resolved by cosmic-ray
measurements solely. Collaboration with experiments at the LHC and fixed target
experiments is mandatory to improve the understanding of multiparticle
production in the (extreme) forward region, thus providing reliable simulation
codes.

Independent and complementary information about the origin of high-energy
cosmic rays can be obtained by measurements of high-energy gamma rays and
neutrinos (multi messenger approach). Recent measurements of TeV gamma rays
from supernova remnants give strong hints for this object class as sources of
galactic cosmic rays. Measurements of neutrino and photon fluxes are very
useful to distinguish between different scenarios for the transition from
galactic to extragalactic cosmic rays. They are important to establish a GZK
feature beyond doubt and provide composition information at the highest
energies.

In the near future new cosmic-ray detectors will provide additional information
on the transition from galactic to extragalactic cosmic rays (KASCADE-Grande
\cite{chiavassapune}, IceTop/IceCube \cite{icecube}, extensions of
Auger-South \cite{Klages:2007x1,Medina:2006my}, extension of the Telescope Array
\cite{Jui05a}), the anisotropy of the arrival directions and composition of cosmic rays at the
highest energies on the whole sky (Auger North
\cite{Nitz:2007ur}) as well as super GZK particles (EUSO \cite{Takahashi:1900zz,*euso}, a
recovery of the cosmic-ray flux beyond the GZK resonance).  A promising new
detection technique, the measurement of radio emission from air showers, is
presently developed in the Pierre Auger Observatory and the LOFAR project. This
technique is expected to allow efficient cosmic-ray measurements in future
large-aperture experiments.

In foreseeable future no significant improvement of hadronic interaction models
is expected. Therefore, the range of direct measurements should be extended to
energies approaching the knee. Ideal would be an ACCESS-type \cite{access}
experiment in outer space with exposure times of several years.  More
experimental efforts are needed in the energy region where the galactic
component is expected to end.  At ultra-high energies the anisotropy studies
are limited by statistics only.  Therefore, large aperture experiments are
needed with full-sky coverage such as the Pierre Auger Project (with
observatories in the southern and northern hemisphere).


\section*{Acknowledgements}

We thank our colleagues from the Pierre Auger and KASCADE-Grande
Collaborations for many fruitful and stimulating discussions and comments on the paper. In particular, we thank Carola Dobrigkeit, Maria Giller and Jim Matthews for reading the manuscript of this article.
We are grateful to D.\ Heck and T.\ Pierog for providing numerical results of air
shower simulations and
S.\ Knurenko,
K.\ Shinozaki,
P.\ Sokolsky,
M.\ Takeda, 
and
G.\ Thomson
for making available data tables of experimental results.
This work has been supported in part by BMBF grant 05A08VK1.


{\small
\singlespacing

}

\end{document}